\documentclass{article}
\usepackage{amsmath}
\usepackage{amsthm}
\usepackage[margin=\parindent]{caption}
\usepackage{dsfont}
\usepackage{enumitem}
\usepackage[T1]{fontenc}
\usepackage[osf]{sourcesanspro}
\usepackage[bitstream-charter,greekuppercase=upright,cal=cmcal]{mathdesign}
\usepackage{mathtools}
\usepackage{subcaption}
\usepackage{xcolor}
\usepackage{placeins}
\usepackage{pgfplots}
\pgfplotsset{compat=1.18}

\usepackage{algpseudocode}
\usepackage{algorithm}
\usepackage{tikz}
\usetikzlibrary{backgrounds,fit,positioning,calc,decorations.markings,decorations.pathreplacing,calligraphy}
\usepackage[paper=a4paper,margin=3cm]{geometry}

\usepackage{sectsty}
\allsectionsfont{\fontseries{sb}\sffamily}

\usepackage[pdfusetitle]{hyperref}
\usepackage[affil-it]{authblk}
\definecolor{dullmagenta}{rgb}{0.5,0,0.4}
\definecolor{dullblue}{rgb}{0.0,0,0.86}
\hypersetup{colorlinks=true,citecolor=dullblue,linkcolor=dullblue,filecolor=dullblue,urlcolor=dullblue,breaklinks=true,bookmarksnumbered=true, bookmarksopen=true,bookmarksopenlevel=1}
\usepackage{bookmark}
\usepackage[capitalize]{cleveref}
\crefname{equation}{}{}
\AddToHook{cmd/appendix/before}{\crefalias{section}{appendix}}
\usepackage[backend=biber,sorting=none,style=phys-jmr,biblabel=brackets,backref=true,bibencoding=utf8,maxnames=20,eprint=true]{biblatex}
\makeatletter
\pretocmd{\blx@head@bibintoc}{\phantomsection}{}{\ddt}
\makeatother
\DefineBibliographyStrings{english}{%
  backrefpage = {page},% originally "cited on page"
  backrefpages = {pages},% originally "cited on pages"
  }
\newtheorem{theorem}{Theorem}[section]
\newtheorem{lemma}{Lemma}[section]
\newtheorem{proposition}{Proposition}[section]
\newtheorem{corollary}{Corollary}[section]

\newtheorem{definition}{Definition}[section]
\newtheorem{problem}{Problem}[section]
\newtheorem{example}{Example}[section]
\newtheorem{procedure}{Procedure}[section]

\newlist{problemdescr}{description}{1} % 1 = max depth
\setlist[problemdescr,1]{font=\textendash\enskip\scshape\bfseries,noitemsep,topsep=0pt}

\tikzset{
  fgfac/.style = {
    shape=rectangle, 
    semithick, 
    fill=white,
    text=black, 
    rounded corners=0.5pt, 
    draw=black, 
    inner sep=0.5mm, 
    outer sep=0.0mm,
    minimum height=4mm, 
    minimum width=4mm,
    text height=1.5ex,
    text depth=.25ex
  },
  fgvar/.style = {
    shape=circle, 
    semithick, 
    fill=white,
    text=black, 
    draw=black,
    inner sep=0.5mm, 
    outer sep=0.0mm,
    minimum height=4mm, 
    minimum width=4mm,
    text height=1.5ex,
    text depth=.25ex
  },
  box/.style = {
    shape=rectangle, 
    semithick, 
    text=black, 
    rounded corners=0.5pt, 
    draw=black, 
    fill=black!5, 
    inner sep=1.5mm, 
    outer sep=0.0mm,
    minimum height=8mm, 
    minimum width=8mm,
    text height=1.5ex,
    text depth=.25ex
  },
  edge/.style = {
    semithick,
    -stealth
  }
 }

\DeclareFontFamily{U}{bbold}{}
\DeclareFontShape{U}{bbold}{m}{n}
 {
  <-5.5> s*[1.069] bbold5
  <5.5-6.5> s*[1.069] bbold6
  <6.5-7.5> s*[1.069] bbold7
  <7.5-8.5> s*[1.069] bbold8
  <8.5-9.5> s*[1.069] bbold9
  <9.5-11> s*[1.069] bbold12 %was 10
  <11-15> s*[1.069] bbold12
  <15-> s*[1.069] bbold17
 }{}

\DeclareRobustCommand{\id}{%
  \text{\usefont{U}{bbold}{m}{n}1}%
}

\newcommand{\cat}{\Vert}

\newcommand{\abs}[1]{\left\lvert#1\right\rvert}
\newcommand{\norm}[1]{\left\lVert #1 \right\rVert}
\providecommand{\ket}[1]{\left\lvert#1\right\rangle}
\providecommand{\bra}[1]{\left\langle#1\right\rvert}
\providecommand{\braket}[2]{\left\langle\smash{#1}\middle\vert\smash{#2}\right\rangle}
\providecommand{\braopket}[3]{\left\langle\smash{#1}\middle\vert\smash{#2}\middle\vert\smash{#3}\right\rangle}
\providecommand{\ketbra}[2]{\left\lvert \smash{#1}\middle\rangle\!\middle\langle\smash{#2}\right\rvert}
\providecommand{\proj}[1]{\ketbra{#1}{#1}}
\DeclareMathOperator{\tr}{Tr}

\def\ff{\mathbb F}
\def\spsc{\ensuremath{\textsc{spsc}}}
\def\bspsc{\ensuremath{\textsc{bspsc}}}
\def\bsc{\ensuremath{\textsc{bsc}}}
\def\mensemble{\ensuremath{\mathcal{E}}}
\def\sdt{\ensuremath{\textsc{sdt}}}
\def\pno{\ensuremath{\textsc{pno}}}

\def\mensemble{\ensuremath{\mathcal{E}}}
\def\blockdec{\ensuremath{\textsc{blockdec}}}
\def\bitdec{\ensuremath{\textsc{bitdec}}}

\def\plusbox{\begin{tikzpicture}\node[fgfac] {$+$};\end{tikzpicture}}
\def\eqbox{\begin{tikzpicture}\node[fgfac] {$=$};\end{tikzpicture}}

\title{\fontseries{m} \LARGE\sffamily  Efficient and optimal quantum state discrimination via quantum belief propagation}
\author[1]{\sffamily Christophe Piveteau}
\author[2]{\sffamily Joseph M.\ Renes}
\affil[1]{\small Inria, France}
\affil[2]{\small Institute for Theoretical Physics, ETH Z\"urich, Switzerland}
\date{}

\begin{document}
\maketitle

\renewcommand{\abstractname}{\vspace{-2.5\baselineskip}} % https://tex.stackexchange.com/a/53175
\begin{abstract}
\noindent 
We present an efficient quantum algorithm for a structured state discrimination problem we call the \emph{subspace decoding task}.
Building on this, we show that the algorithm enables efficient and optimal decoding of certain families of structured classical linear codes transmitted over binary-input classical–quantum pure-state channels.
Such decoders can substantially enhance the performance of quantum algorithms based on Regev’s reduction, such as decoded quantum interferometry.
In particular, we obtain optimal and efficient quantum decoders for all classical codes with efficient trellis representations.
As an application, we design a quantum decoder for turbo codes and, through density evolution, demonstrate decoding thresholds that surpass the Shannon bound and closely approach the Holevo bound.
\end{abstract}
\vspace{3mm}

\section{Introduction}
Quantum state discrimination is a fundamental task in quantum information theory.
It arises in numerous settings, including quantum metrology and learning, quantum communication, and quantum computation.
In some of these applications, the involved families of states are structured in some way, which makes it possible to accomplish the discrimination task with an efficient quantum algorithm.
Understanding when and how such structured state discrimination can be carried out efficiently is of both foundational and practical importance.

In this work, we study a family of state discrimination problems that originate from coding theory.
Specifically, consider an $[n,k]$ binary linear code $\mathcal{C}$ with generator matrix $G\in\ff_2^{k\times n}$.
A uniformly random source string $u\in\ff_2^k$ is encoded as $u^TG$ and then transmitted over a collection of binary-input pure state classical-quantum channels $W_{\omega}^{\otimes n}$ defined as 
\begin{equation}
    W_{\omega}(x) \coloneq \sqrt{1-\omega}\ket{x} + \sqrt{\omega}\ket{1-x}
    \quad\text{for }x\in\ff_2, \omega\in[0,1/2] \, .
\end{equation}
The discrimination task, which we refer to as the \emph{subspace decoding task}, is to estimate a given substring of the input $u$ by measuring the channel outputs. 
Here we develop a new quantum algorithm to efficiently solve subspace decoding problems for certain structured codes and substrings of small size.

When the goal is to estimate the entire bit string $u$, subspace decoding becomes the decoding problem for $\mathcal{C}$ over $W_{\omega}$, sometimes referred to as the \emph{quantum decoding problem}~\cite{chailloux_quantumdecoding_2023}.
We then show how our algorithm can be applied to efficiently and optimally solve the quantum decoding problem for codes that exhibit a certain tree-like structure, such as an efficient trellis representation.
Importantly, this allows us to optimally solve the quantum decoding task on convolutional codes.
Our decoder can be seen as a quantum analogue of both the Viterbi~\cite{forney_viterbi_1973} and BCJR~\cite{bahl_bcjr_1974} algorithms, the standard decoders in classical trellis-based coding theory.
Finally, we utilize our optimal convolutional decoder to construct a quantum decoder for turbo codes.
Using density evolution simulations, we show that this decoder exhibits a threshold which closely approaches the channel capacity, given by the Holevo bound.

The recent development of quantum algorithms based on the technique of \emph{Regev's reduction}~\cite{thomas_reduction_2024,chen_quantumalgo_2022,yamakawa_quantumadvantage_2024,jordan_dqi_2025,chailloux_quantumadvanteage_2025,briaud_quantumadvantage_2025} has kindled interest in finding efficient algorithms to solve the quantum decoding problem.
As a prominent example, the quantum algorithm called \emph{decoded quantum interferometry} (DQI) was recently proposed as a potential candidate for a quantum advantage in solving certain combinatorial optimization problems~\cite{jordan_dqi_2025}.
The central step in DQI consists of solving a specific classical decoding task in coherent superposition.
However, it has been noted that algorithms based on Regev's reduction, including DQI, can be significantly improved by utilizing a \emph{quantum} decoder (i.e. by solving the quantum decoding problem), instead of running a classical decoder in superposition~\cite{chen_quantumalgo_2022,chailloux_quantumdecoding_2023,jordan_dqi_2025}.
The quality of DQI's solution depends directly on the performance of the underlying decoder --- if the decoder can tolerate more noise, then DQI will produce higher quality solutions. 
This motivates the search for efficient quantum decoders tailored to structured codes.

There are few previous results on efficiently solving the quantum decoding problem. 
To our knowledge, the only known algorithm that genuinely exploits the quantum superposition of the error configurations is \emph{belief propagation with quantum messages} (BPQM) introduced in~\cite{renes_belief_2017} and subsequently improved and extended in~\cite{rengaswamy_bpqm_2021,piveteau_bpqm_2022,brandsen_bpqm_2022,delaney_demonstration_2022,mandal_bpqmpolar_2023}.
One can recover BPQM as a special case of our algorithm.
As such, we consider our algorithm to be a generalization of BPQM, and we will also refer to it using the same name.
Unlike the original algorithm, which was fundamentally limited to codes with a tree (or locally tree-like) Tanner graph, our generalized BPQM removes this limitation.
While our improvements immediately impact convolutional and turbo codes, they also lay the groundwork for efficiently tackling quantum decoding for a wider range of structured codes.\footnote{For example, the discussion in \Cref{sec:mpg_oneloop} should allow us to handle codes whith a Tanner graph that has only few cycles locally.}
Expanding the set of codes that can be decoded efficiently and accurately increases the potential for achieving practical quantum advantages using algorithms based on Regev's reduction.

\subsection{Background and motivation}
\paragraph{CQ channel coding with pure-state channels}
Consider the task of achieving reliable communication of classical information over the binary-input pure state CQ channel $W_{\omega}$.
Assume that the $k$-bit data is encoded into $n$ bits using an $[n,k]$ binary linear code $\mathcal{C}$.
If the sender transmits the codeword $x\in\mathcal{C}$, the receiver obtains an $n$-qubit system in the state $\ket{\Psi_x}\coloneqq W_{\omega}^{\otimes n}(x)$.
The quantum decoding problem consists of estimating $x$ given this received quantum system.
It can be understood as the problem of discriminating the family of $2^k$ pure states $\{\ket{\Psi_x}|x\in\mathcal{C}\}$.

\begin{figure}
    \centering
    \includegraphics{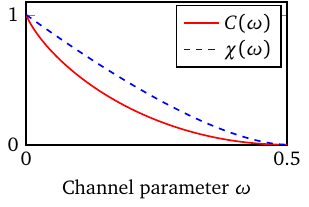}
    \caption{Comparison of the Shannon capacity $C(\omega)$ of the BSC and the Holevo capacity $\chi(\omega)$ of the binary-input pure state CQ channel $W_{\omega}$.}
    \label{fig:capacities}
\end{figure}

A simple strategy to solve the quantum decoding problem is to measure each received qubit in the computational basis, which optimally estimates whether the input bit was more likely to be $0$ or $1$.
This effectively turns the pure-state channel $W_{\omega}$ into a binary symmetric channel (BSC) $\bsc_{\omega}$ with error probability $\omega$, and the decoding can henceforth be achieved with a classical decoder designed for the BSC.
This strategy is fundamentally limited by the Shannon capacity $C(\omega)=1-h(\omega)$ of the induced BSC, where $h$ is the binary entropy function.
However, the celebrated Holevo-Schumacher-Westmoreland theorem states that the maximal possible rate at which classical information can be transmitted over $W_{\omega}$ is not given by $C(\omega)$, but rather by the Holevo capacity $\chi(\omega)=h(\frac{1}{2}-\sqrt{\omega(1-\omega)})$.
The difference between the Shannon capacity of $\bsc_{\omega}$ and the Holevo capacity of $W_{\omega}$ as a function of $\omega$ is depicted in \Cref{fig:capacities}.
Note that the ratio between the capacities $\chi(\omega)/C(\omega)$ diverges as $\omega$ approaches $1/2$.

In order to realize coding schemes which approach the Holevo capacity, it is therefore crucial to design truly quantum decoders which exploit the quantum superposition in the channel output.
In this paper, we will explicitly construct such decoders for various families of codes.

\paragraph{Regev's reduction with quantum decoders}
Regev's reduction is a foundational result in lattice-based cryptography, showing that the \emph{learning with errors} problem is as hard to solve as certain worst-case lattice problems~\cite{regev_reduction_2009}.
In recent years, the reduction has gained renewed attention as a technique to design new quantum algorithms~\cite{thomas_reduction_2024,chen_quantumalgo_2022,yamakawa_quantumadvantage_2024,jordan_dqi_2025,chailloux_quantumadvanteage_2025,briaud_quantumadvantage_2025}.
When phrased in the context of coding theory, Regev's reduction can be viewed as using an efficient decoder for an $[n,k]$ linear code $\mathcal{C}\subset\ff_q^n$ to construct a quantum algorithm that samples low-weight codewords from the shifted dual code $\mathcal{C}^{\perp}+v$ for a given $v\in\ff_q^n$.
For simplicity, we will restrict ourselves to the binary case $q=2$.
Let $f:\ff_2^n\rightarrow \mathbb{R}_{\geq 0}$ be a function such that $\norm{f}_2=1$ and such that the state $\sum_{e\in\ff_2^n}f(e)\ket{e}$ can be efficiently prepared.
Then, Regev's reduction proceeds as follows~\cite{chailloux_quantumdecoding_2023}:
\begin{align}
    \sum_{e\in\ff_2^n}f(e)\ket{e} \otimes \frac{1}{\sqrt{2^k}}\sum_{c\in\mathcal{C}}(-1)^{v\cdot c}\ket{c}
    &\xrightarrow{\quad (1) \quad} \frac{1}{\sqrt{2^k}} \sum_{c\in\mathcal{C}}(-1)^{v\cdot c}\sum_{e\in\ff_2^n}f(e)\ket{c+e}\otimes \ket{c} \\
    &\xrightarrow{\quad (2) \quad} \frac{1}{\sqrt{2^k}} \sum_{c\in\mathcal{C}}(-1)^{v\cdot c}\sum_{e\in\ff_2^n} f(e)\ket{c+e}\otimes \ket{0} \\
    &\xrightarrow{\quad (3) \quad} \sqrt{2^k} \sum_{d\in\mathcal{C}^{\perp}+v} \hat{f}(d)\ket{d}\otimes \ket{0}
    \, .
\end{align}
First, the state $\sum_ef(e)\ket{e}$ is prepared together with a uniform superposition of the codewords of $\mathcal{C}$.
In step (1), the second register is added to the first register.
In step (2), we utilize a decoder for $\mathcal{C}$ in order to uncompute the second register.
Finally, in step (3), we apply a Hadamard transform on the first register.
By measuring the first register, we obtain a random codeword $d\in\mathcal{C}^{\perp}+v$ with a probability proportional to $\hat{f}(d)^2$ where $\hat{f}$ is the Fourier transform of $f$.
Typically, if we choose $f$ to have support on high-weight codewords, then $\hat{f}$ will concentrate on low-weight codewords, effectively allowing us to sample low-weight codewords from the shifted dual code.
Therefore, using a decoder which tolerates stronger noise directly leads to a lower weight of the sampled dual codewords.
A natural choice for the noise function $f$ is to take i.i.d.\ Bernoulli noise~\cite{chailloux_quantumdecoding_2023,chailloux_quantumadvanteage_2025}, i.e., $f(e)^2 = (1-\omega)^{n-w(e)}{\omega}^{w(e)}$ where $\omega\in[0,1/2]$ and $w(e)$ is the Hamming weight of $e\in\ff_2^n$.
The Fourier transform $\hat{f}$ is then also given by Bernoulli noise, but with noise strength $\omega^{\perp}\coloneqq \tfrac{1}{2}-\sqrt{\omega(1-\omega)}$.

The uncomputation in step (2) can be achieved by either executing a classical decoder in superposition, mapping $\ket{c+e}\otimes\ket{c}\mapsto\ket{c+e}\otimes\ket{0}$ for any $c\in\mathcal{C}$ and correctable error $e\in\ff_2^n$, or by a quantum decoder which coherently recovers $c$ from $\sum_ef(e)\ket{c+e}$.
When choosing $f$ as Bernoulli noise, these two decoding problems precisely correspond to decoding $\mathcal{C}$ transmitted over $\bsc_{\omega}$, or over $W_{\omega}$, respectively.
Since a quantum decoder can tolerate stronger noise rates $\omega$ (because it is limited by the Holevo bound of $W_{\omega}$ instead of the Shannon bound of $\bsc_{\omega}$), it can significantly improve the quality of the sampled output from Regev's reduction.
For example, in the context of solving max-XORSAT with DQI, the threshold $\omega_{\star}$ of a (classical or quantum) decoder directly determines the fraction of satisfied equations in the solution, which is precisely $\omega_{\star}^{\perp}$~\cite{chailloux_improved_analysis}.

In summary, any classical code $\mathcal{C}$, combined with a quantum decoder whose threshold exceeds the Shannon bound, provides a promising avenue for finding a quantum advantage within DQI.
Finding such code–decoder pairs is precisely the goal of this paper.
One should note, however, that when $\mathcal{C}$ possesses some strong structure, the associated combinatorial optimization problem solved by DQI may become somewhat artificial and of limited practical relevance.

\subsection{Overview of technical results}
\paragraph{Subspace decoding problem}
Our main technical contribution is an efficient algorithm to solve the subspace decoding problem, which is concretely defined as follows.
Fix some nonnegative integers $n\geq k\geq\ell$, a full-rank matrix $G\in \ff_2^{k\times n}$, and a probability distribution $P$ over $n$ bits.
Define $F_{\ell,G}$ as a random encoding channel which maps the bit string $x\in\ff_2^{\ell}$ to $(x\cat r)^TG$ where $r\in\ff_2^{k-\ell}$ is chosen uniformly at random and $x\cat r$ denotes bit string concatenation.
Furthermore, for an $n$-bit distribution $P$, let $\spsc[P]$ be the symmetric pure state CQ channel which maps $y\in\ff_2^{n}$ to $\spsc[P](y)\coloneqq Z^y\proj{\varphi}Z^y$ where $\ket{\varphi}\coloneqq\sum_{z\in\ff_2^n}\sqrt{P(z)}\ket{z}$ and $Z^y\coloneqq Z^{y_1}\otimes\dots\otimes Z^{y_n}$.
The subspace decoding task $\sdt(n,k,\ell,G,P)$ consists of recovering a uniformly random input $x\in\ff_2^{\ell}$ from an $n$-qubit system prepared in the state $W(x)$, where $W\coloneqq\spsc[P]\circ F_{\ell,G}$.
\begin{figure}
    \centering
    \includegraphics{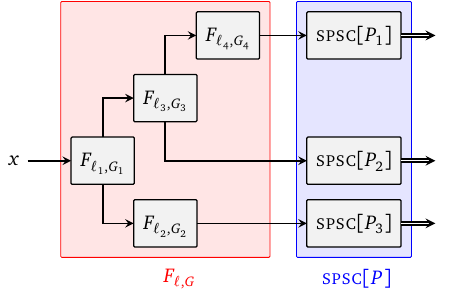}
    \caption{\label{fig:W_decomposition} Decomposition of the channel $W$ in the subspace decoding task.
    In this example, the channel $F_{\ell,G}$, shown in red, decomposes into a concatenation of the four channels $F_{\ell_i,G_i}$ for $i=1,2,3,4$.
    The output of $F_{\ell_1,G_1}$ is split into two bit strings, one of which is then used as input for $F_{\ell_2,G_2}$, and the other as input for $F_{\ell_3,G_3}$.
    Subsequently, the output of $F_{\ell_3,G_3}$ is again split up and one part of the bit string is sent through the channel $F_{\ell_4,G_4}$.
    The channel $\spsc[P]$, shown in blue, decomposes into a product channel $\spsc[P]=\spsc[P_1]\otimes\spsc[P_2]\otimes\spsc[P_3]$.
    Double edges represent quantum systems.
    The nodes and edges contained in the red box make up the MPG representation of $F_{\ell,G}$.
    }
\end{figure}
It is unlikely that every instance $\sdt(n,k,\ell,G,P)$ can be efficiently solved.
However, we show that the task significantly simplifies if a specific instance exhibits some additional structure in the form of the following two requirements, which are depicted in \Cref{fig:W_decomposition}:
\begin{enumerate}
    \item The channel $F_{\ell,G}$ decomposes as a tree-like concatenation of random encodings $F_{\ell_i,G_i}$ of small $[n_i,k_i]$ codes with generator matrix $G_i$.
    The output bit string of each $F_{\ell_i,G_i}$ is split into one or more substrings, and each such substring is passed along an edge to the next node.
    \item $\spsc[P]$ factors into product channels $\spsc[P]=\spsc[P_1]\otimes \spsc[P_2]\otimes\dots$, with one term corresponding to each leaf of the tree representation of $F_{\ell,G}$.
\end{enumerate}
We represent the tree-like decomposition of $F_{\ell,G}$ as a directed tree $\mathcal{G}$, which we call the \emph{message-passing graph} (MPG).
Our main result is an efficient algorithm to solve the subspace decoding problem using this MPG structure.
\begin{theorem}[informal, see \Cref{prop:incoherent_bpqm_complexity} and \Cref{thm:incoherent_bpqm_optimality}]\label{thm:intro_bpqm}
    For any instance $\sdt(n,k,\ell,G,P)$ admitting a tree-like decomposition as described above, we construct a quantum algorithm that solves it with optimal error probability and runtime $\mathcal{O}(\abs{\mathcal{G}}4^{N(\mathcal{G})})$, where $\abs{\mathcal{G}}$ is the number of nodes in the MPG and $N(\mathcal{G})\coloneqq \max_in_i$ is the largest blocklength among the local codes in the MPG.
\end{theorem}
The algorithm operates as a \emph{message-passing procedure} on $\mathcal{G}$: Messages propagate from the leaves towards the root, with each internal node processing the messages from its successors to generate a new message for its predecessor.
Conceptually, this realizes a divide-and-conquer strategy, since each node essentially solves a small instance of $\sdt$ for its associated local code.
The overall runtime is exponential only in the size of the largest local instance, making the algorithm efficient whenever $F_{\ell,G}$ decomposes into local encoding channels of constant or logarithmic size.

\paragraph{Bit-optimal and block-optimal decoding with BPQM}
Our main application of \Cref{thm:intro_bpqm} is to efficiently and optimally solve the quantum decoding problem for certain families of codes.  
We begin with the comparatively simpler task of \emph{bitwise} decoding, i.e., estimating a codeword bit $x_i$ from the channel output $W_{\omega}^{\otimes n}(x)$ for some given $i\in\{1,\dots,n\}$.
We prove a key observation that chaining together bit-optimal decoders yields an optimal block decoder.
\begin{proposition}[informal, see \Cref{prop:bitoptimal_implies_blockoptimal}]\label{prop:intro_bit_to_block}
    Consider an $[n,k]$ binary linear code $\mathcal{C}$ that is transmitted over $W_{\omega}$.
    Suppose for each $i=1,\dots,n$ there exists a two-outcome projective measurement ${\Pi^{(i)}_0,\Pi^{(i)}_1}$ achieving bit-optimal decoding of the $i$-th codeword bit.
    Then, the (sequential) measurement with Kraus operators $\{\tilde\Pi_y\}_{y\in\ff_2^n}$, for $\tilde\Pi_y\coloneqq \Pi^{(n)}_{y_n}\cdots\Pi^{(1)}_{y_1}$, realizes block-optimal decoding.
\end{proposition}
This result is especially remarkable, as the analogous statement does generally not hold in classical coding theory:
It is possible to construct explicit examples where bitwise optimal decoding of a linear code transmitted over the binary symmetric channel does not result in block-optimal decoding.

Thus, block decoding reduces to constructing a quantum algorithm that achieves bit-optimal decoding with two-outcome projective measurements.
Bit decoding is a special case of $\sdt(n,k,\ell,G,P)$ with $\ell=1$, where $G$ is a properly chosen generator matrix of $\mathcal{C}$ and $P$ specifies the pure-state channels.
Thus, \Cref{thm:intro_bpqm} provides an efficient and optimal bit decoder whenever we have an efficient MPG representation of the channel
\begin{equation}
    \ff_2\ni z \mapsto \text{ uniformly random element in } \{x\in\mathcal{C} | x_i=z\}
\end{equation}
for each $i=1,\dots,n$.
We call this channel the $i$-th \emph{bit-transmission channel} of the code $\mathcal{C}$.
\begin{proposition}[see \Cref{prop:trellis_mpg}]
    Let $\mathcal{C}$ be an $[n,k]$ binary linear code that exhibits a trellis representation with maximal state space size $S$.
    Then, for any $i=1,\dots,n$, there exists an MPG representation $\mathcal{G}$ of the $i$-th bit-transmission channel of $\mathcal{C}$ with $N(\mathcal{G})\leq 2\log_2 S$ and $\abs{\mathcal{G}}=n$.
\end{proposition}
Hence, any code with an efficient trellis admits an efficient and optimal bitwise decoder.
To extend this to block decoding, the bit-decoder must implement a two-outcome projective measurement, which is not guaranteed in \Cref{thm:intro_bpqm}. 
We resolve this issue this via the deferred measurement principle, eliminating unnecessary intermediate measurements while preserving runtime complexity by shifting some classical computation into the quantum domain. 
\begin{theorem}[informal, see \Cref{sec:block_optimality}]\label{thm:intro_corollary}
    For any $[n,k]$ binary linear code $\mathcal{C}$ that exhibits a trellis with state space size $S$, we construct a quantum algorithm that solves the quantum decoding task on $\mathcal{C}$ with optimal error probability and runtime $\mathcal{O}(nkS^4)$.
\end{theorem}
This result assumes a computational model where quantum registers can store real numbers in $[0,1]$ with infinite precision at unit cost, analogous to the classical real RAM model. 
In practice, discretization is required.
We prove in \Cref{thm:block_optimal} that our algorithm remains robust under discretization: its runtime stays polynomial, and we pick up an additional $\mathrm{polylog}(1/\epsilon)$ overhead, where $\epsilon$ is the error of the decoder directly caused by the discretization.

\paragraph{BPQM-based decoding of turbo codes}
One particularly important family of codes with efficient trellis representation are convolutional codes.
As such, \Cref{thm:intro_corollary} provides an efficient and optimal quantum algorithm to decode convolutional codes transmitted over a pure-state channel.

We extend these insights to quantum decoding of turbo codes, which are interleaved combinations of two convolutional codes.
In classical channel coding, turbo codes have demonstrated excellent performance as they can operate very close to the Shannon limit~\cite{berrou_turbocodes_1993}.
While turbo codes lack an efficient trellis representation, the convolutional quantum decoder can be adapted to this setting.
Specifically, in the asymptotic limit of large block sizes, windowed decoding with a finite number of iterations yields a tree-like computation graph with probability one~\cite{richardson_modern_2008}, which can be converted into an MPG to decode individual codeword bits.

Since the resulting MPG is not an entirely faithful representation of the bit-transmission channel, the optimality arguments from the previous paragraph do not apply here.
Instead, the asymptotic error rate of BPQM-based bitwise decoding of turbo codes can be numerically characterized using the technique of density evolution.
We numerically estimate these thresholds for the $G=5/7$, $G=13/15$ and $G=23/33$ parallel concatenated turbo code families and depict them in \Cref{tab:turbo_code_results}.
Remarkably, the three code families exhibit thresholds that clearly exceed the Shannon bound, and closely approach the Holevo bound.
\begin{table}
    \centering
      \begin{tabular}{ c | c | c }
        $G$ & $m$ & Threshold (in $\omega$)\\
        \hline
        $5/7$ & $2$ & $0.233$ \\
        $13/15$ & $3$ & $0.243$ \\
        $23/33$ & $4$ & $0.248$ \\
        \hline\hline
        \multicolumn{2}{l|}{$\bsc_{\omega}$ Shannon limit} & $0.17395$ \\
        \multicolumn{2}{l|}{$W_{\omega}$ Holevo limit} & $0.25977$ \\
        \hline
    \end{tabular}
    \caption{BPQM bit decoding thresholds for various rate-$1/3$ parallel concatenated turbo code families transmitted over $W_{\omega}$.
    $G$ denotes the generator polynomials of the constituent convolutional codes in standard octal notation, and $m$ is their memory size (in bits).
    Thresholds are obtained from density evolution numerics (see \Cref{fig:turbo_code_results}).
    }
    \label{tab:turbo_code_results}
\end{table}

\subsection{Relation to prior work}
We show in \Cref{sec:mpg_tanner} that the BPQM algorithm considered in prior work~\cite{renes_belief_2017,rengaswamy_bpqm_2021,piveteau_bpqm_2022,brandsen_bpqm_2022,delaney_demonstration_2022,mandal_bpqmpolar_2023} can be recovered a special case of our algorithm when the local codes in the MPG all correspond to either the $[n_i,1]$ repetition code with generator matrix $G_i=(1 1 \cdots 1)$ or to the trivial $[n_i,n_i]$ code with generator matrix
\begin{equation}
    G_i = \begin{pmatrix}
      1 & 0 & 0 & \dots & 0 \\
      1 & 1 & 0 & \dots & 0 \\
      1 & 0 & 1 & \dots & 0 \\
      \vdots &  &  & \ddots & \vdots \\
      1 & 0 & 0 & \dots & 1  
    \end{pmatrix}
    \, .
\end{equation}
For this reason, we will also refer to our algorithm as ``BPQM''.
We note that previous incarnations of BPQM were fundamentally limited to work with codes that exhibit a tree Tanner graph, or at least, a locally tree-like Tanner graph.\footnote{As an exception, we note that previous did consider adaptations of BPQM for polar codes~\cite{renes_belief_2017,mandal_bpqmpolar_2023}, which do not exhibit a tree Tanner graph. This was possible, because the decoding graph in successive cancellation decoding essentially simplifies to a tree Tanner-like graph.}
As such, it previously remained an open question how to decode codes with small cycles in their Tanner graph.
Our new generalization of BPQM shows that a tree-like Tanner graph is not necessary, as long as another tree-like factor graph structure is provided, such as an efficient trellis.

We note that a special case of the result on block optimality from bit optimality (see \Cref{prop:intro_bit_to_block}) was already implicitly found in~\cite{piveteau_bpqm_2022}, but only for codes $\mathcal{C}$ which exhibit a tree Tanner graph.
Our work generalizes that result to all linear codes $\mathcal{C}$ by using a more refined proof technique.

Our density evolution simulations for the turbo codes are a straightforward adaptation of the BPQM density evolution technique introduced in~\cite{brandsen_bpqm_2022,mandal_bpqmpolar_2023} to our generalized framework.
We note that in~\cite{brandsen_bpqm_2022}, density evolution simulations have demonstrated that BPQM can achieve performance surpassing the Shannon limit on regular LDPC codes.

We note that in literature, the name ``BPQM'' has been used interchangeably to refer to various related algorithms, which differ in some small but important details.
The (inefficient) algorithm described in \cite{renes_belief_2017} corresponds to the ``uniformly-controlled BPQM'' algorithm which we discuss in~\Cref{app:formalization_coherent_bpqm}.
The (inefficient) algorithm called ``BPQM'' in~\cite{piveteau_bpqm_2022} also corresponds to uniformly-controlled BPQM.
The algorithm called ``message-passing BPQM'' in~\cite{piveteau_bpqm_2022} corresponds to the ``coherent BPQM'' algorithm discussed in \Cref{sec:bpqm_coherent}.

\section{Preliminaries}
\subsection{Notation}
\begin{description}
\item[Vectors]
We denote by $\ff_2$ the finite field with $2$ elements.
For two vectors $x\in\ff_2^n$ and $y\in\ff_2^m$, we denote their concatenation by $x\cat y$, an element of $\ff_2^{n+m}$.
Unless stated otherwise, vectors $x\in\ff_2^n$ are understood as column vectors, with $x^T$ denoting their transpose.
The inner product of two vectors $x,y\in\ff_2^n$ is defined as $x\cdot y \coloneqq x^Ty = \sum_{i=1}^n x_iy_i \mod 2$.

\item[Distributions]
For two distributions $P(x)$, $x\in\mathcal{X}$ and $Q(y)$, $y\in\mathcal{Y}$ we denote by $P\times Q$ the product distribution $(P\times Q)(x,y)\coloneqq P(x)Q(y)$.

\item[Quantum states and operators]
For an $n$-qubit system, the computational basis states are denoted by $\ket{x}$ for $x\in\ff_2^n$.
We denote the identity operator and the three Pauli matrices by
\begin{equation}
    \id = \begin{pmatrix}1 & 0 \\ 0 & 1\end{pmatrix}
    \, , \quad
    X = \begin{pmatrix}0 & 1 \\ 1 & 0\end{pmatrix}
    \, , \quad
    Y = \begin{pmatrix}0 & -i \\ i & 0\end{pmatrix}
    \, , \quad
    Z = \begin{pmatrix}1 & 0 \\ 0 & -1\end{pmatrix}
    \, .
\end{equation}
For some $x\in\ff_2^{n}$, we define the $n$-qubit operator $Z^x\coloneq Z^{x_1}\otimes Z^{x_2}\otimes\dots Z^{x_n}$.

\item[Norms and metrics]
For a real vector $x\in\mathbb{R}^n$, we denote its $\ell_p$-norm for $p\geq 1$ by $\norm{x}_p\coloneq \left(\sum_{i=1}^n\abs{x_i}^p\right)^{1/p}$.
For a linear operator $A\in\mathbb{C}^{n\times n}$, we denote its Schatten $p$-norm by $\norm{T}_p\coloneq \tr\left(\abs{A}^p\right)^{1/p}$.
Note that $\norm{A}_{\infty}$ is the spectral norm.
For a pure quantum state $\ket{\psi}\in\mathbb{C}^n$, we denote by $\norm{\ket{\psi}}$ the standard (Euclidean) Hilbert space norm.

For two positive semi-definite operators $\rho,\sigma\in\mathbb{C}^{n\times n}$, we denote their fidelity and trace distance by $F(\rho,\sigma) \coloneq \tr\left[\sqrt{\sqrt{\rho}\sigma\sqrt{\rho}}\right]$ and $\delta(\rho,\sigma) \coloneq \tfrac{1}{2}\norm{\rho-\sigma}_1$.
As a slight abuse of notation, we will write $F(\ket{\psi},\ket{\varphi})=F(\proj{\psi},\proj{\varphi})$ and $\delta(\ket{\psi},\ket{\varphi})=\delta(\proj{\psi},\proj{\varphi})$ for two pure quantum states $\ket{\psi},\ket{\varphi}\in\mathbb{C}^n$.

\item[Logarithm] We denote by $\log$ the logarithm with base $2$.

\item[Circuit complexity]
The circuit complexity of a quantum circuit is defined as the number of single and two-qubit gates that it contains.

\item[Pure-state channels]
Let $P$ be an $n$-bit distribution.
The symmetric pure-state channel $\spsc[P]$ maps $n$ bits to $n$ qubits and is defined as 
\begin{equation}
\label{eq:pscdef}
    \spsc[P](x)\coloneqq Z^x\proj{\varphi}Z^x \quad , \quad \ket{\varphi}\coloneqq\sum_{z\in\mathbb{F}_2^n}\sqrt{P(z)}\ket{z}
    \, .
\end{equation}
The state $\ket{\varphi}$ is sometimes referred to as the channel state. 
The specialization with binary input ($n=1$) will be of particular interest for us, and hence we define the binary-input pure-state channel $\bspsc[p]$ for the channel parameter $p\in[0,1]$ as
\begin{equation}
    \bspsc[p](x) \coloneqq \spsc[(1-p,p)](x) \, .
\end{equation}
\end{description}

\subsection{Forney-style factor graphs}
\begin{figure}
    \centering
    \begin{subfigure}[b]{0.4\textwidth}
        \centering
        \includegraphics{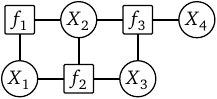}
        \caption{}
        \label{fig:fg_standard}
    \end{subfigure}
    \begin{subfigure}[b]{0.4\textwidth}
        \centering
        \includegraphics{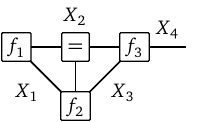}
        \caption{}
        \label{fig:fg_forney}
    \end{subfigure}
    \caption{Standard and Forney-style factor graph representation of the factorization of the function $f(x_1,x_2,x_3,x_4)$ defined in the main text.}
\end{figure}

A factor graph is a graphical representation of how a function of several variables, e.g., a joint probability distribution, factorizes into a product of separate terms each involving only a subset of the variables. 
The standard variant of factor graphs~\cite{kschischang_factor_2001} is a bipartite graph consisting of two types of nodes, called \emph{variable nodes} and \emph{factor nodes}.
Variable nodes are usually represented with circles and factor nodes with rectangular boxes. 
Consider a factorizable function
\begin{equation}
  f(x_1,\dots,x_n)=\prod\limits_{j\in J} f_j(Z_j)
\end{equation}
where $J$ is some finite index set and the $f_j$ are functions taking some subset $Z_j$ of $\{x_1,\dots,x_n\}$ as arguments.
The corresponding factor graph consists of $n$ variable nodes $X_1,\dots,X_n$ as well as a factor node $f_j$ for each $j\in J$.
A variable node $X_i$ and a factor node $f_j$ are connected by an edge if and only if $x_i\in Z_j$.
As an example, \Cref{fig:fg_standard} depicts the factor graph representing the factorization of the function
\begin{equation}
    f(x_1,x_2,x_3,x_4) = f_1(x_1,x_2)f_2(x_1,x_2,x_3)f_3(x_2,x_3,x_4) \, .
\end{equation}

For this work, it will prove advantageous to work with a different graphical representation of factorizable functions, namely the so-called \emph{Forney-style factor graphs}~\cite{forney_codes_2001,loeliger_introduction_2004}.
Here, variables are represented by the edges of the graph whereas the factor terms of the function are represented by the nodes of the graph.
Similarly, an edge and a node of the graph are connected if and only if the corresponding variable is part of the corresponding factor term.

When translating a factor graph to the Forney-style representation, one might encounter the issue that some variables are involved in more than two factors, as depicted in the example in~\Cref{fig:fg_forney}. 
In that case, equality factor nodes must be introduced in the Forney-style factor graph. 
Equality nodes are depicted as $\raisebox{-1mm}{\eqbox}\hspace{.5mm}$ and represent the function that returns $1$ if all connected variables have the same value and $0$ otherwise.
All edges connected to a common equality node essentially represent the same variable.

Consider a Forney-style factor graph representing the factorization of some function in $m+l$ variables $f(V_1,\dots,V_m,H_1,\dots,H_l)$ where one subset of the variables $V_1,\dots,V_m$, called the ``visible'' variables, are represented by half-edges, i.e., they are each connected to exactly one factor node. 
The other variables $H_1,\dots,H_l$ are 'hidden' inside the graph, i.e., they are each connected to exactly two factor nodes.
We define the \emph{exterior function} of the factor graph to be the joint marginal of the visible variables, i.e., the function
\begin{equation}
  v_1,\dots,v_m \mapsto \sum\limits_{h_1,\dots,h_l} f(v_1,\dots,v_m,h_1,\dots,h_l) \, .
\end{equation}

\begin{figure}
    \centering
    \begin{subfigure}[b]{0.49\textwidth}
        \centering
        \includegraphics{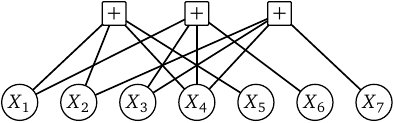}
        \caption{}
        \label{fig:hamming_fg_standard}
    \end{subfigure}
    \begin{subfigure}[b]{0.49\textwidth}
        \centering
        \includegraphics{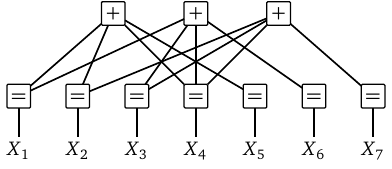}
        \caption{}
        \label{fig:hamming_fg_forney}
    \end{subfigure}
    \caption{Standard and Forney-style representation of the Tanner graph of the Hamming code.}
\end{figure}

We now introduce the \emph{Tanner graph}, a specific type of factor graph that will be of great importance for us.
Consider an $[n,k]$ binary linear code $\mathcal{C}$, i.e., a linear subspace of $\mathbb{F}_2^n$ of dimension $k$.
Let $H\in\ff_2^{(n-k)\times n}$ be a parity-check matrix associated to this code, i.e.,
\begin{equation}
    \mathcal{C} = \{x\in\mathbb{F}_2^n | Hx = 0\} \, .
\end{equation}
The Tanner graph associated to $H$ is a factor graph representation of the factorization of the code inclusion function
\begin{equation}\label{eq:code_inclusion_function}
  I_{\mathcal{C}}:\mathbb{F}_2^n\rightarrow \{0, 1\} \, ,
  \quad
  I_{\mathcal{C}}(x) := \begin{cases}1 \text{ if } x\in \mathcal{C} \\ 0 \text{ else}\end{cases}
  \, .
\end{equation}
In standard factor graph notation, the Tanner graph contains $n$ variables nodes $X_1,\dots,X_n$ and $m$ check nodes (denoted \raisebox{-1mm}{\plusbox}\hspace{.5mm}) which represent the function that is equal to $1$ if the involved variables sum up to $0$ modulo $2$, and $0$ otherwise.
The variable node $X_i$ is connected to the $j$-th check node if and only if $H_{ji}=1$.
Hence, each check node corresponds to a parity check encoded in one of the rows of $H$.
\Cref{fig:hamming_fg_standard} depicts the Tanner graph for the Hamming code based on the parity check matrix
\begin{equation}
    H = 
    \begin{pmatrix}
        1 & 1 & 0 & 1 & 1 & 0 & 0 \\
        1 & 0 & 1 & 1 & 0 & 1 & 0 \\
        0 & 1 & 1 & 1 & 0 & 0 & 1
    \end{pmatrix}
    \, .
\end{equation}
In the Forney-style representation of the Tanner graph, we simply replace each variable node with an equality node.
As an example, the Forney-style Tanner graph of the Hamming code is depicted in \Cref{fig:hamming_fg_forney}.
As a convention, we will always choose the exterior function of a Forney-style Tanner graph to be the code inclusion function $I_{\mathcal{C}}(x_1,\dots,x_n)$.
This is achieved by connecting a half-edge to every equality node.

\subsection{Trellis representation of codes}
\label{sec:trellis}
Here, we briefly introduce the trellis representation of binary codes.
Compared to the Tanner graph, the trellis representation can be viewed as an alternative technique to graphically represent a linear code.
We refer interested readers to the excellent review in~\cite{vardy_trellisstructure_1998} for an exhaustive treatment.
In the context of classical coding theory, trellis representations have been extensively studied as they form the basis for the Viterbi~\cite{forney_viterbi_1973} and BCJR~\cite{bahl_bcjr_1974} decoding algorithms, which respectively realize the blockwise and bitwise optimal decoders over a classical channel. 
The runtime of these algorithms scales with the size of the trellis, which implies that the classical decoding problem can be optimally and efficiently solved for any code that exhibits an efficient (i.e.\ small) trellis representation.
Perhaps the most well-known family of codes with efficient trellises are the convolutional codes.
By extension, trellis representations also play an important role in the decoding of turbo codes~\cite{berrou_turbocodes_1993}, which are built from convolutional codes.
Turbo codes can operate very close to the Shannon limit of a classical channel.

\begin{figure}
    \centering
    \includegraphics{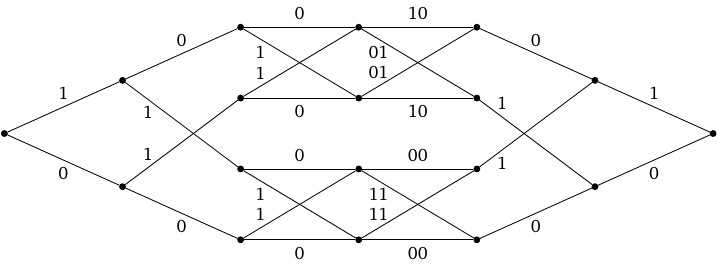}
    \caption{Trellis representation of the $[7,4]$ Hamming code. The codewords can be read off by considering paths from the leftmost node (source) to the rightmost node (sink).}
    \label{fig:trellis_hamming}
\end{figure}

We now formally define a trellis representation of a code.
In \Cref{fig:trellis_hamming}, we depict a trellis representation of the $[7,4]$ Hamming code with generator matrix
\begin{equation}
    G = \begin{pmatrix}
        1 & 0 & 0 & 0 & 1 & 1 & 0 \\
        0 & 1 & 0 & 0 & 1 & 0 & 1 \\
        0 & 0 & 1 & 0 & 0 & 1 & 1 \\
        0 & 0 & 0 & 1 & 1 & 1 & 1 
    \end{pmatrix}
    \, .
\end{equation}
A trellis $T$ is an acyclic directed graph $T=(V,E)$ where $V$ denotes the set of vertices and $E$ the set of edges.
The vertices are partitioned into $m+1$ disjoint subsets (called ``state sets'')
\begin{equation}
    V = V_0 \cup V_1 \cup \dots \cup V_m
\end{equation}
such that the first and last state set contain only one element each $V_0=\{v_0\}$, $V_m=\{v_{\star}\}$.
We refer to $v_0$ and $v_{\star}$ as the ``source'' and the ``sink'' of the trellis.
The edges connect neighboring state sets in increasing direction, i.e., every edge $e$ goes from some element in $V_{i-1}$ to some element in $V_{i}$ for some $i\in\{1,\dots,m\}$.
Similar to the vertices, we can group the edges into disjoint layers
\begin{equation}
    E = E_1 \cup E_2 \cup \dots \cup E_m
\end{equation}
where $E_i$ contains all edges going from $V_{i-1}$ to $V_i$.
To any edge $e\in E_i$, we associate some label $\lambda(e)\in\ff_2^{\ell_i}$ where $\ell_i\in\mathbb{N}$ is some fixed number for every $i\in\{1,\dots,m\}$.

For any path $p=(e_1,\dots,e_m)$ from the source to the sink (which must fulfill $e_i\in E_i$ for all $i$), we define the path label to be the concatenation of the labels of the individual edges
\begin{equation}
    \lambda(p) \coloneqq \lambda(e_1)\cat \lambda(e_2) \cat \dots \cat \lambda(e_m) \, .
\end{equation}
We say that a trellis $T=(V,E)$ represents a binary linear code $\mathcal{C}\subset\ff_2^n$ if the path labels precisely correspond to the codewords, i.e.,
\begin{equation}
    \mathcal{C} = \left\{ \lambda(p) \middle\vert p \text{ is a path in $T$ from the source to the sink} \right\} \, .
\end{equation}
The \emph{maximal state space size} $S(T)$ for some trellis $T$ is defined as
\begin{equation}
    S(T) \coloneqq \max\limits_{i\in \{0,\dots,m\}} \abs{V_i}
    \, .
\end{equation}

\subsection{Pretty good measurement}
Consider a quantum system whose state is described by an ensemble $\{P(x), \rho_x\}_x$, i.e., it is in the state $\rho_x$ where the random variable $X$ is randomly sampled according to the distribution $P(x)$.
The task of quantum hypothesis testing is to estimate $X$ given access to this quantum system.
One strategy is to perform the so-called \emph{pretty good measurement} (PGM) described by the POVM operators
\begin{equation}
    \Lambda_x \coloneq  P(x) \rho^{-1/2}\rho_x\rho^{-1/2} \,,
\end{equation}
where $\rho\coloneq\sum_x P(x)\rho_x$.
In case that $\sum_x\Lambda_x\leq \id$, the POVM is completed with an additional operator $\id-\sum_x\Lambda_x$.
In certain circumstances, the success probability of solving the quantum hypothesis testing task $\sum_x P(x)\tr[\rho_x\Lambda_x]$ can be shown to be maximal.
\begin{proposition}[Optimality of the PGM~\cite{ban_optimum_1997,sasaki_quantum_1998,eldar_2001}]\label{prop:PGMopt}
    Consider a set of linearly independent pure states $\mathcal{S}=\{\ket{\phi(x)}\}_x$ which form a geometrically uniform state set, i.e.,
    \begin{equation}
      \mathcal{S} = \{U\ket{\phi} | U\in \mathcal{G} \}
    \end{equation}
    for some state $\ket{\phi}$ and some finite Abelian group of unitary matrices $\mathcal{G}$ .
    The PGM distinguishes the uniform mixture of $\mathcal{S}$ with the optimal success probability.
\end{proposition}
When the PGM is applied to an ensemble of pure states $\{P(x),\ket{\phi_x}\}_x$, as for example in \Cref{prop:PGMopt}, then the associated POVM is realized by a projective measurement w.r.t.\ the orthonormal basis $\{\ket{f_x}\}_x$ defined as
\begin{equation}
  \ket{f_x} \coloneqq \sqrt{P(x)}\rho^{-1/2}\ket{\phi_x} \, .
\end{equation}

\begin{lemma}\label{lem:optimal_mmt}
Let $\ket{\varphi}$ be an arbitrary $\ell$-qubit state. Given access to a quantum system in a uniformly randomly-chosen state $\{Z^x\ket{\varphi}\}_{x\in \mathbb F_2^\ell}$, the optimal measurement to determine $x\in \mathbb F_2^\ell$ is given by the POVM $\Lambda_x\coloneq H^{\otimes \ell}\proj{x}H^{\otimes \ell}$.
\end{lemma}
\begin{proof}
We expand the state $\ket{\varphi}=\sum_{y\in \mathbb F_2^\ell}\sqrt{P_Y(y)}\ket{y}$ where $\{\ket{y}\}_y$ are the computational basis states, up to some complex phase.
Due to the Abelian symmetry of our state ensemble, \Cref{prop:PGMopt} ensures that the PGM achieves the optimal distinguishing probability.
The PGM realizes a projective measurement w.r.t. the orthonormal basis $\ket{f_x}\coloneqq \tfrac{1}{\sqrt{2^{\ell}}}\rho^{-1/2}Z^x\ket{\varphi}$, where $\rho\coloneq\sum_{x\in \mathbb F_2^{\ell}}\tfrac{1}{2^{\ell}}Z^x\proj{\varphi}Z^x$. 
Evidently, $\rho=\sum_{y\in\mathbb F_2^{\ell}}P_Y(y)\proj{y}$, since the sum completely dephases the state $\proj{\varphi}$. 
Hence, 
\begin{equation}
\tfrac{1}{\sqrt{2^\ell}}\rho^{-1/2}Z^x \ket{\varphi}=\tfrac{1}{\sqrt{2^\ell}} Z^x\rho^{-1/2}\ket{\varphi}=\tfrac{1}{\sqrt{2^\ell}} Z^x\sum_{y\in\mathbb F_2^\ell}\ket{y}
    \, .
\end{equation}
This is equal to $Z^x\ket{+}^{\otimes \ell}$ up to some complex phase.
Therefore, $\Lambda_x=H^{\otimes \ell}\proj{x}H^{\otimes \ell}=Z^x\proj{+}^{\otimes \ell}Z^x$ is the optimal measurement. 
\end{proof}

\section{Subspace decoding task}\label{sec:node_op}
\begin{figure}[h]
\begin{center}
\includegraphics{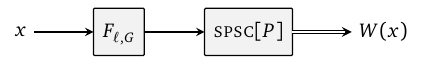}
\caption{\label{fig:subspacedecoding} The CQ channel in \Cref{eq:Wbar}. Double edges represent quantum systems.}
\end{center}
\end{figure}

This section is devoted to a problem which we call the ``subspace decoding task''.
The goal is to extract certain classical information that has been encoded into the output of a symmetric pure state quantum channel.
\begin{problem}[Subspace decoding task $\sdt(n,k,\ell,G,P)$]
\hfill
\begin{problemdescr}
    \item[Instance parameters:] Nonnegative integers $n\geq k\geq\ell$, full-rank matrix $G\in \ff_2^{k\times n}$, probability distribution $P$ over $n$ bits.
    \item[Input:] An $n$-qubit system in the state $W(x)$ for $x\in \ff_2^\ell$ sampled uniformly at random, where $W$ is the channel defined as
    \begin{equation}\label{eq:Wbar}
        x\mapsto W(x)\coloneqq \frac{1}{2^{k-\ell}}\sum_{r\in\ff_2^{k-\ell}} \spsc[P]\left( (x\cat r)^TG \right)\,. 
    \end{equation}   
    \item[Goal:] Given the input quantum system, determine $x$.
\end{problemdescr}
\end{problem}
As depicted in \Cref{fig:subspacedecoding}, the channel $W$ defining this task can be seen as the concatenation of a classical channel $F_{\ell,G}$ and the CQ channel $\spsc[P]$, where $F_{\ell,G}$ is defined as follows.
\begin{definition}
    For a full-rank matrix $G\in\ff_2^{k\times n}$ and some positive integer $\ell\leq k$, we define $F_{\ell,G}$ to be the channel which maps a bit string $x\in\ff_2^{\ell}$ to $(x\cat r)^TG$ where $r\in\ff_2^{k-\ell}$ is chosen uniformly at random.
\end{definition}
Intuitively, $F_{\ell,G}$ extends the input $x$ to a $k$-bit source bit string, and then encodes it through $G$.
The subspaces in the $\sdt$ problem are precisely the images of $x$ under $F_{\ell,G}$, and the difficulty of determining $x$ is that one only has access to the image of the subspace under $\spsc[P]$. 

In the following, we show that $\sdt(n,k,\ell,G,P)$ can be optimally solved by first performing some specific $n$-qubit unitary followed by a measurement of a particular subset of $\ell$ qubits.
This unitary operation will be a central ingredient for the BPQM algorithm later, so for convenience we denote it by $\pno(n,k,\ell,G,P)$.
We call it a ``primitive node operation'', since it will essentially constitute the node operation for the message-passing procedure of BPQM.
\begin{definition}[Primitive node operation $\pno(n,k,\ell,G,P)$]\label{def:pno}
Let $n,k,\ell,G,P$ be the parameters of some instance of the subspace decoding task.
Then $\pno(n,k,\ell,G,P)$ is the $n$-qubit unitary $\pno(n,k,\ell,G,P)\coloneqq U_2U_1$, where the $n$-qubit unitaries $U_1,U_2$ are constructed as follows:
\begin{enumerate}[noitemsep]
    \item Let $M=\left(\begin{smallmatrix}G \\ K\end{smallmatrix}\right)$ for some $K\in \ff_2^{(n-k)\times n}$ such that $M$ is invertible.
    \item Define $U_1\coloneqq\sum_{z\in \ff_2^n}\ket{Mz}\!\bra{z}$.
    \item Let $Y$, $S$ and $A$ be $\ell$-bit, $(k-\ell)$-bit and $(n-k)$-bit random variables, respectively, with the joint distribution
    \begin{equation}\label{eq:PYSA}
        P_{Y,S,A}(y,s,a)\coloneqq P\left( M^{-1}(y\cat s\cat a) \right).
    \end{equation}
    \item For any $s\in \ff_2^{k-\ell}$ and $y\in \ff_2^\ell$, define the $(n-k)$-qubit state
    \begin{equation}
    \ket{\xi_{y,s}}\coloneqq \sum_{a\in \mathbb F_{2}^{n-k}}\sqrt{P_{A|Y=y,S=s}(a)}\ket{a}\,,
    \end{equation} and let $V(y,s)$ be an $(n-k)$-qubit unitary which maps $\ket{\xi_{y,s}}$ to $\ket{0}^{\otimes (n-k)}$.
    \item Define $U_2\coloneqq \sum_{y\in \mathbb F_2^{\ell},s\in \mathbb F_2^{k-\ell}} \proj{y} \otimes \proj{s} \otimes V(y,s)$.
\end{enumerate}
\end{definition}
Note that the definition of $\pno$ is not unique: there may exist many valid choices for the matrix $M$ and the unitaries $V(y,s)$.
Unless otherwise specified, any such choice will suffice for our purposes.
Thus, whenever we refer to the unitary $\pno(n,k,\ell,G,P)$, we are implicitly picking one valid choice. % out of the set of all valid unitaries.

Following the \pno\ by a measurement of the first $\ell$ qubits in the conjugate basis is optimal for solving \sdt. 
\begin{theorem}\label{thm:primitiveNode}
    Consider an instance $T=\sdt(n,k,\ell,G,P)$ and let $U=\pno(n,k,\ell,G,P)$. 
    Define the projective measurement with projectors $\Lambda_x$ for $x\in \ff_2^{\ell}$ by 
    \begin{equation}\label{eq:optimalPOVM}
        \Lambda_x\coloneq U^\dagger(H^{\otimes \ell}\proj{x}H^{\otimes \ell}\otimes \id^{\otimes (n-\ell)})U\,.
    \end{equation}
   Then $\{\Lambda_x\}_{x\in \ff_2^\ell}$ determines the channel input $x$ in $T$ with optimal (i.e., lowest possible) error probability. 
\end{theorem}

The proof makes use of the fact that $\pno(n,k,\ell,G,P)$ effectively compresses the information about $x$ contained in the $n$ qubits into just $\ell$ qubits and $k-\ell$ classical bits. 
Moreover, the state of the $\ell$ qubits can be understood as the output of a heralded symmetric pure-state channel given input $x$.
More formally, we have the following result. 
\begin{lemma}\label{lem:node_operation}
    Consider an instance  $\sdt(n,k,\ell,G,P)$ and let $U=\pno(n,k,\ell,G,P)$.
    For all $x\in \ff_2^\ell$, 
    \begin{equation}\label{eq:endstate}
       U{W}(x)U^{\dagger} = \sum_{s\in\ff_2^{k-\ell}} P_S(s) \, \spsc[P_{Y|S=s}](x) \otimes \proj{s}\otimes  \proj{0}^{\otimes (n-k)}\,,
    \end{equation}
    where $P_S$ and $P_{Y|S=s}$ are marginal and conditional marginal distributions defined from \Cref{eq:PYSA}.
\end{lemma}
\begin{proof}
    The proof proceeds by using $U_1$ to bring $W(x)$ into a CQ form and then $U_2$ to ``decouple'' the final $n-k$ qubits from the rest and leave them in the all-zero state. 

    First observe that $U_1 Z^{v^TM}=Z^vU_1$ for any vector $v\in \mathbb F_2^n$:
      \begin{equation}
      \label{eq:U1action}
      \begin{aligned}
        U_1 Z^{v^TM} U_1^{\dagger}
        &= \sum_{z,z'\in \mathbb F_2^n} \ketbra{Mz}{z} Z^{v^TM} \ketbra{z'}{Mz'} \\
        &= \sum_{z\in \mathbb F_2^n} \ketbra{Mz}{Mz} (-1)^{v^TMz} = \sum_{z \in \mathbb F_2^n} \ketbra{z}{z} (-1)^{v\cdot z} 
           = Z^{v} \, .
      \end{aligned}
    \end{equation}
    Furthermore, $U_1$ acts on the channel state $\ket\varphi=\sum_z\sqrt{P(z)}\ket{z}$ as follows:
    \begin{equation}
      \begin{aligned}
        U_1\ket{\varphi}
        &= \sum_{z \in \ff_2^n} \sqrt{P(z)} \ket{Mz} = \sum_{z \in \ff_2^n} \sqrt{P(M^{-1}z)} \ket{z} \\
        &= \sum_{y\in \ff_2^\ell,s\in \ff_2^{k-\ell},a\in \ff_2^{n-k}} \sqrt{P_{Y,S,A}(y,s,a)} \ket{y}\otimes\ket{s}\otimes\ket{a}\\
        &=\sum_{y\in \ff_2^\ell,s\in \ff_2^{k-\ell}} \sqrt{P_{Y,S}(y,s)} \ket{y}\otimes\ket{s}\otimes\ket{\xi_{y,s}}\,.
      \end{aligned}
    \end{equation}
    Let us now consider the action of $U$ on the channel output for a fixed bit string $r\in\ff_2^{k-\ell}$.
    By \Cref{eq:U1action}, we have 
    \begin{equation}
        U_1 Z^{(x\cat r\cat 0)^TM}\ket{\varphi}=\sum_{y\in \mathbb F_2^\ell,s\in \mathbb F_2^{k-\ell}} \sqrt{P_{Y,S}(y,s)} Z^x\ket{y}\otimes Z^r\ket{s}\otimes \ket{\xi_{y,s}}\,.
    \end{equation}
    Applying $U_2$ decouples the last subsystem, producing
    \begin{equation}\label{eq:nodeop_purestate}
    U_2U_1 Z^{(x\cat r\cat 0)^TM}\ket{\varphi}=\sum_{s\in \mathbb F_2^{k-\ell}} \sqrt{P_{S}(s)} Z^x\ket{\varphi'_s}\otimes Z^r\ket{s}\otimes \ket{0}^{\otimes (n-k)}\,,
    \end{equation}
    where $\ket{\varphi_s'}=\sum_{y\in \ff_2^{\ell}} \sqrt{P_{Y|S=s}(y)}\ket{y}$. 
    Mixing these pure states uniformly over values of $r$ pinches the qubits in the second register, removing all off-diagonal elements. 
    Hence $UW(x)U^\dagger$ has the form claimed in \Cref{eq:endstate}. 
\end{proof}
Observe that the purity of the output of $W$ is crucial to the result. 
The case of mixed states can be easily described by having an additional purification register $R$ such that $\ket{\varphi}\in (\mathbb {C}^2)^{\otimes n}\otimes \mathcal H_R$. The operation $U_1$ will work as before, but now $U_2$ will not generally be able to decouple $\ket{\xi_{y,s}}\in (\mathbb{C}^2)^{\otimes (n-k)}\otimes \mathcal H_R$ from the rest of the state without access to $R$.\footnote{One case that does work is when all the $\ket{\xi_{y,s}}$ are different maximally-entangled states, or, more generally, when all the states have the same Schmidt coefficients.}

\begin{proof}[Proof of \Cref{thm:primitiveNode}]
It is equivalent to distinguish the states $\{W(x)|x\in\ff_2^{\ell}\}$ and $\{UW(x)U^{\dagger}|x\in\ff_2^\ell\}$.
\Cref{lem:node_operation} implies that by observing the classical bit $s$, this latter task reduces to the problem of distinguishing the states $\{\spsc[P_{Y|S=s}](x) | x\in\ff_2^\ell\}$.
By \Cref{lem:optimal_mmt}, this is optimally achieved by a measurement in the conjugate basis.
Importantly, the same conjugate-basis measurement is optimal for every value of $s$.
\end{proof}

\section{General description of BPQM}\label{sec:general_bpqm}
The prototypical node operation of the previous section performs optimal subspace decoding. 
However, the circuit complexity of realizing the associated unitary will generally grow exponentially with $n$.
In this section, we will see how subspace decoding can still be solved efficiently when the code associated with $G$ has some additional structure that allows the problem to be split into a sequence of smaller instances with small $n$ each.

Recall that the channel $W$ in $\sdt(n,k,\ell,G,P)$ can be written as $W=\spsc[P]\circ F_{\ell,G}$.
There are two requirements for a given subspace decoding task to decompose into smaller instances, which are depicted in \Cref{fig:W_decomposition}:
\begin{enumerate}
  \item The channel $F_{\ell,G}$ can be written as a tree-like concatenation of random encodings $F_{\ell_i,G_i}$ of smaller $[n_i,k_i]$ codes $G_i$.
  The output bit string of each $F_{\ell_i,G_i}$ is split into one or more substrings, and each such substring is passed along an edge to the next node.
  \item $\spsc[P]$ factors into product channels $\spsc[P]=\spsc[P_1]\otimes \spsc[P_2]\otimes\dots$, with one term corresponding to each leaf of the tree representation of $F_{\ell,G}$.
\end{enumerate}

The tree representation of $F_{\ell,G}$ will be central for us, so we give an explicit name and call it a \emph{message-passing graph} (MPG) representation of $F_{\ell,G}$.
In the remainder of this section, we first formally define MPGs and then introduce the incoherent BPQM algorithm, which can be considered a message-passing algorithm operating on the MPG (the reason for the adjective \emph{incoherent} will become clear in \Cref{sec:bpqm_decoding}).
We then show that the algorithm achieves the optimal error probability. 
Finally, we show that incoherent BPQM is robust to discretization errors, which are inevitable in practical implementations of the algorithm.

\subsection{Message-passing graphs}
Formally, an MPG $\mathcal{G}=(V,E)$ is a directed tree, where $V$ and $E$ denote the sets of nodes and edges respectively.
The MPG has both normal edges, connecting two nodes, and half-edges, connected to only one node. 
There are at least $2$ half-edges: one serves as the ``root edge'' and the rest serve as ``leaf edges''.
Each node has degree at least two and each edge is directed away from the root.
Hence, each node $v\in V$ in an MPG has one incoming edge and $m_v\geq 1$ outgoing edges, which we denote by $e^{\mathrm{in}}(v)$ and $e^{\mathrm{out}}_1(v),\dots,e^{\mathrm{out}}_{m_v}(v)$ respectively.

To each edge $e\in E$, we associate a positive integer $n_e$, which represents the number of bits associated to the represented intermediate step in the concatenated encoding procedure.
We then define $n_v\coloneq \sum_{i=1}^{m_v} n_{e^{\mathrm{out}}_i(v)}$ and $\ell_v\coloneqq n_{e^{\mathrm{in}}(v)}$. 
Additionally, to each node $v\in V$ we associate a full-rank generator matrix $G_v\in \ff_2^{k_v\times n_v}$.
Furthermore, it must hold that $n_v\geq k_v\geq \ell_v$. 

As depicted in \Cref{fig:W_decomposition}, every MPG represents a certain classical channel, which we formalize as follows.
\begin{definition}
    Consider an MPG $\mathcal{G}$ with root edge $e^{\star}$ and leaf edges $e_1,\dots,e_m$.
    Let $n_{\mathrm{out}}\coloneqq\sum_{i=1}^mn_{e_i}$.
    The channel induced by $\mathcal{G}$, denoted $F[\mathcal{G}]$, is an $n_{e^{\star}}$-bit to $n_{\mathrm{out}}$-bit channel.
    We define it by giving an explicit procedure to sample an output $F[\mathcal G](x)$:
    \begin{enumerate}
        \item Assign the message $x$ to the root edge $e^{\star}$.
        \item Perform message-passing from the root towards the edges. Upon encountering node $v$, append a random string $r\in \ff_2^{k_v-\ell_v}$ to the message $z\in \ff_2^{\ell_v}$, and transform the result via $G_v$ to $(z\cat r)^TG_v$.
        Then, partition the output bits and send them along the corresponding output edges to subsequent nodes.
        Repeat until messages have been generated for all leaf edges.
        \item Output the concatenation of the messages over all leaf edges.
    \end{enumerate}
\end{definition}

It can often be convenient to think of the MPG $\mathcal{G}$ as a Forney-style factor graph, by having the node $v$ represent the channel $F_{\ell_v,G_v}$.
More formally, the node $v$ represents the indicator function
\begin{equation}\label{eq:mpg_factors}
  x^{\mathrm{in}},x^{\mathrm{out}}_1,\dots,x^{\mathrm{out}}_{m_v} \mapsto 
  \left\{\begin{array}{cl} 
  1 & \text{if }\exists r\in\ff_2^{k_v-\ell_v} : (x^{\mathrm{in}}\cat r)^TG_v = x^{\mathrm{out}}_1\cat\dots\cat x^{\mathrm{out}}_{m_v}\\ 
  0 & \text{else}\end{array}\right.\,,
\end{equation}
where $x^{\mathrm{in}}$ is the variable associated with the edge $e^{\mathrm{in}}(v)$ and $x^{\mathrm{out}}_i$ the variable associated to $e^{\mathrm{out}}_i(v)$.
If $\mathcal{G}$ now represents the channel $F_{\ell,G}$, then the outer function of the factor graph exactly represents $F_{\ell,G}$ in the sense of \Cref{eq:mpg_factors}.
In \Cref{sec:codes}, we will discuss concrete MPG constructions for interesting channels $F_{\ell,G}$.

\begin{figure}
    \centering
    \includegraphics{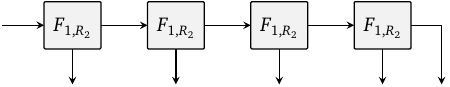}
    \caption{MPG representing the encoding channel $F_{1,R_5}$ of the $5$-bit repetition code.}
    \label{fig:mpg_example_repetition}
\end{figure}

\begin{figure}
    \centering
    \begin{subfigure}[T]{0.49\linewidth}
    \centering
    \includegraphics{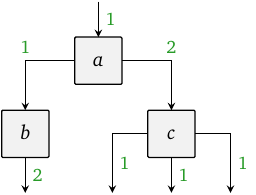}
    \end{subfigure}
    \begin{subfigure}[T]{0.49\linewidth}
    \centering
    \begin{equation*}
    \begin{array}{c|c|c}
    \text{Node} &  n_v,k_v,\ell_v &  G_v\\ \hline\hline
    a & 3,3,1 & \begin{pmatrix}0&0&1 \\ 1&1&0 \\ 0&1&1\end{pmatrix} \\ \hline
    b & 2,1,1 & \begin{pmatrix}1&1\end{pmatrix} \\ \hline
    c & 3,2,2 & \begin{pmatrix}1&0&1 \\ 0&1&1 \end{pmatrix}
    \end{array}
    \end{equation*}
    \end{subfigure}
    \caption{MPG representing the encoding channel $F_{\ell,G}$ introduced in \Cref{ex:mpg_example_2}. The green numbers depict the number of bits $n_e$ associated to each edge $e$.}
    \label{fig:mpg_example_2}
\end{figure}

Given an appropriate product distribution, we can now formalize the channel associated to the tree subspace decoding task, as depicted in \Cref{fig:W_decomposition}. 
\begin{definition}
    Consider an MPG $\mathcal{G}$ with $m$ leaves $e_1,\dots,e_m$ and let $\mathcal{P}=(P_i)_{i=1,\dots,m}$ be a sequence of $n_{e_i}$-bit distributions.
    We define the channel induced by $\mathcal{G}$ and $\mathcal{P}$ as
    \begin{equation}
        \mathcal{W}[\mathcal{G},\mathcal{P}] \coloneqq \left( \bigotimes_{i=1}^m \spsc[P_i] \right) \circ F[\mathcal{G}]\,,
    \end{equation}
    where the concatenation should be understood as passing the output of the leaf edge $e_i$ through the channel $\spsc[P_i]$.
\end{definition}
To end this section, we provide two illustrative examples.
\begin{example}\label{ex:mpg_example_repetition}
    Consider the $[n,1]$ repetition code given by the generator matrix $G=R_n\coloneq \begin{pmatrix}1 & 1 & \cdots & 1\end{pmatrix}\in\ff_2^{1\times n}$.
    The $n$-bit repetition code can be seen as a repeated concatenation of $2$-bit repetition codes.
    As depicted in \Cref{fig:mpg_example_repetition} for $n=5$, this allows us to express $F_{1,R_n}=F[\mathcal{G}]$ with an MPG $\mathcal{G}$ containing $n-1$ nodes.
    For every edge $e$ in the MPG, we have $n_e=1$, and for every node $v$ in the MPG, we have $n_v=2$, $k_v=\ell_v=1$ and $G_v=R_2$.
\end{example}
\begin{example}\label{ex:mpg_example_2}
    Consider the encoding channel $F_{\ell,G}$ for $\ell=1$ and the generator matrix
    \begin{equation}
        G = \begin{pmatrix}
        0 & 0 & 0 & 1 & 1 \\
        1 & 1 & 1 & 0 & 1 \\
        0 & 0 & 1 & 1 & 0 
        \end{pmatrix}
        \, .
    \end{equation}
    In \Cref{fig:mpg_example_2}, we depict an MPG $\mathcal{G}$ which represents $F_{\ell,G}$, i.e., $F_{\ell,G}=F[\mathcal{G}]$.
    We now verify that this MPG is indeed a valid representation.
    Consider $x$ to be the input bit.
    The node $a$ maps this bit to the vector
    \begin{equation}
        x \xlongrightarrow{a}
        x \left(\begin{array}{@{}c@{}} 0 \\ \hline 0 \\ 1 \end{array}\right) +
        r_1 \left(\begin{array}{@{}c@{}} 1 \\ \hline 1 \\ 0 \end{array}\right) +
        r_2 \left(\begin{array}{@{}c@{}} 0 \\ \hline 1 \\ 1 \end{array}\right)
    \end{equation}
    where $r_1,r_2$ are uniformly random bits.
    The horizontal line in the vectors depicts how the three bits get split up, i.e., the first bit is then sent to $b$ and the last two bits are sent to $c$.
    The nodes $b$ and $c$ act on the vector as follows
    \begin{equation}
        \xlongrightarrow{b}
        x \left(\begin{array}{@{}c@{}} 0 \\ 0 \\ \hline 0 \\ 1 \end{array}\right) +
        r_1 \left(\begin{array}{@{}c@{}} 1 \\ 1 \\ \hline 1 \\ 0 \end{array}\right) +
        r_2 \left(\begin{array}{@{}c@{}} 0 \\ 0 \\ \hline 1 \\ 1 \end{array}\right)
        \xlongrightarrow{c}
        x \left(\begin{array}{@{}c@{}} 0 \\ 0 \\ \hline 0 \\ \hline 1 \\ \hline 1 \end{array}\right) +
        r_1 \left(\begin{array}{@{}c@{}} 1 \\ 1 \\ \hline 1 \\ \hline 0 \\ \hline 1 \end{array}\right) +
        r_2 \left(\begin{array}{@{}c@{}} 0 \\ 0 \\ \hline 1 \\ \hline 1 \\ \hline 0 \end{array}\right)
        \, .
    \end{equation}
    The three vectors in the last expression exactly correspond to the rows of $G$, so the combined action of $a$, $b$ and $c$ precisely corresponds to $F_{\ell,G}$.
\end{example}

\subsection{Incoherent BPQM algorithm}\label{sec:incoherent_bpqm}
Incoherent BPQM is a quantum algorithm in which messages are passed along the edges of an MPG $\mathcal{G}=(V,E)$. 
Messages travel from the leaves towards the root, i.e., against the direction of the edges. 
At each node, the outgoing edges correspond to input messages and the incoming edge corresponds to the output message. 
The message passed over the edge $e\in E$ consists of two parts: an $n_e$-qubit quantum system $Q_e$ and a (classical) description of an $n_e$-bit probability distribution $P_e$ (e.g.\ a vector of $2^{n_e}$ probabilities $P_e(x)$ for $x\in\mathbb{F}_2^{n_e}$).
Each node processes its output message once all incoming messages have been received.

\begin{figure}
    \centering
    \includegraphics{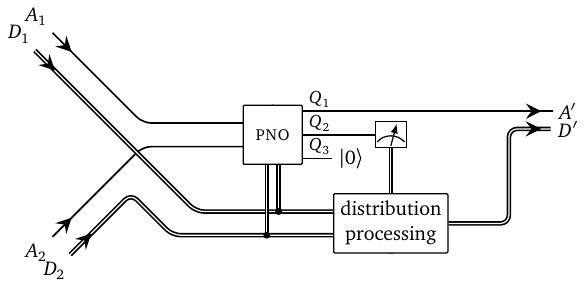}
    \caption{Quantum circuit representation of the incoherent BPQM node operation for some node $v$ wich $m_v=2$.
    Here, double-edged lines represent classical registers and single-edged lines quantum registers.
    The depicted node receives two input messages (with quantum part $A_i$ and classical part $D_i$ for $i=1,2$ respectively).
    The joint quantum parts are passed through the primitive node operation (the $\pno$ unitary), which depends on the values of $D_1$ and $D_2$.
    The resulting quantum system $Q_3$ is always in the $\ket{0}$ state (see proof of \Cref{lem:incoherent_bpqm_intermediate_state})  and the system $Q_2$ is measured in the computational basis.
    Given the measurement outcome, the classical part of the input messages is processed into a new distribution $D'$.
    The output message consists of the quantum system $A'=Q_1$ and the classical distribution $D'$.
    }
    \label{fig:incoherent_node_op}
\end{figure}

So to define incoherent BPQM, we first specify how each node $v\in V$ processes its input messages coming from the outgoing edges in order to generate a message that is sent over its incoming edge.
This node operation essentially performs the primitive node operation from \Cref{def:pno} with the generator matrix $G=G_v$. \Cref{fig:incoherent_node_op} depicts the steps of the node operation, including measurement and classical registers. 
\begin{definition}[Node operation]\label{def:nodeop}
Consider a node $v$ of some MPG.
\begin{problemdescr}
    \item[Input:] For each outgoing edge $e^{\mathrm{out}}_i(v)$, $i=1,\dots,m_v$: a message consisting of an $n_{e^{\mathrm{out}}_i(v)}$-qubit system $Q_{e^{\mathrm{out}}_i(v)}$ and an $n_{e^{\mathrm{out}}_i(v)}$-bit distribution $P_{e^{\mathrm{out}}_i(v)}$.
    \item[Output:] An $n_{e^{\mathrm{in}}(v)}$-qubit system $Q_{e^{\mathrm{in}}(v)}$ and an $n_{e^{\mathrm{in}}(v)}$-bit distribution $P_{e^{\mathrm{in}}(v)}$.
    \item[Algorithm:]
    Denote the joint $n_v$-qubit quantum system $\bar Q\coloneqq\bigotimes_{i=1}^{m_v}Q_{e^{\mathrm{out}}_i(v)}$ and the joint $n_v$-bit distribution $\bar P \coloneqq P_{e_1^{\mathrm{out}}(v)}\times P_{e_2^{\mathrm{out}}(v)} \times \dots \times P_{e_{m_v}^{\mathrm{out}}(v)}$
    \begin{enumerate}
         \item Set $U=\pno(n_v,k_v,\ell_v,G_v,\bar P)$ and apply $U$ to the quantum system $\bar Q$.
        Partition the resulting quantum system into three parts $Q_1\otimes Q_2\otimes Q_3$ consisting of $\ell_v,(k_v-\ell_v),(n_v-k_v)$ qubits each.
        \item Measure $Q_2$ in the computational basis, obtaining the result $s\in\ff_2^{k_v-\ell_v}$.
        \item Set the output message to be the quantum system $Q_1$ together with the distribution $P_{Y|S=s}$, where $P_{Y,S,A}(y,s,a)$ is defined as in \Cref{eq:PYSA} using the $n_v\times n_v$ matrix $M$ from $U$.
    \end{enumerate}
\end{problemdescr}
\end{definition}

The incoherent BPQM algorithm simply performs this node operation on each node of the MPG, followed by a measurement of the final quantum message. 
\begin{definition}[Incoherent BPQM algorithm]\label{def:incoherent_bpqm}
Consider an instance $\sdt(n,k,\ell,G,P)$ where the associated channel $W$ can be written as $W=\mathcal{W}[\mathcal{G},\mathcal{P}]$ for some MPG $\mathcal{G}$ and a sequence  $\mathcal{P}=(P_i)_{i=1,\dots,m}$ of probability distributions.
\begin{problemdescr}
    \item[Input:] An $n$-qubit quantum system denoted by $Q$ in the state $W(x)$, where $x\in\ff_2^\ell$.
    \item[Output:] An estimate $\hat x\in \ff_2^{\ell}$ for $x$. 
    \item[Algorithm:]
    Denote the root edge of $\mathcal{G}$ by $e^{\star}$.
    Partition the quantum system into subsystems $Q=\bigotimes_{i=1}^mQ_i$ where $Q_i$ corresponds to the output of $\spsc[P_i]$.
    Accordingly denote the leaf edges of $\mathcal{G}$ by $e_1,\dots,e_m$.
    \begin{enumerate}
        \item For $i=1,\dots,m$, fix the message associated with the edge $e_i$ to be the quantum system $Q_i$ together with the distribution $P_i$.
        \item Proceeding from leaves to root, perform node operations according to \Cref{def:nodeop} until all edges have an associated message.
        \item Apply the Hadamard transform $H^{\otimes n_{e^{\star}}}$ on the quantum system $Q_{e^{\star}}$ of the final message.
        \item Measure $Q_{e^{\star}}$ in the computational basis and set $\hat x$ to the resulting measurement outcome. 
    \end{enumerate}
\end{problemdescr}
\end{definition}

\subsection{Efficiency and optimality of incoherent BPQM}
The runtime of incoherent BPQM is governed by two factors: The number of nodes in the MPG as well as the following quantity.
\begin{definition}[Maximum variable dimension]
  The maximum variable dimension of an MPG $\mathcal{G}=(V,E)$ is defined as $N(\mathcal{G}) \coloneqq \max_{v\in V}n_v$.
\end{definition}
\begin{proposition}\label{prop:incoherent_bpqm_complexity}
    The incoherent BPQM algorithm operating on some MPG $\mathcal{G}=(V,E)$ has a time complexity of $\mathcal{O}(\abs{V}4^{N(\mathcal{G})})$.
\end{proposition}
\begin{proof}
    The number of performed node operations is $\abs{V}$ and the complexity of the node operation associated with $v\in V$ is generally $\mathcal{O}(4^{n_v})$, as this is the circuit complexity to implement a general $n_v$-qubit unitary~\cite{mottonen_multiqubitgates_2004}.
    The time complexity associated with the classical computation scales as $\mathcal{O}(2^{n_v})$.
\end{proof}
We note that if the MPG is well balanced, and node operations are allowed to be performed in parallel, then the $\abs{V}$ term in the time complexity of the algorithm can potentially be reduced to a sub-linear term.

We can already appreciate the efficiency of incoherent BPQM by considering the $n$-bit repetition code we encountered in \Cref{ex:mpg_example_repetition}.
Naively implementing the primitive node operation solving the entire subspace decoding task would entail a complexity of $\mathcal{O}(4^n)$.
By using incoherent BPQM on the proposed MPG, the complexity instead reduces to $\mathcal{O}(n)$ since $\abs{V}=n-1$ and $N(\mathcal{G})=2$.

Now we turn to our main result, the optimality of incoherent BPQM.
\begin{theorem}\label{thm:incoherent_bpqm_optimality}
    Consider an instance $T=\sdt(n,k,\ell,G,P)$ where the associated channel $W$ can be written as $W=\mathcal{W}[\mathcal{G},\mathcal{P}]$ for some MPG $\mathcal{G}$ with $m$ leaves and a sequence of probability distributions $\mathcal{P}=(P_i)_{i=1,\dots,m}$.
    Incoherent BPQM determines the input $x$ in $T$ with optimal (i.e., the lowest possible) error probability. 
\end{theorem}

To prove this theorem, we need to characterize the messages that are sent over the various edges of the MPG.
Since the measurement of $Q_2$ makes the incoherent BPQM node operations inherently stochastic, the values of the messages are not deterministic and must instead be described by a distribution over their possible values.
We need some technical groundwork before we can arrive to such a characterization, which will be stated in \Cref{lem:incoherent_bpqm_intermediate_state} below.

We define an \emph{$\ell$-bit distribution ensemble} $\mathcal{E}$ to be a list of pairs $\mathcal{E}=(p_i,D_i)_{i=1,\dots,m}$ where $p_i\geq 0$ are probabilities, so that $\sum_{i=1}^mp_i=1$, and $D_i$ are $\ell$-bit probability distributions. 
Next, we define the product $\mathcal{E}_1\otimes\mathcal{E}_2$ of an $\ell_1$- and $\ell_2$-bit ensemble $\mathcal{E}_1=(p_i^{(1)},D_i^{(1)})_i,\mathcal{E}_2=(p_j^{(2)},D_j^{(2)})_j$ to be the $(\ell_1+\ell_2)$-bit distribution ensemble $(p_i^{(1)}p_j^{(2)},D_i^{(1)}\times D_j^{(2)})_{i,j}$. 
A distribution ensemble of particular importance is the one which describes the possible residual channels that occur in a primitive node operation.
\begin{definition}[Node message ensemble]
    Consider a primitive node operation $U=\pno(n,k,\ell,G,P)$ and its associated $n\times n$ matrix $M$. 
    We define the distribution ensemble
    \begin{equation}
        f_{U}(D) \coloneqq (P_S(s),P_{Y|S=s})_{s\in\ff_2^{k-\ell}}\,,
    \end{equation}
    where $P_{Y,S,A}$ is defined as in \Cref{def:pno}, i.e., $P_{Y,S,A}(y,s,a)\coloneqq D(M^{-1}(y\cat s\cat a))$.
\end{definition}
Combining the node message ensembles yields the message ensemble for the root edge of the BPQM algorithm.
\begin{definition}[BPQM message ensemble]
\label{def:bpqm_message_ensemble}
    Consider an MPG $\mathcal{G}=(V,E)$ with $m$ leaves $e_1,\dots,e_m$ and a sequence $\mathcal{P}=(P_i)_{i=1,\dots,m}$ of $n_{e_i}$-bit probability distributions.
    The \emph{message ensemble} of $\mathcal{G}$ w.r.t.\ $\mathcal{P}$, denoted by $\mensemble_{\mathcal{G}}^{\mathcal{P}}$, is the $(k-\ell)$-bit distribution ensemble defined inductively as follows. 
    \begin{itemize}
        \item If $\mathcal{G}$ has zero nodes (and thus $\mathcal{P}=(P_1)$), then $\mensemble_{\mathcal{G}}^{\mathcal{P}}$ is given by the list with only one element $(1,P_1)$.
        \item If $\mathcal{G}$ has at least one node, then let $v$ be the node of $\mathcal{G}$ connected to the root edge.
        Each outgoing edge $e^{\mathrm{out}}_i(v)$ for $i=1,\dots,m_v$ can be considered to be the root of a smaller MPG $\mathcal{G}_i$ which only includes $e^{\mathrm{out}}_i(v)$ and all of its successor nodes and edges in $\mathcal{G}$.
        Define the ensemble $\mathcal{E}_i\coloneqq\mensemble_{\mathcal{G}_i}^{\mathcal{P}_i}$ where $\mathcal{P}_i$ only contains the distributions associated to the leaves of $\mathcal{G}_i$.
        Let us write the product ensemble as $\bigotimes_{i=1}^{m_v}\mathcal{E}_i = (\bar p_j,\bar D_j)_j$.
        For each $\bar D_j$, we write $f_{U}(\bar D_j)=(p_{k|j}',D_{k|j}')_k$ where $U=\pno(n_v,k_v,\ell_v,G_v,\bar{D}_j)$\footnote{Note that technically speaking $f_U$ depends on the choice of the matrix $M$. The exact choice if of no further consequence, except that the same choice needs to be used for each $j$.}.
        Then, define
        \begin{equation}
            \mensemble_{\mathcal{G}}^{\mathcal{P}}\coloneqq (\bar p_j p_{k|j}', D_{k|j}')_{j,k} \ .
        \end{equation}
    \end{itemize}
\end{definition}
\begin{lemma}\label{lem:incoherent_bpqm_intermediate_state}
    Consider the setup of \Cref{thm:incoherent_bpqm_optimality}. 
    Denote the root of $\mathcal{G}$ by $e^{\star}$ and let $\mensemble_{\mathcal{G}}^{\mathcal{P}}=(p_s,D_s)_s$ be the BPQM message ensemble from \Cref{def:bpqm_message_ensemble}. 
    Applying incoherent BPQM to $W(x)$ results in a final message sent over $e^{\star}$ which is given by a quantum system in the state $\spsc[D_s](x)$ and the classical distribution $D_s$ with probability $p_s$.
\end{lemma}
\begin{proof}
The proof proceeds by induction over the number of nodes in the MPG $\mathcal{G}$, as in the construction of $\mensemble_{\mathcal{G}}^{\mathcal{P}}$ from \Cref{def:bpqm_message_ensemble}. 
    If $\mathcal{G}$ has no nodes, then the statement is true by definition.
    For the induction step, consider an MPG $\mathcal{G}$ with at least one node, and let $v$ denote the node connected to the root and $F_{\ell_v,G_v}$ the associated randomized encoding.
    For an input $x\in \ff_2^{n_{e^\star}}$, the output of $F_{\ell_v,G_v}$ is $(x\cat r)G_v$ for a randomly-chosen $r\in \ff_2^{k_v-\ell_v}$. 
    Denote the portion of this string associated to outgoing edge $e^{\mathrm{out}}_i(v)$ by $x_i$. 
    Each outgoing edge $e^{\mathrm{out}}_i(v)$ for $i=1,\dots,m_v$ can be considered the root of a smaller MPG $\mathcal{G}_i$ which includes only its successor nodes and child edges in the original MPG. 
    The input to $\mathcal G_i$ is $x_i$.

    We apply the induction hypothesis on these smaller sub-MPGs. 
    Define the ensemble $\mathcal{E}_i\coloneqq\mensemble_{\mathcal{G}_i}^{\mathcal{P}_i}$ where $\mathcal{P}_i$ only contains the distributions associated to the leaves contained in $\mathcal{G}_i$.
    By assumption, the message sent over the edge $e_i^{\mathrm{out}}$ is precisely characterized by the ensemble $\mathcal{E}_i=(p_{\ell|i},D_{\ell|i})_\ell$ in that the classical part of the message is $D_{\ell|i}$ and the quantum part $\spsc[D_{\ell|i}](x_i)$ with probability $p_{\ell|i}$. 
    Let us write the product ensemble as $\bigotimes_{i=1}^{m_v}\mathcal{E}_i = (\bar p_j,\bar D_j)_j$.
    The combined quantum message going into the node is hence given by the state
    \begin{equation}
        \frac{1}{2^{k_v-\ell_v}} \sum_{r\in\ff_2^{k_v-\ell_v}} \spsc[\bar D_j]((x\cat r)^TG)
    \end{equation}
    with probability $\bar p_j$.

    The first step of the node operation is application of the unitary $U=\pno(n_v,k_v,\ell,G_v,\bar D_j)$ to this state.
    By \Cref{lem:node_operation}, the resulting state is precisely given by
    \begin{equation}\label{eq:incoherent_bpqm_cq_state}
        \sum_{k\in\ff_2^{k_v-\ell_v}}  p_{k|j}' \spsc[D_{k|j}'](x)_{Q_1} \otimes \proj{k}_{Q_2} \otimes \proj{0}^{\otimes n_v-k_v}_{Q_3}\,,
    \end{equation}
    where we write $f_U(\bar D_j)=(p_{j|j}',D_{k|j}')_k$.
    In the next step, $Q_2$ is measured to obtain the value of $k$.
    As such, the output message is precisely given by $\spsc[D_{k|j}'](x)$ and $D_{k|j}'$ with probability $\bar p_jp_{k|j}'$. Thus the induction step is proven.
\end{proof}
\begin{proof}[Proof of \Cref{thm:incoherent_bpqm_optimality}]
    From \Cref{eq:incoherent_bpqm_cq_state}, we can observe that for each node operation, the state of the system $Q_1Q_2$ is given by a quantum-classical state.
    As such, the measurement of the system $Q_2$ in the standard basis is a reversible (i.e. it does not disturb the state) and no information is lost through its action.
    
    Together with the statement of \Cref{lem:optimal_mmt}, we conclude that the sequence of all node operations reversibly maps $W(x)$ to the quantum-classical state
    \begin{equation}
        \sum_{(p_s,D_s)\in\mensemble_{\mathcal{G}}^{\mathcal{P}}} p_s \spsc[D_s](x)\otimes \proj{s}\otimes \proj{0}^{\otimes n-k} \, .
    \end{equation}
    Therefore we may just as well determine $x$ from this state as from $W(x)$.
    By \Cref{lem:optimal_mmt} the optimal strategy to guess $x$ from this quantum-classical state is simply to perform a measurement of the first $k$ qubits in the conjugate basis.

    We briefly note that the reversibility of the node operations generally only holds for input states of the form $W(x)$.
    However, we will see later in \Cref{sec:bpqm_decoding} that it is possible to modify BPQM so as to realize the node operations reversibly for any input state.
\end{proof}
From the proof of \Cref{thm:incoherent_bpqm_optimality}, we can also directly find a closed-form expression for the success probability of incoherent BPQM.
\begin{corollary}
    Consider an instance $\sdt(n,k,\ell,G,P)$ where the associated channel $W$ can be written as $W=\mathcal{W}[\mathcal{G},\mathcal{P}]$ for some MPG $\mathcal{G}$ and a sequence of probability distributions $\mathcal{P}=(P_i)_{i=1,\dots,m}$.
    The optimal success probability of solving the subspace decoding task is given by
    \begin{equation}\label{eq:optimal_succ_prob}
        p_{\mathrm{succ}} = \sum_{(p,D)\in \mensemble_{\mathcal{G}}^{\mathcal{P}}} p2^{H_{1/2}(D) - \ell}
    \end{equation}
    where $H_{1/2}(D)\coloneqq 2\log\sum_{z\in\ff_2^{\ell}}\sqrt{D(z)}$ is the Rényi entropy of order $1/2$.
\end{corollary}
\begin{proof}
    The probability of correctly obtaining $x$ from $\spsc[D](x)$ is given by
    \begin{equation}
        \Big|\bra{+}^{\otimes\ell} \sum_{z\in\ff_2^{\ell}} \sqrt{D(z)} \ket{z}\Big|^2 = \frac{1}{2^\ell} \Big(\sum_{z,z'}\sqrt{D(z)}\braket{z'}{z} \Big)^2 = 2^{H_{1/2}(D)-\ell}
        \, .
    \end{equation}
    Averaging over the choice of $D$ in $\mathcal E_{\mathcal G}^{\mathcal P}$ gives the desired statement. 
\end{proof}

We briefly note that while incoherent BPQM is a quantum algorithm with quantum input, and thus requires an actual quantum device to execute, it is possible to efficiently evaluate its success probability on a classical machine.
Since BPQM only operates on non-entangled quantum systems of size at most $N(\mathcal{G})$ qubits, the classical cost to evaluate BPQM's performance scales as $\mathcal{O}(\abs{V}2^{N(\mathcal{G})})$, which is essentially identical to the runtime of BPQM itself on a quantum computer.

\subsection{Robustness to discretization}\label{sec:discretization}
In the previous discussion of incoherent BPQM, we made the implicit assumption that the classical part of the computation, which involves real numbers, can be realized with infinite accuracy.
Such assumptions are common in the study of message passing algorithms, though one must ensure that the algorithm remains robust when these real numbers are approximated with a discretized representation.
It is especially important to formally define and analyze a properly discretized version of incoherent BPQM, since, as we will see later in \Cref{sec:bpqm_decoding}, it can be useful to perform the classical computation coherently on a quantum register.

There are essentially two aspects of the incoherent BPQM algorithm which are idealized and need some discretization procedure to work in practice.
Firstly, the classical part of the messages are real-valued, and would in principle require an infinite number of classical bits to represent.
Secondly, as part of every primitive node operation, a circuit for the unitary $U_2$ must be explicitly (classically) constructed, which, as opposed to $U_1$, generally also involves real numbers.

We address these issues by approximating the classical part of the message in a discretized fashion.
We consider the arguably simplest discretization scheme to simplify the analysis: Store each number in the probability vector using a fixed-point representation that utilizes $B$ bits to represent a real number in $[0,1]$.
The $\ell_1$-norm error caused by the discretized representation of the distribution is at most $2^{N(\mathcal{G})}/2^B$, since $2^{N(\mathcal{G})}$ is the size of the largest probability vector that ever occurs in the incoherent BPQM algorithm and each entry is off by at most $2^{-B}$.
Therefore, to achieve an $\ell_1$-norm error of $\Delta$ on all distributions, we require a discretized representation with a total number of $B=N(\mathcal G)+\log(1/\Delta)$ bits for each probability. 

The full details of the discretized incoherent BPQM algorithm are rather technical and not of particular importance for most considerations.
We therefore refer the interested reader to \Cref{app:discretized_incoherent_bpqm}, and for now, the following informal definition will suffice.
\begin{definition}[discretized incoherent BPQM, informal]\label{def:discretized_incoherent_bpqm_informal}
    The discretized incoherent BPQM algorithm is defined identically to \Cref{def:incoherent_bpqm}, except that any real number occurring in the classical computation is represented in a discretized fashion.
\end{definition}
We now briefly discuss the impact that discretization errors have on the accuracy and runtime of BPQM.
Roughly speaking, the analysis of the discretization errors has to consider two aspects.
First, since we store the classical messages in a quantized fashion, they will accumulate more and more errors during the execution of the algorithm.
In turn, these errors also affect the quantum part of the message, since each node operation applies a unitary $\pno(n,k,\ell,G,P)$ with an erroneous value of $P$.
Fortunately, it can be shown that these errors can be efficiently suppressed by choosing $B$ to be large enough.
\begin{theorem}\label{thm:bit_optimal_discretized}
    Consider an instance $\sdt(n,k,\ell,G,P)$ where the associated channel $W$ can be written as $W=\mathcal{W}[\mathcal{G},\mathcal{P}]$ for some MPG $\mathcal{G}=(V,E)$ and a sequence of probability distributions $\mathcal{P}=(P_i)_{i=1,\dots,m}$.
    For any $\epsilon >0$, discretized incoherent BPQM solves this problem with a success probability that is at most $\epsilon$ smaller than the optimal probability when choosing $B=\mathcal{O}(N(\mathcal{G})+\log(1/\epsilon)+\abs{V}\log(N(\mathcal{G})))$.
    The runtime of the algorithm is $\mathcal{O}(4^{N(\mathcal{G})}N(\mathcal{G})^2\abs{V}\mathrm{poly}(B))$.
\end{theorem}
The proof is delegated to \Cref{app:discretization}.
We briefly note that the exact polynomial in the complexity term depends on implementation details of the discretization scheme, such as the precise asymptotic cost of performing basic operations (addition, division, square root, arcsin) on discretized $B$-bit fixed-point numbers.
By appropriately choosing the algorithms to realize these basic operations, the polynomial can be quadratic~\cite{burge_quantum_2024}.
We also expect our estimate of the required number of bits $B$ to be overly pessimistic and that improved techniques and a more refined analysis could lower it.

\section{Decoding the pure-state channel with BPQM}\label{sec:bpqm_decoding}
In this section, we will show how BPQM can be utilized to solve the following block decoding task.
Recall that for $p\in[0,1/2]$, we define $\bspsc[p]\coloneqq\spsc[(1-p,p)]$.
\begin{problem}[Block decoding problem $\blockdec(\mathcal{C},p)$]\label{prob:block_decoding}
\hfill
\begin{problemdescr}
    \item[Instance parameters:] An $[n,k]$ binary linear code $\mathcal{C}$ and a vector $p\in[0,1/2]^n$.
    \item[Input:] A codeword $x\in\mathcal{C}$ is chosen uniformly at random and then transmitted over a collection of independent binary-input pure-state channels, yielding the state
    \begin{equation}
      \ket{\Psi_x} \coloneq \bspsc[p_1](x_1)\otimes \bspsc[p_2](x_2)\otimes\dots\otimes \bspsc[p_n](x_n) \, .
    \end{equation}
    The input is an $n$-qubit quantum system in the state $\ket{\Psi_x}$.
    \item[Goal:] Given the input quantum system, determine $x$.
\end{problemdescr}
\end{problem}
Before solving $\blockdec$, let us first discuss the comparatively simpler task of bit decoding.
\begin{problem}[Bit decoding problem $\bitdec(\mathcal{C},p,i)$]
\hfill
\begin{problemdescr}
    \item[Instance parameters:] An $[n,k]$ binary linear code $\mathcal{C}$, a vector $p\in[0,1/2]^n$ and $i\in\{1,\dots,n\}$.
    \item[Input:] identical to the input of \Cref{prob:block_decoding}.
    \item[Goal:] Given the input quantum system, determine $x_i$.
\end{problemdescr}
\end{problem}
Clearly, $\bitdec(\mathcal{C},p,i)$ is an instance of the subspace decoding problem $\sdt(n,k,\ell,G,P)$ considered in \Cref{sec:node_op} with $\ell=1$, $P=(1-p_1,p_1)\times\dots\times(1-p_n,p_n)$ and with an appropriately chosen generator matrix $G$ of $\mathcal{C}$ where only the first row has a non-zero entry at the $i$-th position.\footnote{I.e., $G_{1i}=1$ and $G_{ji}=0$ for $j>1$. This can be achieved by appropriately applying elementary row operations on any given generator matrix of $\mathcal{C}$.}
As such, we can optimally solve this problem with incoherent BPQM as long as we are given an MPG representation of the following channel.

\begin{definition}[Bit-transmission channel]
    Let $\mathcal{C}$ be an $[n,k]$ binary linear code and $i\in\{1,\dots,n\}$.
    We define the bit-transmission channel $\mathcal{F}_{\mathcal{C},i}$ as the $1$-bit to $n$-bit channel, which maps the bit $z\in\{0,1\}$ to a uniformly random element in $\{x\in\mathcal{C} | x_i=z\}$.
\end{definition}
\begin{corollary}
    Consider an instance $\bitdec(\mathcal{C},p,i)$.
    If $\mathcal{G}$ is an MPG such that $F[\mathcal{G}]=\mathcal{F}_{\mathcal{C},i}$, then incoherent BPQM on $\mathcal{G}$ optimally solves the $\bitdec(\mathcal{C},p,i)$ problem.
\end{corollary}
Recall that the runtime complexity of incoherent BPQM is $\mathcal{O}(4^{N(\mathcal{G})}\abs{V})$ where $\mathcal{G}=(V,E)$.
Hence, to efficiently and optimally solve the $\bitdec$ problem, it suffices to find efficient MPG representations of the bit-transmission channel.
In \Cref{sec:codes}, we will see how this can be achieved for different families of codes.

We now turn our attention back to block decoding.
In principle, the $\blockdec(\mathcal{C},p)$ problem is also an instance of the subspace decoding task with $\ell =k$, and could thus be solved by incoherent BPQM.
However, since the maximum variable dimension of the involved MPG is necessarily at least $k$ (because the root edge $e^{\star}$ of the MPG must have $n_{e^{\star}}=k$), incoherent BPQM has an exponentially scaling runtime when $k=\Omega(n)$.

To decode codes with non-vanishing rate, we must therefore take another approach.
Taking inspiration from classical BP, we instead decode each codeword bit separately.
We will therefore assume that for every codeword bit $x_i$ (or at least for $k$ linearly independent codeword bits) we know an efficient MPG representation of the bit-transmission channel $\mathcal{F}_{\mathcal{C},i}$, such that we can efficiently decode said codeword bit.

For block decoding with BPQM, there are two important distinctions from classical BP that stem from the quantum nature of the problem.
First, in contrast to classical BP, it is not possible to decode all codeword bits in parallel. 
That would require duplication of the channel outputs, which is forbidden by the no-cloning theorem.
Instead we decode the individual codeword bits sequentially one at a time.
Second, we cannot naively apply the incoherent BPQM algorithm sequentially. 
Measurements in the node operations of incoherent BPQM may significantly disturb the quantum state, potentially impeding the sufccessful decoding of subsequent bits.

For this reason, we need to modify incoherent BPQM to be as minimally destructive as possible.
We cannot eliminate the measurement of the final message, since we need to extract one classical bit out of the decoding procedure. 
But all node operations can be performed completely coherently by utilizing the deferred measurement principle.
We call this modification of BPQM with reversible node operations \emph{coherent BPQM} to highlight its difference with incoherent BPQM.

After decoding a single codeword bit with coherent BPQM, the node operations can be reversed in order to return to a state that is as close as possible to the original state.
Remarkably, it turns out that sequentially decoding in this fashion is block-optimal, i.e., it allows us to estimate the codeword with the maximal success probability.
This will be proven in \Cref{sec:block_optimality}.

\subsection{Coherent BPQM algorithm}\label{sec:bpqm_coherent}
The incoherent BPQM algorithm described in \Cref{sec:incoherent_bpqm} involves two types of measurements: The conjugate basis measurement on the final message as well as the measurements of the register $Q_2$ in each node operation. 
The measurements of $Q_2$ can be avoided by using the deferred measurement principle.
To do so, we first need to upgrade the classical registers storing the distributions ($D_1$, $D_2$ and $D'$ in \Cref{fig:incoherent_node_op}) to quantum registers.
Then, the primitive node operation ($\pno$ in the \Cref{fig:incoherent_node_op}) must be coherently controlled on the values of $D_1$ and $D_2$.
Furthermore, the classical computation (the ``distribution processing'' box in the \Cref{fig:incoherent_node_op}) needs to be translated from a classical circuit to a quantum circuit.
Finally, instead of measuring the $Q_2$ register coming out from $\pno$, the distribution processing step must be coherently controlled on $Q_2$.

We call the resulting algorithm \emph{coherent BPQM}.
Due to the deferred measurement principle, coherent BPQM solves the subspace decoding task with the same probability as incoherent BPQM.
It can hence realize optimal bit-wise decoding.
The drawback is that now we offload some of the classical computation to the quantum domain, so additional quantum memory and computation is required.
However, the overall (classical and quantum) asymptotic complexity of the algorithm can be shown to remain unchanged.

In the discussion of incoherent BPQM, we distinguished between ideal incoherent BPQM and discretized incoherent BPQM, where in the latter the classical part of the message is stored in a discretized fashion using only a finite number of bits.
It is important to note that, strictly speaking, coherent BPQM is only a well-defined algorithm when based on the discretized incoherent BPQM algorithm.
Indeed, without discretization, the register containing the description of the distribution ($D_i$ in \Cref{fig:incoherent_node_op}) would need to be infinite dimensional to reliably represent the infinite number of admissible distributions\footnote{It could be argued that incoherent BPQM without discretization is also an ill-defined algorithm in the first place, as it requires infinite precision to represent the classical part of the message. However, making this assumption of a real RAM computation model is generally more common for classical algorithms.}.
\begin{definition}[Coherent BPQM algorithm]\label{def:coherent_bpqm}
    The coherent BPQM algorithm is identical to the discretized incoherent BPQM algorithm, except that all measurements, besides the final one, are avoided by using the deferred measurement principle.
\end{definition}
For a more detailed step-by-step description of coherent BPQM, as well as a discussion of its circuit complexity, we refer the reader to \Cref{app:coherent_bpqm}.

This situation is somewhat cumbersome, as it leaves us without a precise notion of an ``idealized coherent BPQM'' algorithm, of which the discretized version would be an approximation.
We address this in \Cref{app:formalization_coherent_bpqm} by introducing another algorithm, called \emph{uniformly-controlled BPQM} (UC-BPQM).
UC-BPQM is a generalization of the algorithm initially introduced in \cite{renes_belief_2017}. 
It is an inefficient algorithm, but does formalize the notion of what an ideal coherent BPQM computation (without discretization errors) would be.
The main idea of UC-BPQM is to not keep track of the distribution information of the state on-the-fly, but instead simply condition the primitive node unitaries on all (classical) bits produced by preceding node operations.

\subsection{Block-optimality of sequential BPQM decoding}\label{sec:block_optimality}
Consider some instance $\blockdec(\mathcal{C},p)$.
To estimate $x$ from $\ket{\Psi_x}$, it suffices to correctly decode $k$ linearly independent codeword bits, which then fully determine $x$.
For convenience of notation, we will assume that the first $k$ codeword bits $x_1,\dots,x_k$ are linearly independent, which can always be achieved by appropriate relabeling.
We assume that for each $i=1,\dots,k$ an MPG $\mathcal{G}_i$ describing the bit-transmission channel $\mathcal{F}_{\mathcal{C},i}$ is given.

Notice that coherent BPQM w.r.t. $\mathcal{G}_i$ utilizes ancilla scratch registers in the zero state during its execution.
\footnote{For example, these ancilla qubits are used to store the distribution part of the messages.}
Denote the total number of such ancilla qubits by $A_i$.
Therefore, the action of the coherent BPQM algorithm w.r.t. $\mathcal{G}_i$ on the system $Q$ in state $\ket{\Psi_x}$ can be written as a unitary evolution
\begin{equation}
    \ket{\Psi_x}_Q\otimes\ket{0^{A_i}}_{E_i} \mapsto U_{\mathrm{cBPQM},i} \left( \ket{\Psi_x}_Q\otimes\ket{0^{A_i}}_{E_i} \right)
\end{equation}
for some ${A_i}$-qubit ancilla system $E_i$, followed by some projective measurement $\{\Pi_i,\id-\Pi_i\}$ acting on $QE$.
The $(n+A_i)$-qubit unitary $U_{\mathrm{cBPQM},i}$ captures all of the node operations performed by the algorithm, including initialization of the classical part of the message at the leaves into the ancilla. 
Correspondingly, the projection $\Pi_i$ captures the measurement of the final qubit in the conjugate basis.

To decode the complete codeword with BPQM, we sequentially apply the coherent BPQM algorithm for the codeword bits $1,\dots,k$.
After each codeword bit $x_i$ has been decoded, the node operations are ``undone'' by reversing the quantum circuit in order to return to a state that is as close as possible to the original channel output.
\begin{definition}[BPQM block decoding]\label{def:bpqm_block_decoding}
Consider an instance $\blockdec(\mathcal{C},p)$ for some $[n,k]$ binary linear code where $x_1,\dots,x_k$ are linearly independent codeword bits.
Let $\mathcal{G}_1,\dots,\mathcal{G}_k$ be MPG representations of $\mathcal{F}_{\mathcal{C},1},\dots,\mathcal{F}_{\mathcal{C},k}$.
\begin{problemdescr}
    \item[Input:] $n$-qubit quantum system $Q$ in the state $\ket{\Psi_x}$ for some $x\in\mathcal{C}$.
    \item[Output:] An estimate $\hat{x}$ of $x$.
    \item[Algorithm:] For $i=1,\dots,k$ perform the following actions
    \begin{enumerate}
        \item Prepare $A_i$-qubit ancilla register $E_i$ in the all-zero state
        \item Apply the coherent BPQM circuit $U_{\mathrm{cBPQM},i}$ to $QE_i$
        \item Set $\hat x_i$ to the result of measuring $\{\Pi_i,\id-\Pi_i\}$ on $QE_i$
        \item Apply the reversed coherent BPQM circuit $U_{\mathrm{cBPQM},i}^{\dagger}$ to $QE_i$
        \item Discard the register $E_i$
    \end{enumerate}
    Finally, complete $\hat{x}_1,\dots,\hat{x}_k$ to a valid codeword in $\mathcal{C}$.
\end{problemdescr}
\end{definition}

We observe that, somewhat surprisingly, BPQM not only realizes bit-optimal decoding, but also block-optimal decoding.
This is remarkable, since classical BP on the binary symmetric channel does not necessarily realize block-optimal decoding, even on codes with a tree factor graph.
\begin{theorem}\label{thm:block_optimal}
    Consider an instance $\blockdec(\mathcal{C},p)$ for an $[n,k]$ binary linear code $\mathcal{C}$.
    For $k$ linearly independent codeword bits $x_i$, let $\mathcal{G}_i=(V_i,E_i)$ be an MPG representation of $\mathcal{F}_{\mathcal{C},i}$.
	Define $N\coloneq\max_{i}N(\mathcal{G}_i)$, $M\coloneq\max_{i}\abs{V_i}$ and $\bar p\coloneqq\min_{j=1,\dots,n}p_j$.

    For any $\epsilon >0$, the algorithm in \Cref{def:bpqm_block_decoding} with $B$-bit fixed point representation solves $\blockdec(\mathcal{C},p)$ with a success probability that is at most $\epsilon$ smaller than the optimal probability when choosing $B=\mathcal{O}(N + M\log N + \log k + \log(1/\epsilon) + n\log(1 / \bar p))$.
	The circuit complexity of the algorithm is $\mathcal{O}(4^NN^2kM\,\mathrm{poly}(B))$.
\end{theorem}
We briefly note that the exact polynomial in the complexity term depends on implementation details of the discretization scheme, such as the precise asymptotic cost of performing basic operations (addition, division, square root, arcsin) on discretized $B$-bit fixed-point numbers.
By appropriately choosing the algorithms to realize these basic operations, the polynomial can be quadratic~\cite{burge_quantum_2024}.
We also expect our estimate of the required number of bits $B$ to be overly pessimistic and that improved techniques and a more refined analysis could lower it.
In practice, it is likely sufficient for most applications to simply fix a number of discretization bits $B$ that is large enough, as is commonplace for most classical numerical computing.
In that case, the considerations about the asymptotics simplify significantly, and the overhead reduces to $\mathcal{O}(4^NMk)$ (up to polynomial terms in $N$).

There are two main steps involved in the proof of this theorem.
First, we will argue that in the absence of discretization errors, the algorithm would be block optimal.
Second, we show that it is possible to efficiently approximate this ideal computation up to an error $\epsilon$.
We briefly sketch out the proof for the first step here, since some of the involved ideas are interesting on their own.
For the second step, we delegate the full details to \Cref{app:discretization}, since they are more technical and less insightful.

Due to the Abelian structure, the PGM realizes block-optimal decoding of binary linear codes sent over pure-state channels (recall \Cref{prop:PGMopt}).
In this case, the PGM is given by an orthogonal measurement into the orthonormal basis $\{\ket{f_x}\}_{x\in\mathcal{C}}$ where
\begin{equation}
    \ket{f_x} \coloneqq \frac{1}{\sqrt{2^k}}\rho^{-1/2}\ket{\Psi_x} \quad \text{ and }\quad  \rho \coloneqq \frac{1}{2^k}\sum_{x\in\mathcal{C}}\proj{\Psi_x} \, .
\end{equation}
To prove that a decoding scheme is block-optimal, it thus suffices to show that it acts equivalently to a projective measurement in the basis $\{\ket{f_x}\}_x$ on the subspace $V\coloneqq \mathrm{span}\{\ket{\Psi_x} | x\in\mathcal{C}\}$.

The following proposition shows that any optimal bit-wise decoding algorithm, which in some sense doesn't measure more than it actually needs to, can be chained into a block-wise optimal decoder which realizes the PGM.
\begin{proposition}\label{prop:bitoptimal_implies_blockoptimal}
    Consider an instance $\blockdec(\mathcal{C},p)$ for some $[n,k]$ binary linear code $\mathcal{C}$, and let $i\in\{1,\dots,n\}$.
    Define
    \begin{equation}
        \rho_0 \coloneqq \frac{1}{2^{k-1}}\sum_{x\in\mathcal{C}: x_i=0}\proj{\Psi_x}
        \, , \qquad
        \rho_1 \coloneqq \frac{1}{2^{k-1}}\sum_{x\in\mathcal{C}: x_i=1}\proj{\Psi_x}
        \, .
    \end{equation}
    Consider an orthogonal two-outcome projective $n$-qubit measurement $\{\Pi,\id-\Pi\}$ that optimally decodes the $i$-th codeword bit, i.e., it maximizes $\tr[\Pi \rho_0] + \tr[(\id -\Pi) \rho_1]$.
    Then, for $P$ the projection onto $V$,
    %the subspace $\mathrm{span}\{\ket{\Psi_x} | x\in\mathcal{C}\}$,
    \begin{equation}
    \label{eq:bitoptgoal}
        \Pi P = \sum_{x\in\mathcal{C}: x_i=0}\proj{f_x}\,. 
    \end{equation} 
\end{proposition}
\begin{proof}
    First, we consider the spectral decomposition of $(\rho_0-\rho_1)$.
    The orthogonal complement of $V$ is clearly an eigenspace with eigenvalue zero.
    The eigenvalues corresponding to eigenvectors in $V\coloneqq \mathrm{span}\{\ket{\Psi_x} | x\in\mathcal{C}\}$ are all non-zero, because for all $\ket{\phi}\in V$
    \begin{equation}
        \left( \rho_0-\rho_1 \right)\ket{\phi} = \frac{1}{2^{k-1}}\sum\limits_{x\in\mathcal{C}} (-1)^{x_i} \braket{\Psi_x}{\phi} \ket{\Psi_x}
    \end{equation}
    which can only be zero if $\ket{\phi}=0$ (due to the linear independence of the $\ket{\Psi_x}$).

    By \cite[Equation 2.41]{holevo_quantum_2019} every optimal measurement has the form $\Pi=\{\rho_0-\rho_1>0\}+\Lambda$, where $\{A>0\}$ denotes the projection onto the subspace of strictly positive eigenvalues of $A$ and $\Lambda$ is an operator supported on the kernel of $\rho_0-\rho_1$ such that $0\leq \Lambda\leq \id$. 
    Since the support of $\rho_0-\rho_1$ is $V$, we have $\{\rho_0-\rho_1>0\}P=\{\rho_0-\rho_1>0\}$, and similarly, $\Lambda P=0$.
    Hence, $\Pi P=\{\rho_0-\rho_1>0\}$ for all optimal $\Pi$. 

    Therefore, to complete the proof, it suffices to show that $\Pi'=\{\rho_0-\rho_1>0\}$ for $\Pi'\coloneq\sum_{x\in\mathcal{C}: x_i=0}\proj{f_x}$. 
    Let $G$ be a generator matrix for $\mathcal C$ adapted to the $\bitdec(\mathcal{C},p,i)$ problem, as described above. 
    Then define $U=\pno(n,k,1,G,P)$ for $P$ the distribution corresponding to the $n$ independent binary pure-state channels with parameters $p_1,\dots,p_n$.
    We will verify the equality $\Pi'=\{\rho_0-\rho_1>0\}$ under the basis transformation given by $U$.

    By \Cref{lem:node_operation}, the unitary $U$ transforms $\rho_i$ to the state 
    \begin{equation}
        U\rho_i U^\dagger = \sum_{s\in \ff_2^{k-1}} P_S(s) \spsc[P_{Y|S=s}](i)\otimes \proj{s}\otimes \proj{0}^{\otimes (n-k)}
        \,. 
    \end{equation}
    Therefore, we have
    \begin{equation}\label{eq:bit_block_proof_1}
        U\{\rho_0-\rho_1>0\}U^\dagger = \{U\rho_0 U^\dagger-U\rho_1 U^\dagger>0\}=\proj{+}\otimes \id^{\otimes (k-1)}\otimes \proj{0}^{\otimes (n-k)}\,. 
    \end{equation}
    Next, notice that we can write $\Pi'$ as follows
    \begin{equation}
        \Pi' = \sum_{x\in\mathcal{C}, x_i=0} \proj{f_x} = \frac{1}{2^k}\rho^{-1/2} \left( \sum_{x\in\mathcal{C}, x_i=0} \proj{\Psi_x} \right) \rho^{-1/2} = \frac{1}{2}\rho^{-1/2}\rho_0\rho^{-1/2}
        \, .
    \end{equation}
    Moreover, $U\rho U^\dagger=U(\tfrac{\rho_0+\rho_1}{2})U^{\dagger}$ is diagonal:  
    \begin{equation}
        U\rho U^\dagger =\sum_{s\in \ff_2^{k-1}}\sum_{y\in \ff_2} P_S(s)P_{Y|S=s}(y) \proj{y}\otimes \proj{s}\otimes \proj{0}^{\otimes (n-k)}
        \,. 
    \end{equation}
    Therefore, we observe
    \begin{equation}\label{eq:bit_block_proof_2}
        U\Pi'U^{\dagger} = \frac{1}{2}(U\rho U^{\dagger})^{-1/2} (U\rho_0U^{\dagger}) (U\rho U^{\dagger})^{-1/2} = \proj{+}\otimes \id^{\otimes (k-1)}\otimes \proj{0}^{\otimes (n-k)}
        \, .
    \end{equation}
    Comparing \Cref{eq:bit_block_proof_1} with \Cref{eq:bit_block_proof_2}, we get the desired statement.
\end{proof}
Therefore, if we have a sequence of bit-optimal decoders for the codeword bits $i=1,\dots,k$, each described by the projective measurement $\{\Pi_i^0,\Pi_i^1\}$, then this proposition implies
\begin{equation}
    \Pi_i^0 P = \sum_{x\in\mathcal{C}: x_i=0}\proj{f_x}
    \quad\text{and}\quad
    \Pi_i^1 P = \sum_{x\in\mathcal{C}: x_i=1}\proj{f_x}\,.
\end{equation}
and thus
\begin{equation}
    \forall\hat{x}\in\mathcal{C} : \Pi_k^{\hat{x}_k}\cdots \Pi_2^{\hat{x}_2}\Pi_1^{\hat{x}_1} P = \proj{f_{\hat{x}}} \, .
\end{equation}
because $P\Pi_i^{\hat{x}_i}P=\Pi_i^{\hat{x}_i}P$.
Therefore, sequentially chaining the $k$ bit-wise decoders together provides a block-optimal decoder.
We note that, instead of decoding $k$ linearly independent bits and then completing them to a codeword, it would also be possible to instead decode all $n$ bits bit-optimally.
We also highlight that the bitwise projective measurements commute on $V$.

Ideally, we would like to apply \Cref{prop:bitoptimal_implies_blockoptimal} to the coherent BPQM algorithm to prove \Cref{thm:block_optimal}.
Unfortunately, this is not possible for two reasons: Firstly, it is a priori unclear if coherent BPQM realizes a projective measurement, since it acts on an extended space (i.e., it is described by an isometry and not a unitary).
Secondly, coherent BPQM is only $\epsilon$-close to a bit-optimal decoder.
These two problems can be solved by considering the UC-BPQM algorithm mentioned previously. 
It requires no ancilla qubits, and therefore \Cref{prop:bitoptimal_implies_blockoptimal} can be applied to UC-BPQM.
We refer to \Cref{app:formalization_coherent_bpqm} for a detailed discussion.
So to prove \Cref{thm:block_optimal}, it suffices to show that coherent BPQM can efficiently approximate UC-BPQM.
The details of this are technical and can be found in \Cref{app:discretization}.
We also briefly highlight that coherent BPQM does indeed realize a projective measurement.
\begin{lemma}\label{lem:coherent_bpqm_projection}
    For all $i=1,\dots,k$, the five steps in \Cref{def:bpqm_block_decoding} act like a two-outcome projective measurement on the system $Q$.
\end{lemma}
The proof can also be found in \Cref{app:discretization}.

\section{MPG construction for bit-transmission channel}\label{sec:codes}
In \Cref{sec:bpqm_decoding}, we showed that BPQM can achieve optimal bitwise and blockwise decoding of a binary linear code $\mathcal{C}$ under the condition that we are given an efficient MPG representation of the involved bit-transmission channels $\mathcal{F}_{\mathcal{C},i}$.
In this section, we will show how such MPGs can be explicitly constructed for different classes of codes.
The first construction, presented in \Cref{sec:mpg_tanner}, works for codes that exhibit a tree Tanner graph.
This construction essentially recovers the BPQM description that was presented in previous works \cite{renes_belief_2017,piveteau_bpqm_2022}, and we add it here for completeness.
The subsequent constructions in \Cref{sec:mpg_oneloop,sec:mpg_trellis} are novel.
They apply to certain families of codes that do not have a tree Tanner graph, but still exhibit some alternative tree factor graph representation.

\begin{figure}
    \centering
    \includegraphics{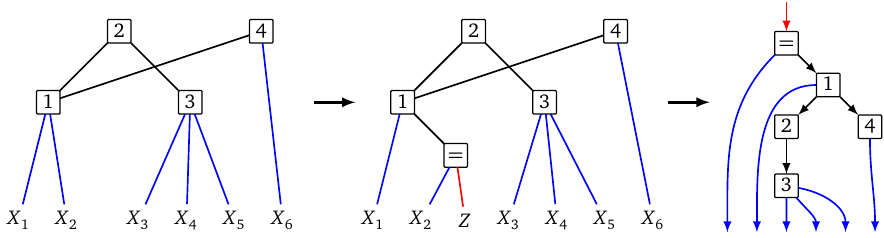}    
    \caption{Illustration of \Cref{proc:fg_to_mpg} which constructs an MPG for the bit-transmission channel $\mathcal{F}_{\mathcal{C},i}$ from a tree Forney-style factor graph representation of the code inclusion function, here for some $6$-bit code $\mathcal{C}$ and $i=2$.
    In the original factor graph (left), the external edge $X_i$ is copied using an equality node (center).
    We denote the copy by $Z$, depicted in red.
    The remaining external edges $X_1,\dots,X_n$ are depicted in blue.
    The desired MPG (right) is obtained by fixing the red edge to be the root.
    }
    \label{fig:mpg_from_tree}
\end{figure}

We begin with a brief overview of the general strategy that underlies all the constructions before addressing each one in turn.
As an illustrative example, we depict in \Cref{fig:mpg_from_tree} how this general procedure constructs an MPG representation $\mathcal{F}_{\mathcal{C},i}$ for some $6$-bit code $\mathcal{C}$ and $i=2$.
\begin{procedure}[Procedure to construct MPG representation of $\mathcal{F}_{\mathcal{C},i}$]
\label{proc:fg_to_mpg}
Assume we know some tree Forney-style factor graph representation of the code inclusion function $I_{\mathcal{C}}$ (recall \Cref{eq:code_inclusion_function}).
More concretely, this Forney-style factor graph has $n$ external edges representing the variables $X_1,\dots,X_n$ and the exterior function of the factor graph is precisely $I_{\mathcal{C}}$.
Then, copy the external leg $X_i$ using an equality node, and call the copy $Z$.
The resulting factor graph has the exterior function
\begin{equation}
    x_1,\dots,x_n,z \mapsto I_{\mathcal{C}}(x_1,\dots,x_n) \cdot \begin{cases} 1 &\text{ if } x_i=z \\ 0 &\text{ otherwise}\end{cases}
    \, ,
\end{equation}
and therefore, can be understood as a representation of the bit-transmission channel $\mathcal{F}_{\mathcal{C},i}$ in the sense of \Cref{eq:mpg_factors}.
This new factor graph is still a tree, and we can induce an order on all its edges by choosing $Z$ to be the root and the $X_1,\dots,X_n$ to be leaves (and choosing the direction from root to leaves).
In order for this graph to represent a valid MPG, we additionally need all internal edges to have an alphabet size that is a power of $2$ and also each node $v$ to represent a channel of the form $F_{\ell_v,G_v}$ for some $\ell_v,G_v$.
\end{procedure}

We emphasize that this procedure is generally not applicable to any arbitrary tree factor graph representation of $I_{\mathcal{C}}$, since the individual factor nodes may not necessarily represent a channel $F_{\ell_v,G_v}$.
Furthermore, notice that the (logarithm of the) alphabet sizes of the factor graph edges directly translates to the variable dimensions $n_e$ in the MPG.
Therefore, the main difficulty in this section will be to construct factor graph representations of the code inclusion function $I_{\mathcal{C}}$ that
\begin{enumerate}
    \item are trees,
    \item have a small (polynomial in code size) number of nodes,
    \item have small (polynomial in code size) alphabet sizes of the internal variables and
    \item only contain nodes representing channels of the form $F_{\ell,G}$.
\end{enumerate}

\subsection{Construction from tree Tanner graph}\label{sec:mpg_tanner}
If a certain $[n,k]$ binary linear code $\mathcal{C}$ exhibits a Tanner graph which is a tree, then \Cref{proc:fg_to_mpg} goes through quite directly.
All nodes $v$ in the Forney representation of the Tanner graph are either equality nodes or check nodes, and in both cases, one can interpret the node as representing a channel $F_{1,G_v}$ for some code $\mathcal{C}_v$ of blocksize $\mathrm{deg}(v)-1$ and generator matrix $G_v$, where $\mathrm{deg}(v)$ denotes the degree of $v$.
More concretely, for an equality node, $\mathcal{C}_v$ is the $[\mathrm{deg}(v)-1,1]$ repetition code with generator matrix $G_v=(1 1 \dots 1)$, and for a check node, $\mathcal{C}_v$ is the trivial $[\mathrm{deg}(v)-1,\mathrm{deg}(v)-1]$ code with generator matrix
\begin{equation}
    G_v = \begin{pmatrix}
      1 & 0 & 0 & \dots & 0 \\
      1 & 1 & 0 & \dots & 0 \\
      1 & 0 & 1 & \dots & 0 \\
      \vdots &  &  & \ddots & \vdots \\
      1 & 0 & 0 & \dots & 1  
    \end{pmatrix}
    \, .
\end{equation}

\begin{figure}
    \centering
    \includegraphics{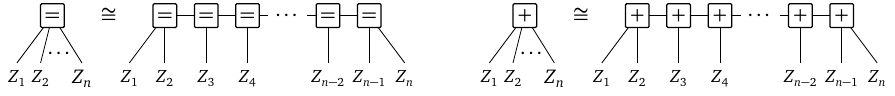}
    \caption{In a Forney-style factor graph, an equality or check node of degree $n>3$ can be decomposed into smaller nodes of degree $3$.
    The equality here should be understood as saying that the exterior function of both factor graphs are identical.}
    \label{fig:eq_check_decomp}
\end{figure}

While this construction does provide us with a valid MPG $\mathcal{G}$ for the bit-transmission channel, the resulting maximal variable dimension $N(\mathcal{G})$ could potentially be large, as it is essentially given by the maximum node degree in the original Tanner graph.
To avoid this issue, we can do a simple pre-processing step on the Tanner graph before converting it into an MPG.
The crucial insight here is that any check or equality node of degree larger than $3$ in the Forney-style representation of the Tanner graph can be split up into multiple smaller nodes of degree $3$ by iteratively applying the identities in \Cref{fig:eq_check_decomp}.
As such, the maximal degree of the pre-processed factor graph is at most $3$, and hence, $N(\mathcal{G})\leq 2$.

\begin{figure}
    \centering
    \includegraphics[width=\textwidth]{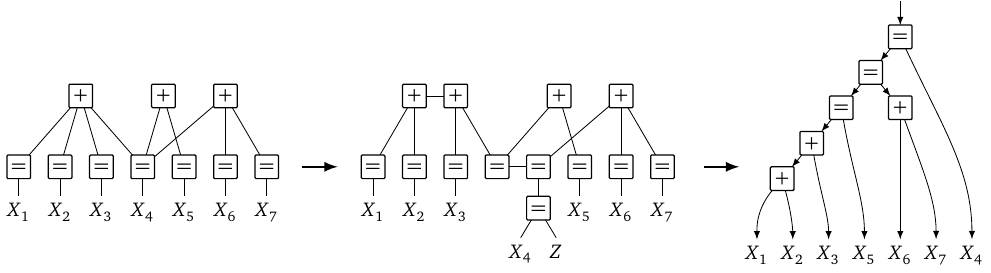}
    \caption{Construction of MPG representation of bit-transmission channel $\mathcal{F}_{\mathcal{C},4}$ of some $7$-bit code with tree Tanner graph (left). In the first step, the nodes with degree $>3$ are split up and the bit $X_4$ is copied with an additional equality node (middle). Unrolling this graph around the root $Z$ provides us with the desired MPG (right). The nodes with degree 2 are removed from the MPG, since they correspond to the trivial trivial channel $F_{\ell,G}$ with $\ell=1$ and $G=(1)$. }
    \label{fig:mpt_tree_tanner_example}
\end{figure}

In \Cref{fig:mpt_tree_tanner_example}, we exemplify the generation of the efficient MPG representation of $\mathcal{F}_{\mathcal{C},i}$ where $i=4$ and $\mathcal{C}$ is the $7$-bit code with parity check matrix
\begin{equation}
    H = 
    \begin{pmatrix}
        1&1&1&1&0&0&0 \\
        0&0&0&1&1&0&0 \\
        0&0&0&1&0&1&1
    \end{pmatrix}
    \, .
\end{equation}

While the pre-processing step guarantees a small constant alphabet size, it increases the number of nodes in the resulting MPG.
More concretely, for each (equality or check) node of degree $d>3$, it is split into $d-2$ nodes of degree $3$ each.
If we denote by $d^{(c)}_i$ and $d^{(v)}_j$ the degree of the $i$-th variable node and $j$-th check node in the original (Forney-style) Tanner graph, we see that the total number of nodes in the pre-processed factor graph is
\begin{equation}
    \sum_{i=1}^n \begin{cases} d_i^{(v)}-2\text{ if } d_i^{(v)}>2\\ 1 \text{ else}\end{cases}
    + \sum_{j=1}^{m} \begin{cases} d_j^{(c)}-2\text{ if } d_j^{(c)}>2\\ 1 \text{ else}\end{cases}
    \leq \sum_{i=1}^n d_i^{(v)} + \sum_{j=1}^{m} d_j^{(c)} = 3n + 2m -2
\end{equation}
where $n$ is the code size, $m$ the number of parity checks in the Tanner graph and we used that the number of internal edges is precisely $n+m-1$.
We summarize our discussion as follows.
\begin{proposition}
    Let $\mathcal{C}$ be an $[n,k]$ binary linear code with a tree Tanner graph.
    For any $i\in\{1,\dots,n\}$, there exists an MPG representation $\mathcal{G}$ of the bit-transmission channel $\mathcal{F}_{\mathcal{C},i}$ that has maximum variable dimension $N(\mathcal{G})\leq 2$ and $\mathcal{O}(n)$ nodes.
\end{proposition}

\subsection{Construction from Tanner graph with one cycle}\label{sec:mpg_oneloop}
\begin{figure}
    \centering
    \includegraphics{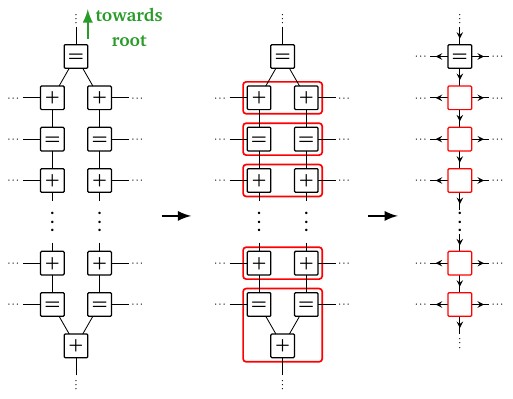}
    \caption{Procedure to obtain an efficient MPG from a Tanner graph with one loop.
    The loopy tanner graph is depicted on the left (only the nodes involved in the loop are shown).
    With appropriate pre-processing, it can be assumed that all involved nodes have degree three.
    By pairing up the nodes and merging each pair into one node (see middle), we obtain nodes that are valid for an MPG (see right), i.e., they represent some channel of the form $F_{\ell,G}$.}
    \label{fig:loop_merging}
\end{figure}

In some situations, an efficient MPG can also be obtained from a Tanner graph that contains only a few cycles.
We illustrate this procedure on the simplest case: A Tanner graph with a single cycle.
The central idea is that the cycle can be effectively removed from the factor graph by merging certain groups of check and/or equality nodes into a larger node\footnote{Merging two or more nodes $v_1,\dots,v_m$ is achieved by replacing them with one node $\bar v$ representing the exterior function of the sub-factor graph containing only $v_1,\dots,v_m$ and their connected edges. This merging operation is sometimes called ``closing the box'' in the Forney-style factor graph nomenclature~\cite{loeliger_introduction_2004}. It can also be understood as a tensor contraction by interpreting the factor graph as a tensor network. The exterior function of a Forney-style factor graph remains unchanged when some of its nodes are merged.}.
We depict the procedure in \Cref{fig:loop_merging}.
Two nodes (or three nodes at the bottom) are each merged into one single node, depicted in red.

\begin{figure}
    \centering
    \includegraphics{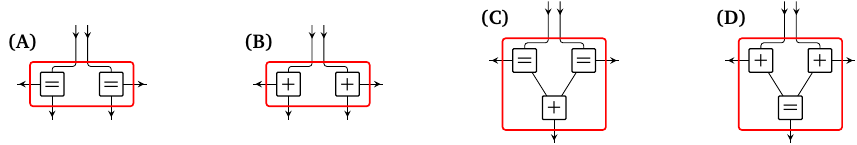}
    \caption{The four different types of merged nodes that can occur in \Cref{fig:loop_merging}.}
    \label{fig:loop_merging_nodes}
\end{figure}

More precisely, we turn the (non-tree) Tanner graph into a tree factor graph representation of the code inclusion function by performing following steps:
\begin{enumerate}
    \item Every node of degree-$2$ in the factor graph can be removed and the two neighboring nodes merged using the identities
    \begin{center}
        \includegraphics{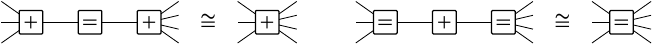} \, .
    \end{center}
    Note that the remaining factor graph remains a bipartite graph of equality and check nodes.
    
    \item Ensure that all remaining nodes in the graph are of degree $3$, by splitting up high-degree nodes using the identity in \Cref{fig:eq_check_decomp}.
    For nodes that are involved in the cycle, this has to be done specifically in a way such that the cycle remains alternating between equality nodes and check nodes.
    \item Merge the nodes involved in the cycle as depicted in \Cref{fig:loop_merging}.
\end{enumerate}
Steps 1 and 2 ensure that the cycle in the factor graph only contains degree-$3$ nodes, alternating between equality and check nodes.
After the merging in step 3, the resulting factor graph now contains new node types (besides check and equality nodes).
We exhaustively enumerate the types of possible nodes in \Cref{fig:loop_merging_nodes}.
To verify that the resulting tree graph is a valid MPG, we simply have to verify that each of these possible nodes does represent a channel of the form $F_{\ell,G}$ for some $\ell,G$.\footnote{Note that all other nodes in the graph are equality and check nodes, and therefore already covered by the discussion in \Cref{sec:mpg_tanner}.}
We show this for each node type separately.
In the following, we choose the order of bits to be left-to-right according to \Cref{fig:loop_merging_nodes}.
\begin{enumerate}[label=(\Alph*)]
    \item This node encodes $\ell=2$ bits $z_1,z_2$ into $(z_1\cat z_2)^TG_A$ where
    \begin{equation}
        G_A = 
        \begin{pmatrix}
            1&1&0&0 \\
            0&0&1&1
        \end{pmatrix}
        \, .
    \end{equation}
    \item This node encodes $\ell=2$ bits $z_1,z_2$ into $(z_1\cat z_2\cat r)^TG_B$ where $r\in\mathbb{F}_2^2$ is uniformly random and 
    \begin{equation}
        G_B = 
        \begin{pmatrix}
            1&0&0&0 \\
            0&0&1&0 \\
            1&1&0&0 \\
            0&0&1&1
        \end{pmatrix}
        \, .
    \end{equation}
    \item This node encodes $\ell=2$ bits $z_1,z_2$ into $(z_1\cat z_2)^TG_C$ where
    \begin{equation}
        G_C = 
        \begin{pmatrix}
            1&1&0 \\
            0&1&1
        \end{pmatrix}
        \, .
    \end{equation}
    \item This node encodes $\ell=2$ bits $z_1,z_2$ into $(z_1\cat z_2\cat r)^TG_D$ where $r\in\mathbb{F}_2$ is uniformly random and 
    \begin{equation}
        G_D = 
        \begin{pmatrix}
            1&0&0 \\
            0&0&1 \\
            1&1&1
        \end{pmatrix}
        \, .
    \end{equation}
\end{enumerate}
Our result can be summarized as follows.
\begin{proposition}
    Let $\mathcal{C}$ be an $[n,k]$ linear code with a Tanner graph containing one cycle.
    For any $i\in\{1,\dots,n\}$, there exists an MPG representation $\mathcal{G}$ of the bit-transmission channel $\mathcal{F}_{\mathcal{C},i}$ that has maximum variable dimension $N(\mathcal{G})\leq 4$ and $\mathcal{O}(n)$ nodes.
\end{proposition}
We note that for Tanner graphs with multiple cycles, the ``cycle removal'' procedure could in principle be repeated multiple times to obtain a tree factor graph representation of the code inclusion function.
However, proving that the fused nodes remain valid MPG nodes while controlling the maximum variable dimension is more involved.
Characterizing the cases where this remains possible is an interesting open research direction for future work.

\subsection{Construction from trellis representation}\label{sec:mpg_trellis}
We now discuss how to construct MPGs from classical linear codes that exhibit an efficient trellis representation.
For now, we will restrict our attention to trellises that have only binary labels (i.e. either $0$ or $1$) on their edges, as opposed to possibly longer bit strings. 
In the notation of \Cref{sec:trellis}, this means $\ell_i=1$ for all $i$.
We will come back to the general case towards the end of this section.

\begin{figure}
    \centering
    \includegraphics{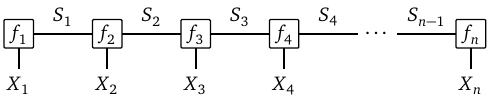}
    \caption{Structure of a Forney-style factor graph which was derived from a trellis and which represents the code inclusion function $I_{\mathcal{C}}$.}
    \label{fig:trellis_fg}
\end{figure}

Consider some trellis representation $T=(V,E)$ of some binary linear code $\mathcal{C}\subset\ff_2^n$ with binary labels.
There exists a well-known procedure to generate a factor graph from $T$ of the form of \Cref{fig:trellis_fg} that represents the code inclusion function $I_{\mathcal{C}}$ \cite{loeliger_introduction_2004}.
Each external edge $X_i$ is connected to a distinct factor $f_i$ for $i=1,\dots,n$.
The factor graph also contains internal variables $S_j$ for $j=1,\dots,n-1$ connecting the factors $f_j$ and $f_{j+1}$.
The alphabet size of $S_j$ is given by state space size at the $j$-th level, i.e., $\abs{V_j}$ in the notation of \Cref{sec:trellis}.
The factor nodes are defined as
\begin{equation}
    f_i(s_{i-1},x_i,s_i) \coloneqq
    \left\{
    \begin{array}{@{}l@{\quad}p{8cm}@{}}
    1 & if there exists an edge $e\in E$ connecting the $s_{i-1}$-th node in $V_{i-1}$ to the $s_{i}$-th node in $V_{i}$ with label $\lambda(e)=x_i$, \\
    0 & otherwise.
    \end{array}
    \right.
\end{equation}
A detailed discussion of this construction,  including a detailed explanation why the resulting factor graph describes the code inclusion function $I_{\mathcal{C}}$, is not necessary for our purposes, and we refer interested readers to \cite{loeliger_introduction_2004}.
The Viterbi~\cite{forney_viterbi_1973} and BCJR~\cite{bahl_bcjr_1974} algorithms, the standard trellis-based decoders in classical coding theory, can both be seen as some variation of belief propagation on this factor graph~\cite{wiberg_codes_1996,loeliger_introduction_2004}.
Naturally, one may attempt to attempt to base the MPG for BPQM off of the same factor graph.
Unfortunately, this is generally not possible, as the not every trellis-induced factor graph representing $I_{\mathcal{C}}$ will be suitable for our procedure to generate MPGs, since the individual nodes generally do not represent channels of the form $F_{\ell,G}$.

However, similar to factor graph representations, any one code has a multitude of trellis representations, some being more efficient than others.
It turns out that the construction of an MPG is always possible if we focus on one specific family of trellis representations, namely, trellises constructed from a ``minimal-span generator matrix'' (MSGM, also called ``trellis-oriented generator matrix'').
These MSGM-derived trellises have been extensively studied in literature~\cite{forney_cosetcodes_1988,kschischang_trellis_1995,mceliece_bcjrtrellis_1996}, since they are optimal in a multitude of notions.
For example, they minimize the state space size of the trellis at every position~\cite[Theorem 5.1]{mceliece_bcjrtrellis_1996} (in fact, they are the unique optimum up to isomorphism).
It should also be noted that an MSGM of a code can always be found efficiently (see e.g. \cite[Section IV.C]{kschischang_trellis_1995} or \cite[Section VI]{mceliece_bcjrtrellis_1996}).

In the following, we will now explain how a MPG for the bit-transmission channel can be constructed from an MSGM of a code.
In order to simplify and shorten the exposition as much as possible, we will not explain the construction of the MSGM-derived trellis from the MSGM itself; instead we will directly show how the MPG can be obtained from the MSGM, skipping the intermediate  trellis.
We start by summarizing the definition and a few results about MSGMs.

\begin{definition}[Span]
    Consider a nonzero vector $x\in\mathbb{F}_2^n$.
    The left index $L(x)$ is defined as the smallest index $i\in\{1,\dots,n\}$ such that $x_i\neq 0$, and similarly, the right index $R(x)$ is defined as the largest index $j\in\{1,\dots,n\}$ such that $x_j\neq 0$.
    The span of $x$ is the discrete interval $(L(x),L(x)+1,\dots,R(x))$.
    The span length of $x$ is the number of elements in its span.
\end{definition}
\begin{definition}[Minimal-span generator matrix]
    The span length of a generator matrix $G\in\ff_2^{k\times n}$ for some $[n,k]$ binary linear code is defined as the sum of the span lengths of its rows.
    A generator matrix of a code that achieves the minimal spanlength is called a minimal-span generator matrix (MSGM).
\end{definition}
\begin{lemma}{\cite[Theorem 6.11]{mceliece_bcjrtrellis_1996}}
    A generator matrix $G$ is an MSGM if and only if it fulfills the LR property, i.e., the $i$-th and $j$-th rows $g_i,g_j$ of $G$ fulfill $L(g_i)\neq L(g_j)$ and $R(g_i)\neq R(g_j)$ when $i\neq j$.
\end{lemma}

\begin{figure}
    \centering
    \includegraphics[width=\textwidth]{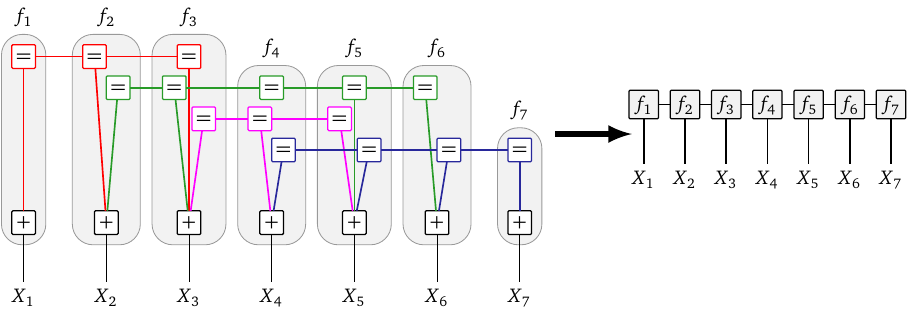}
    \caption{
    Construction of Forney-style factor graph describing the code inclusion $I_{\mathcal{C}}$ for the $[7,4]$ Hamming code.
    The color indicates to which row of the MSGM an equality node is associated to.
    The gray boxes show which nodes are combined into the factors $f_1,\dots,f_7$.
    }
    \label{fig:hamming_trellis_mpg}
\end{figure}

Given a generator matrix $G$ of some $[n,k]$ binary linear code $\mathcal{C}$ fulfilling the LR property, we can construct a factor graph representation of $I_{\mathcal{C}}$ using the following procedure below.
As an example, the procedure is illustrated in \Cref{fig:hamming_trellis_mpg} for the Hamming code with MSGM
\begin{equation}
    G = \begin{pmatrix}
        1 & 1 & 1 & 0 & 0 & 0 & 0 \\
        0 & 1 & 1 & 0 & 1 & 1 & 0 \\
        0 & 0 & 1 & 1 & 1 & 0 & 0 \\
        0 & 0 & 0 & 1 & 1 & 1 & 1 \\
    \end{pmatrix}
    \, .
\end{equation}
\begin{enumerate}
    \item Introduce a parity check node for every bit, denoted by $c_1,\dots,c_n$.
    For each $i$, connect $c_i$ to an external edge corresponding to $X_i$.
    \item For $j=1,\dots,k$:
    \begin{itemize}
        \item Let $g_j$ be the $j$-th row of $G$.
        \item For $i=L(g_j),\dots,R(g_j)$: Introduce an equality node $e_{j,i}$.
        If $G_{j,i}=1$, connect this node to $c_i$.
        \item For $i=L(g_j),\dots,R(g_j)-1$: connect the nodes $e_{j,i}$ and $e_{j,i+1}$.
    \end{itemize}
    \item For every $j=1,\dots,k$: Combine $c_j$ and all nodes $e_{j,i}$ for any $i$ into one single node.
\end{enumerate}
It is easy to verify that this constructed factor graph has the exterior function 
\begin{equation}
    x_1,\dots,x_n \mapsto \sum_{u_1,\dots,u_k=0,1} I_{u^TG=x^T}
\end{equation}
where $I$ is the indicator function, as each row of connected equality nodes $(e_{j,i})_i$ represents a source bit $u_j$.
This map is equal to the code inclusion function $I_{\mathcal{C}}$.

\begin{figure}
    \centering
    \includegraphics{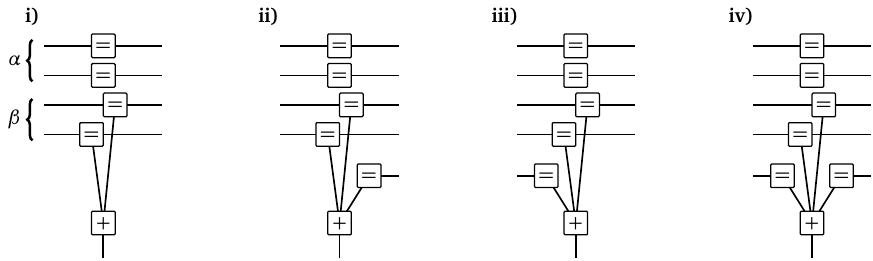}
    \caption{In an MSGM-derived trellis, we can categorize the factor nodes into four different categories, depending on whether there are zero or one ``terminating'' bits coming from the left and from the right.
    We denote these four cases i) ii) iii) and iv).
    Furthermore, we denote by $\alpha$ and $\beta$ the number of equality nodes in that are not connected, or respectively are connected, to the check node.
    In the depicted examples, $\alpha=\beta=2$.}
    \label{fig:trellis_cases_1}
\end{figure}

\begin{figure}
    \centering
    \includegraphics{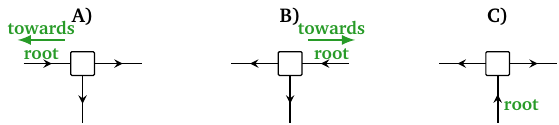}
    \caption{When translating a trellis-induced factor graph, as depicted in \Cref{fig:trellis_fg}, into an MPG, the orientation of the edges connected to some factor node can only be one of three possible configurations, which we denote A), B) and C).}
    \label{fig:trellis_cases_2}
\end{figure}

To verify that our factor graph induces a valid MPG, we have to check that the individual factor nodes represent channels of the form $F_{\ell,G}$.
To this end, the LR property of $G$ is very important: It implies that at every codeword bit check node, there may be at most one source bit from the left and at most one source bit from the right ``terminating'' at that node.
Hence, we can categorize four different configurations (according to the number of ``terminating bits''),  which are depicted in \Cref{fig:trellis_cases_1}.
There is also another difficulty to consider: Contrary to the Tanner graph-derived MPG, the factor nodes here are not symmetric under permutation of the edges.
As such, we have to make sure that the nodes represent valid encoding channels for \emph{all} relevant configurations of edge directions.
Concretely, there are three possible configurations, which are depicted in \Cref{fig:trellis_cases_2}.


In total, there are thus $4\cdot 3=12$ different cases, and for each one, we have to verify that the factor represents a valid channel $F_{\ell,G}$ for some $\ell$ and $G$.
We now do this by enumerating all these possible cases.
Note that situations A) and B) are symmetric to each other, so it will suffice to only consider A) and C).
Furthermore, as in \Cref{fig:trellis_cases_1}, let $\alpha$ and $\beta$ denote the number of equality nodes that are involved in a factor nodes, which are not connected, respectively connected, to the check node.
In the following, $I_n$ denotes the $n\times n$ identity matrix and $\vec{1}_n$ denotes the column vector of size $n$ containing only the entry $1$.
Furthermore, in reference to \Cref{fig:trellis_cases_1}, the order of the bits is always chosen as follows: First, the bits corresponding to open edges on the left (top-to-bottom), then the bits corresponding to open edges on the right (top-to-bottom), and then the bit corresponding to the open edge at the bottom
\begin{enumerate}
    \item[\textbf{Ai)}]
    This node encodes $l=\alpha+\beta$ bits $z\in\ff_2^{\alpha+\beta}$ into $z^TG$ where 
    \begin{equation}
        G = 
        \begin{pmatrix}
            I_{\alpha} & 0 & 0 \\
            0 & I_{\beta} & \vec{1}_{\beta}
        \end{pmatrix}
        \, .
    \end{equation}
    
    \item[\textbf{Aii)}]
    This node encodes $l=\alpha+\beta$ bits $z\in\ff_2^{\alpha+\beta}$ into $(z\cat r)^TG$ where $r\in\mathbb{F}_2$ is uniformly random and 
    \begin{equation}
        G = 
        \begin{pmatrix}
            I_{\alpha} & 0 & 0 \\
            0 & I_{\beta+1} & \vec{1}_{\beta+1}
        \end{pmatrix}
        \, .
    \end{equation}
    
    \item[\textbf{Aiii)}]
    This node encodes $l=\alpha+\beta+1$ bits $z\in\ff_2^{\alpha+\beta+1}$ into $z^TG$ where 
    \begin{equation}
        G = 
        \begin{pmatrix}
            I_{\alpha} & 0 & 0 \\
            0 & I_{\beta} & \vec{1}_{\beta} \\
            0 & 0 & 1
        \end{pmatrix}
        \, .
    \end{equation}
    
    \item[\textbf{Aiv)}]
    This node encodes $l=\alpha+\beta+1$ bits $z\in\ff_2^{\alpha+\beta+1}$ into $(z\cat r)^TG$ where $r\in\mathbb{F}_2$ is uniformly random and 
    \begin{equation}
        G = 
        \begin{pmatrix}
            I_{\alpha} & 0 & 0 & 0 \\
            0 & I_{\beta} & 0 & \vec{1}_{\beta} \\
            0 & 0 & 0 & 1 \\
            0 & 0 & 1 & 1
        \end{pmatrix}
        \, .
    \end{equation}
    
    \item[\textbf{Ci)}]
    This node encodes $l=1$ bit $z$ into $(z\cat r)^TG$ where $r\in\mathbb{F}_2^{\alpha+\beta-1}$ is uniformly random and 
    \begin{equation}
        G = 
        \begin{pmatrix}
            0 & 1 & 0 & 0 & 1 & 0 \\
            0 & \vec{1}_{\beta-1} & I_{\beta-1} & 0 & \vec{1}_{\beta-1} & I_{\beta-1} \\
            I_{\alpha} & 0 & 0 & I_{\alpha} & 0 & 0
        \end{pmatrix}
        \, .
    \end{equation}
    
    \item[\textbf{Cii)}]
    This node encodes $l=1$ bit $z$ into $(z\cat r)^TG$ where $r\in\mathbb{F}_2^{\alpha+\beta}$ is uniformly random and 
    \begin{equation}
        G = 
        \begin{pmatrix}
            0 & 1 & 0 & 0 & 1 & 0 & 0 \\
            0 & \vec{1}_{\beta-1} & I_{\beta-1} & 0 & \vec{1}_{\beta-1} & I_{\beta-1} & 0 \\
            I_{\alpha} & 0 & 0 & I_{\alpha} & 0 & 0 & 0 \\
            0 & 1 & 0 & 0 & 1 & 0 & 1
        \end{pmatrix}
        \, .
    \end{equation}
    
    \item[\textbf{Ciii)}]
    This node encodes $l=1$ bit $z$ into $(z\cat r)^TG$ where $r\in\mathbb{F}_2^{\alpha+\beta}$ is uniformly random and 
    \begin{equation}
        G = 
        \begin{pmatrix}
            0 & 1 & 0 & 0 & 0 & 1 & 0 \\
            0 & \vec{1}_{\beta-1} & I_{\beta-1} & 0 & 0 & \vec{1}_{\beta-1} & I_{\beta-1} \\
            I_{\alpha} & 0 & 0 & 0 & I_{\alpha} & 0 & 0 \\
            0 & 1 & 0 & 1 & 0 & 1 & 0
        \end{pmatrix}
        \, .
    \end{equation}
     
    \item[\textbf{Civ)}]
    This node encodes $l=1$ bit $z$ into $(z\cat r)^TG$ where $r\in\mathbb{F}_2^{\alpha+\beta+1}$ is uniformly random and 
    \begin{equation}
        G = 
        \begin{pmatrix}
            0 & 1 & 0 & 0 & 0 & 1 & 0 & 0 \\
            0 & \vec{1}_{\beta-1} & I_{\beta-1} & 0 & 0 & \vec{1}_{\beta-1} & I_{\beta-1} & 0\\
            I_{\alpha} & 0 & 0 & 0 & I_{\alpha} & 0 & 0 & 0\\
            0 & 1 & 0 & 1 & 0 & 1 & 0 & 0 \\
            0 & 1 & 0 & 0 & 0 & 1 & 0 & 1 \\
        \end{pmatrix}
        \, .
    \end{equation}
\end{enumerate}

We summarize the result of our construction as follows.
\begin{proposition}\label{prop:trellis_mpg}
    Let $\mathcal{C}$ be an $[n,k]$ binary linear code that exhibits a trellis representation $T$ with binary labels and maximal state space size\footnote{In the notation of \Cref{sec:trellis}, $S\coloneq\max_i \abs{V_i}$.} $S$.
    For any $i\in\{1,\dots,n\}$, there exists an MPG representation $\mathcal{G}$ of the bit-transmission channel $\mathcal{F}_{\mathcal{C},i}$ that has maximum variable dimension $N(\mathcal{G})\leq 2\log S$ and $n$ nodes.
\end{proposition}
\begin{proof}
    Consider an MSGM $G'$ of some arbitrary binary linear $[n',k']$ code $\mathcal{C}'$.
    Let's denote the rows of $G'$ by $g_1,\dots,g_{k'}$ and define $A_i(G')\coloneqq \{j\in \{1,\dots,k'\} | i\in \mathrm{span}(g_j) \}$ for $i=1,\dots,n'$.
    Then, define $\beta_i(G)$ to be the cardinality of the set $A_i(G')\cap A_{i+1}(G')$.
    We use following two facts and refer to literature for the proofs:
    \begin{itemize}
        \item The trellis derived from $G'$ has state space $2^{\beta_i(G')}$ in the $i$-th layer of the trellis~\cite[Theorems 7.2 and 7.4]{mceliece_bcjrtrellis_1996}.
        \item No trellis representation of $\mathcal{C}'$ has a state space size at the $i$-th layer which is smaller than $2^{\beta_i(G')}$~\cite[Theorem 5.1]{mceliece_bcjrtrellis_1996}.
    \end{itemize}
    Now, let $G$ be an MSGM for the given code $\mathcal{C}$.
    We can conclude that $\beta_i(G)\leq \log S$ for all $i$.

    Consider the factor graph constructed from $G$, as described by \Cref{fig:hamming_trellis_mpg}.
    Notice that the resulting external edges have alphabet size $2$, whereas the $i$-th internal edge has alphabet size $2^{\beta_i(G)}$.
    By the previous considerations, we can utilize this factor graph to construct an MPG by \Cref{proc:fg_to_mpg}.
    The largest possible value of $n_v$ can be achieved by setting C) in \Cref{fig:trellis_cases_2}.
    We conclude that $N(\mathcal{G})\leq 2\max_i\beta_i(G)$.
\end{proof}

Up to now, we have only considered trellises with binary labels on each edge, and therefore, each node in the factor graph (as in \Cref{fig:trellis_fg}) was connected to only one single external (binary) edge.
However, it is sometimes possible to obtain more efficient trellises by allowing edge labels with multiple bits.
This is nicely exemplified in \Cref{fig:trellis_hamming}, where combining the fourth and fifth bits reduces the maximal state space from $S=8$ to $S=4$.

An optimal trellis (in terms of maximal state space size) with longer edge labels can always be obtained from the MSGM-derived trellis by merging subsequent trellis stages together.\footnote{This follows from the fact that the optimal trellis state space size at any given layer only depends on the ``past'' and ``future'' codes at this point --- see~\cite[Appendix A]{forney_cosetcodes_1988} for a detailed discussion.}
In terms of the factor graph picture in \Cref{fig:trellis_hamming}, this corresponds to merging neighboring nodes in the factor graph.
The resulting factor graph can therefore have more than one external edge connected to any node.
It is easy to verify that the resulting factor graph still produces a valid MPG when processed with \Cref{proc:fg_to_mpg}.
Indeed, this follows from the fact that the concatenation of $F_{\ell_2,G_2}\circ F_{\ell_1,G_1}$, where $F_{\ell_2,G_2}$ acts on a subset of the output bits of $F_{\ell_1,G_1}$, is again a channel of the form $F_{\ell',G'}$ for some $\ell'$ and $G'$.
\begin{corollary}
    Let $\mathcal{C}$ be an $[n,k]$ binary linear code and $T$ a trellis representation of $\mathcal{C}$ with maximum state space size $S$ and where the length of the label bit strings is $\leq L$.
    For any $i\in\{1,\dots,n\}$ there exists an MPG representation $\mathcal{G}$ of the bit-transmission channel $\mathcal{F}_{\mathcal{C},i}$ that has maximum variable dimension $N(\mathcal{G})\leq 2\log S + (L-1)$ and $n$ nodes.
\end{corollary}
To illustrate the trellis-based MPG construction, we now consider a simple convolutional codes as example.

\paragraph{Example: $G=13/15$ binary systematic recursive convolutional code}
We illustrate the trellis-based MPG construction on the $G=13/15$ binary systematic recursive convolutional code.
For this example, it will prove a bit simpler and less verbose to not explicitly work out the MSGM of the code, and instead directly construct the MSGM-derived trellis as in \Cref{fig:hamming_trellis_mpg}.
The MSGM could then be inferred from that factor graph, if desired.

\begin{figure}
    \centering
    \includegraphics{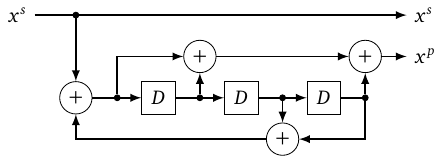}
    \caption{Encoding diagram for the $G=13/15$ binary systematic recursive convolutional code.
    $x^s$ and $x^p$ denote the systematic bit and parity bit.
    The square boxes depict delay elements and correspond to the $m=3$ bits of memory in the encoder.
    }
    \label{fig:convo_example_encoder}
\end{figure}

\begin{figure}
    \centering
    \begin{subfigure}[b]{0.49\textwidth}
        \centering
        \includegraphics{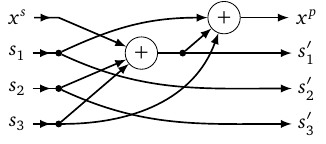}
        \caption{}
        \label{fig:convo_example_transition}
    \end{subfigure}
    \begin{subfigure}[b]{0.49\textwidth}
        \centering
        \includegraphics{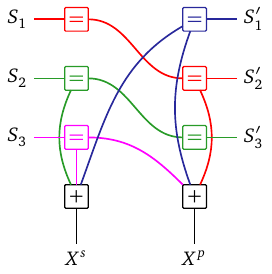}
        \caption{}
        \label{fig:convo_example_transition_fg}
    \end{subfigure}
    \caption{Graphical representation of the state transition map $T$ of the $G=13/15$ convolutional code, as a circuit diagram (a) and as a Forney-style factor graph (b).}
\end{figure}

\begin{figure}
    \centering
    \includegraphics{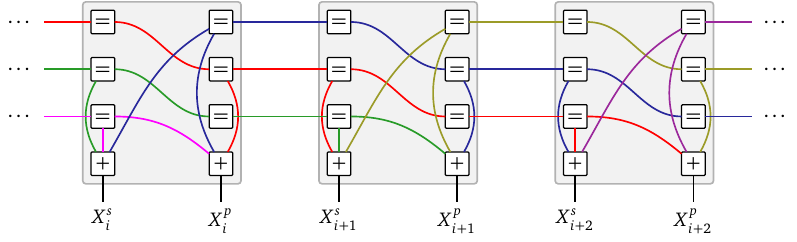}
    \caption{Forney-style factor graph representation of the $G=13/15$ binary systematic recursive convolutional code.
    Here, $X_i^s$ an $X_i^p$ are the variables representing the $i$-th systematic bit and the $i$-th parity bit.
    Each color corresponds to a row of the MSGM.
    By grouping together the nodes of each gray box, one can obtain an MPG representation of the bit-transmission channels with maximum variable dimension of $3$.}
    \label{fig:convo_example_fg}
\end{figure}

For this example, we will assume basic familiarity with the theory of convolutional codes.
The $13$ corresponds to the feedforward polynomial $p(D)=1+D+D^3$ and $15$ corresponds to the feedback polynomial $q(D)=1+D^2+D^3$.
The encoding diagram for this convolutional code is depicted in \Cref{fig:convo_example_encoder}.
This convolutional code has $m=3$ bits of memory, and therefore, it exhibits a trellis representation with a state space size of $2^3=8$.
The convolutional encoder implicitly describes the state transition map $T$
\begin{equation}
    (x^s,s) \mapsto T(x^s,s) = (x^p,s')
\end{equation}
which describes how the input systematic bit $x^s$ at time $t$ and the state $s$ at time $t-1$ are mapped to the output parity bit $x^p$ at time $t$ and the state $s'$ at time $t$.
Since we have three memory bits, we have $s,s'\in\ff_2^3$.
A circuit representation of the transition map $T$ is given in \Cref{fig:convo_example_transition}.
In \Cref{fig:convo_example_transition_fg}, we translate this circuit into a factor graph representation of this transition function, i.e., a Forney-style factor graph with exterior function
\begin{equation}
    x^s,s,x^p,s' \mapsto I_{T(x^s,s)=(x^p,s')}=\begin{cases} 1 \text{ if } T(x^s,s)=(x^p,s') \\ 0 \text{ otherwise}\end{cases}
    \, .
\end{equation}
The code inclusion function of the convolutional code $\mathcal{C}$ can be written as
\begin{equation}
    I_{\mathcal{C}}(x_1^s,x_1^p,x_2^s,x_2^p,x_3^s,x_3^p,\dots)
    =
    \sum_{s_1,s_2,s_3,\dots\in\ff_2^3}
    I_{T(x_1^s,0)=(x_1^p,s_1)}
    I_{T(x_2^s,s_1)=(x_2^p,s_2)}
    I_{T(x_3^s,s_2)=(x_3^p,s_3)}
    \dots
\end{equation}
which implies that by ``gluing'' together repeated copies of the factor graph in \Cref{fig:convo_example_transition_fg}, we can obtain a factor graph representation of the code inclusion function $I_{\mathcal{C}}$, which is depicted in \Cref{fig:convo_example_fg}.
If we pick the order of the codeword bits to be $x_1^sx_1^px_2^sx_2^px_3^sx_3^p\dots$ where $x_i^s,x_i^p$ are the $i$-th systematic and parity bit, this factor graph takes exactly the form of an MSGM-derived factor graph, i.e., at each check node there is at most one ``terminating'' edge coming from either side.
By running the construction in reverse, one could also retrieve a valid MSGM from this factor graph.
More importantly, because of the previous discussion in this section, we can turn this factor graph into an MPG for each bit-transmission channel using \Cref{proc:fg_to_mpg} such that the resulting MPG has a maximum variable dimension of $4$.
If we group the two neighboring factors corresponding to $x_i^s$ and $x_i^p$ together for each $i$, as depicted by the gray boxes in \Cref{fig:convo_example_fg}, then the maximum variable dimension can be reduced to $3$.

We note that \Cref{fig:convo_example_fg} only depicts the ``bulk'' of the factor graph which exhibits the depicted repeated pattern.
Special care has to be taken when constructing the MPG nodes of the first few and the last few codeword bits, as the exact behavior of the MSGM will depend on the exact details of the initialization and termination procedure of the convolutional code.

\section{BPQM decoding of turbo codes}
Previously, in \Cref{sec:bpqm_decoding}, we saw that BPQM achieves bit-optimal and block-optimal decoding for codes where an efficient MPG representation of the bit-transmission channels is available.
Later, in \Cref{sec:codes}, we saw that such efficient MPG representations can be constructed for various code families which exhibit some efficient tree factor graph representation, such as a tree Tanner graph or a trellis with small state space size.
Unfortunately, many of the best performing classical codes do not exhibit such a tree-like structure.
Fortunately, this doesn't necessarily imply that BPQM cannot be effective in decoding these codes --- some simple modifications to the decoding procedure can be sufficient in order to successfully apply BPQM.
However, as a consequence of these modifications, the bit- and block-optimality arguments from \Cref{sec:bpqm_decoding} are not applicable in these settings and the performance of BPQM must be characterized on a case-by-case basis.

In previous works, such BPQM-based decoding beyond tree factor graphs was considered for polar codes~\cite{renes_belief_2017} and LDPC codes~\cite{brandsen_bpqm_2022}.
The main trick to apply BPQM to polar codes is the realization that the factor graph induced by successive cancellation decoding is a tree, which can therefore be translated into an MPG.
A quantum version of the polarization argument~\cite{wilde_polarcodes_2013}, together with the quantum union bound~\cite{gao_unionbound_2015}, then suffice to prove that BPQM-based decoding can achieve the Holevo capacity.

The BPQM-based decoding procedure for LDPC codes is more closely related to the techniques we will later consider in this section.
While LDPC codes do not exhibit a tree-like structure, they can be seen as being \emph{locally} tree-like.
More formally, in the limit of large block sizes, the environment of a codeword bit in the Tanner graph is almost certainly a tree.
As such, the computation graph associated to a message-passing decoder that is truncated to a finite number of steps will not encounter any loops in the asymptotic limit, which enables us to translate this computation graph to an MPG and run BPQM on it.
The asymptotic performance of BPQM on regular LDPC codes was investigated using density evolution in \cite{brandsen_bpqm_2022}.

\begin{figure}
    \centering
    \includegraphics{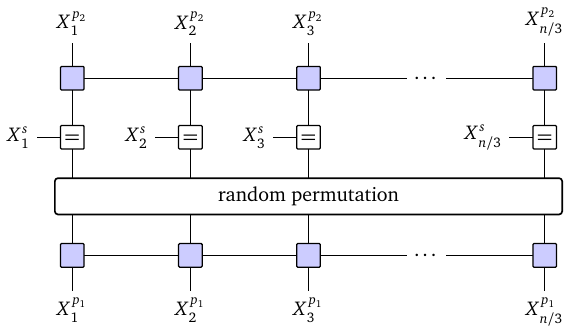}
    \caption{Forney-style factor graph representation of the code inclusion function of a parallel concatenated rate-$1/3$ turbo code of blocksize $n$.
    Here $X_i^s$ denotes the $i$-th systematic bit and $X_i^{p_1},X_i^{p_2}$ denote the $i$-th parity bits.
    The blue nodes at the top and at the bottom each correspond to the factor graph representation of the constituent convolutional code.
    }
    \label{fig:turbo_fg}
\end{figure}

In this section, we will extend the applicability of BPQM to another family of binary linear codes which do not exhibit a tree factor graph structure: turbo codes.
These codes exhibit excellent performance for classical channel coding and they closely approach the Shannon limit~\cite{berrou_turbocodes_1993}.
In this section, we will show that they also approach the Holevo limit in classical-quantum channel coding when paired with BPQM.
For simplicity, we will restrict our attention to parallel concatenated turbo codes without puncturing, i.e., with rate $1/3$.
We refer the reader to~\cite{richardson_modern_2008} for a precise definition of these codes, and from now on, we will assume basic familiarity with turbo codes.
We depict the Forney-style factor graph structure of such a code in \Cref{fig:turbo_fg}.
It essentially consists of two copies of the factor graph of the constituent convolutional code which are interleaved with a random permutation.
Furthermore, we will only focus on the task of bit decoding on these turbo codes.
To lift the results from bit error rates to block error rates, one can use the quantum union bound~\cite{gao_unionbound_2015} (which is applicable because of \Cref{lem:coherent_bpqm_projection}), provided that the bit error rates are sufficiently small.

The general strategy for decoding turbo codes with BPQM is similar as for LDPC codes, namely, a truncated message-passing decoder is executed for a finite number of iterations.
More concretely, we employ a \emph{windowed} message passing schedule, parametrized by a window size $w\in\mathbb{N}$ and the number of steps $l\in\mathbb{N}$.
The schedule operates as follows: To decode a given systematic bit, messages are required from both the upper and lower constituent convolutional codes.
Each of these messages is computed by performing BPQM on the corresponding trellis within a symmetric window of size $w$, that is, by restricting the message passing to the $w$ trellis sections to the left as well as to the right of the target bit.
The resulting messages for the neighboring $2w$ systematic bits are obtained in the same way, by running local windowed decoding on the opposite constituent code. This process is iterated $l$ times, enabling information to propagate gradually between the two trellises through the interleaver while keeping the decoding complexity bounded.
For a more detailed explanation of windowed decoding, we refer to~\cite[Sections 6.4]{richardson_modern_2008}.
We note that for each trellis section, we need to choose some initial message at the boundaries of the window.
We choose the value of that message to be associated with the completely information-destroying channel $\spsc[P_0]$ where $P_0$ is the $m$-bit distribution with $P_0(0^m)=1$ and $m$ is the number of memory cells in the convolutional code.
This mirrors the classical strategy, where these messages are typically initialized as the uniform distribution.

The computation tree resulting from such a windowed decoder is asymptotically tree-like, i.e., the probability that it contains a loop converges to zero as the block size of the code goes to infinity.
As such, the computation tree can be converted to a MPG by the considerations in \Cref{sec:trellis}.
One can characterize the asymptotic performance of BPQM by density evolution, i.e., by tracking the distribution of the classical part of the messages that are transmitted by incoherent BPQM over this MPG.
For a more detailed explanation of density evolution on turbo codes, we refer to~\cite[Sections 6.4]{richardson_modern_2008}.

We numerically implement density evolution for various parallel concatenated rate-$1/3$ turbo codes with the constituent convolutional codes parametrized by the rational function $G(D)=p(D)/q(D)$.
More concretely, we consider the three code families $G=5/7$, $G=13/15$ and $G=23/33$ where the polynomials $p,q$ are denoted in standard octal notation.
These codes have memory size $m=2$, $m=3$ and $m=4$ respectively.
We will assume that the codeword bits are transmitted over independent copies of the binary-input pure-state channel $W_{\omega}\coloneq\bspsc[\frac{1}{2}-\sqrt{\omega(1-\omega)}]$ for some parameter $\omega\in[0,1/2]$.
This parametrization is natural, because $W_{\omega}$ acts, up to a unitary basis change on its output, as
\begin{equation}
    W_{\omega}(x) = \sqrt{1-\omega}\ket{x} + \sqrt{\omega}\ket{1-x}
\end{equation}
for $x\in\ff_2$.
Therefore, by performing the optimal distinguishing measurement on the output of $W_{\omega}(x)$, it degrades to the binary symmetric channel $\bsc_{\omega}$ with error probability of $\omega$.
The amount of information that can be transmitted over $W_{\omega}$ using a classical decoding strategy is precisely characterized by the Shannon bound of $\bsc_{\omega}$, whereas a truly quantum decoder (such as BPQM) can surpass that Shannon bound and achieve rates up to the Holevo bound of $W_{\omega}$.

\begin{figure}
    \centering
    \includegraphics{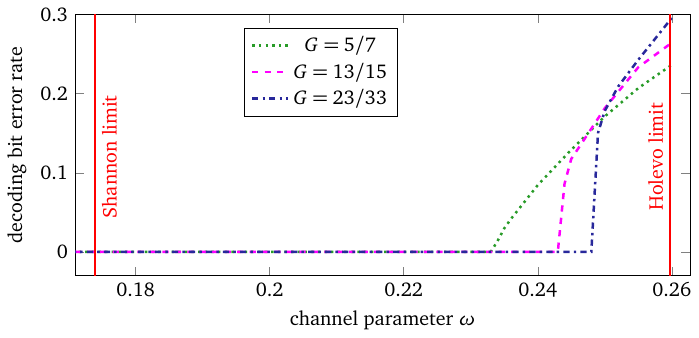}
    \caption{Asymptotic bit error rates of BPQM for various rate-$1/3$ parallel concatenated turbo codes obtained from density evolution.
    The red vertical lines depict the Shannon and Holevo bounds which characterize the maximal tolerable noise rates $\omega$ at which reliable transmission is possible with rate $1/3$ over $\bsc_{\omega}$ and $W_{\omega}$ respectively.
    }
    \label{fig:turbo_code_results}
\end{figure}

To numerically perform density evolution, we utilize the Monte Carlo density evolution technique described in~\cite{brandsen_bpqm_2022}.
We fix the parameters $l=200$ and $w=200$ and utilize a population size of $10000$.
In \Cref{fig:turbo_code_results}, we depict the expected bit error rate associated with the decoding of a systematic bit for each of our three chosen turbo code families.
The points where the lines reach $0$ indicate the location of the BPQM thresholds on the corresponding code families.
We summarize the estimated thresholds in \Cref{tab:turbo_code_results}.
Observe that all three code families exhibit a threshold that clearly exceeds the Shannon bound.
Furthermore, the thresholds lie close to the Holevo bound.

\section*{Acknowledgements}
We thank André Chailloux, Anthony Leverrier, Avijit Mandal, Henry Pfister, and Jean-Pierre Tillich for insightful discussions.
CP acknowledges financial support from the the Swiss National Science Foundation (SNSF) Postdoc.Mobility fellowship P500PT\_230724. JMR acknowledges support from the SNF under grant number 20QU-1\_225224 as well as the ETH Quantum Center and the SNF NCCR SwissMAP.  

\FloatBarrier
\appendix
\section{Formal definition of discretized incoherent BPQM}\label{app:discretized_incoherent_bpqm}
Here, we provide a formal version of \Cref{def:discretized_incoherent_bpqm_informal}.
In order to obtain a discretized BPQM algorithm valid in a model of computation without the capability to store arbitrary real numbers with infinite accuracy, we modify incoherent BPQM in two ways.

First, the classical part of the messages is replaced with an approximate discretized representation thereof, as described in the main text.
For convenience, define the values that are representable with a $B$-bit fixed-point representation:
\begin{equation}
    \mathfrak{Q}_B \coloneqq \left\{ \frac{i}{2^B} \middle\vert i=0,\dots,2^B-1 \right\} \subset [0,1) \, .
\end{equation}
We will assume that basic operations (addition, multiplication, division, square root, arcsin) of elements in $\mathfrak{Q}_B$ can be performed in $\mathrm{poly}(B)$ complexity.
Then, we denote the set of discretized $m$-bit distributions by
\begin{equation}
  \mathfrak{D}_{B,m} \coloneqq \left\{ p:\ff_2^m\rightarrow\mathfrak{Q}_B \middle\vert \sum_{x\in\ff_2^m}p(x)=1 \right\}
  \, .
\end{equation}
Hence, an element $p\in\mathfrak{D}_{B,m}$ can be represented using a $2^mB$-bit string.
For convenience, we will often treat elements of $\mathfrak{D}_{B,m}$ interchangeably as elements of $\ff_2^{2^mB}$, e.g., $\ket{p}$ is understood as a computational basis state of a $2^mB$-qubit register.
For convenience, we will also define a rounding function $\mathfrak{r}_{B,m}$ which maps an $m$-bit distribution $p$ to the (or one of the) closest element in $\mathfrak{D}_{B,m}$.
It generally holds that
\begin{equation}
    \norm{ \mathfrak{r}_{B,m}(q) - q}_1 \leq 2^{m-B}
\end{equation}
for any $m$-bit distribution $q$~\cite[Lemma 3]{bocherer_optimal_2016}.

Second, for the primitive node operation applied within each node operation, we need an efficient procedure to construct an efficient circuit that is a good approximation to the desired unitary.
This is possible, as captured by the following lemma.
\begin{lemma}\label{lem:discretized_pno}
    Consider some instance $\sdt(n,k,\ell,G,P)$ and $B\in\mathbb{N}$.
    Let $Q$ be an $n$-qubit register and $E$ a $2^nB$-qubit register.
    We construct an explicit quantum circuit $C$ on $QE$ such that
    \begin{itemize}
        \item $C$ only contains unitary operations,
        \item $C$ implements the operation
        \begin{equation}
            \sum\limits_{\tilde{P}\in\mathfrak{D}_{B,n}} \pno(n,k,\ell,G,P'(\tilde{P}))_Q \otimes \proj{\tilde{P}}_E\,,
        \end{equation}
        where for any $\tilde{P}\in\mathfrak{D}_{B,n}$, $P'(\tilde{P})$ is some $n$-bit distribution such that $\norm{P'(\tilde{P}) - \tilde{P}}_1\leq (\pi+1)2^{n-B}$,
        \item $C$ has circuit complexity $\mathcal{O}(\mathrm{poly}(B)4^nn^2)$.
    \end{itemize}
\end{lemma}
Here, we perform the operations conditional on the classical distribution coherently. 
This will be useful in \Cref{app:coherent_bpqm}. 
We note this statement is not entirely trivial, as a naive implementation of the gate $\sum_{\tilde{P}}\pno(n,k,\ell,G,\tilde{P})\otimes\proj{\tilde{P}}$ would have a complexity exponential in $2^nB$.
The proof, including the explicit circuit construction, of this lemma is provided at the end of this section.
We now formally define discretized incoherent BPQM in full analogy to \Cref{def:nodeop} and \Cref{def:incoherent_bpqm}.
\begin{definition}[Discretized node operation]\label{def:discretized_nodeop}
Consider a node $v$ of some MPG and some fixed $B\in\mathbb{N}$.
\begin{problemdescr}
    \item[Input:] For each outgoing edge $e^{\mathrm{out}}_i(v)$, $i=1,\dots,m_v$: a message consisting of an $n_{e^{\mathrm{out}}_i(v)}$-qubit system $Q_{e^{\mathrm{out}}_i(v)}$ and a discretized distribution $P_{e^{\mathrm{out}}_i(v)}\in\mathfrak{D}_{B,n_{e^{\mathrm{out}}_i(v)}}$.
    \item[Output:] An $n_{e^{\mathrm{in}}(v)}$-qubit system $Q_{e^{\mathrm{in}}(v)}$ and a discretized distribution $P_{e^{\mathrm{in}}(v)} \in \mathfrak{D}_{B,n_{e^{\mathrm{in}}(v)}}$.
    \item[Algorithm:]
    Denote the joint $n_v$-qubit quantum system $\bar Q\coloneqq\bigotimes_{i=1}^{m_v}Q_{e^{\mathrm{out}}_i(v)}$.
    \begin{enumerate}
        \item Compute and store the joint distribution $\bar P\coloneqq \mathfrak{r}_{B,n_v}(P_{e_1^{\mathrm{out}}(v)}\times P_{e_2^{\mathrm{out}}(v)} \times \dots \times P_{e_{m_v}^{\mathrm{out}}}(v))$.
        \item Prepare a $2^{n_v}B$-qubit quantum register $F$ in the state $\ket{{\bar P}}_F$ and apply the quantum circuit from \Cref{lem:discretized_pno} on $\bar{Q}$ and $F$.
        Partition the resulting quantum system $\bar{Q}$ into three parts $Q_1\otimes Q_2\otimes Q_3$ consisting of $\ell_v,(k_v-\ell_v),(n_v-k_v)$ qubits each.
        \item Measure $Q_2$ in the computational basis, obtaining the result $s$.
        \item Define $P_{Y,S,A}(y,s,a)\coloneqq \bar P(M^{-1}(y\cat s\cat a))$ using the $n_v\times n_v$ matrix $M$ from the primitive node operation.
        Compute and store the distribution $P'\coloneqq \mathfrak{r}_{B,\ell_v}(P_{Y|S=s})$.
        \item Output the quantum system $Q_1$ and the distribution $P'$.
    \end{enumerate}
\end{problemdescr}
\end{definition}
\begin{definition}[Discretized incoherent BPQM algorithm]\label{def:discretized_bpqm}
Consider an instance $\sdt(n,k,\ell,G,P)$ where the associated channel $W$ can be written as $W=\mathcal{W}[\mathcal{G},\mathcal{P}]$ for some MPG $\mathcal{G}$ and a sequence of probability distributions $\mathcal{P}=(P_i)_{i=1,\dots,m}$.
Also, let $B\in\mathbb{N}$.
\begin{problemdescr}
    \item[Input:] An $n$-qubit quantum system denoted $Q$ in the state $W(x)$, where $x\in\ff_2^{\ell}$.
    \item[Output:] An estimate $\hat x\in \ff_2^{\ell}$ for $x$. 
    \item[Algorithm:]
    Let us denote the root edge of $\mathcal{G}$ by $e^{\star}$.
    Partition the quantum system into subsystems $Q=\bigotimes_{i=1}^mQ_i$ where $Q_i$ corresponds to the output of $\spsc[P_i]$.
    Accordingly denote the leaf edges of $\mathcal{G}$ by $e_1,\dots,e_m$.
    \begin{enumerate}
        \item For $i=1,\dots,m$, fix the message associated with the edge $e_i$ to be the quantum system $Q_i$ together with the discretized distribution $\mathfrak{r}_{B,n_{e_i}}(P_i)$.
        \item Proceeding from leaves to root, perform node operations according to \Cref{def:discretized_nodeop}, with $B$, until all edges have an associated message.
        \item Apply the Hadamard transform $H^{\otimes n_{e^{\star}}}$ on the quantum system $Q_{e^{\star}}$ of the final message.
        \item Measure $Q_{e^{\star}}$ in the computational basis and set $\hat x$ to the resulting measurement outcome. 
    \end{enumerate}
\end{problemdescr}
\end{definition}
The complexity of discretized incoherent BPQM is dominated by the node operations.
All of the five steps in \Cref{def:discretized_nodeop} each have complexity of $\mathcal{O}(\mathrm{poly}(B)4^{n_v}n_v^2)$
\footnote{
The complexity of computing $\bar P$ is at most $2^{n_v}\cdot m_v \cdot \mathrm{poly}(m_vB) = \mathcal{O}(4^{n_v}\mathrm{poly}(B))$ since we are computing the product of $m_v$ $m_vB$-bit numbers $2^{n_v}$ times.
The vector $M^{-1}(y\cat s\cat a)$ can be computed (exactly) in time $n_v^2$ (respectively $n_v^3$ if the matrix inverse is not computed offline).
Therefore, $P_{Y,S,A}$ can be exactly computed and stored in the time $n_v^2$.
The computation and storage of $P_S(s)$ and the vector $P_{Y,S}(\cdot,s)$ (for the given $s$)
can be done in time $2^{n_v}\mathrm{poly}(B)$ (where the $\mathrm{poly}(B)$ term comes from addition), and importantly, without any discretization errors, since $P_{Y,S}\in\mathfrak{D}_{B,k_v}$ and $P_{S}\in\mathfrak{D}_{B,k_v-\ell_v}$.
To compute $P'$, it remains to perform $2^{\ell_v}$ divisions, which has complexity $\mathcal{O}(2^{n_v}\mathrm{poly}(B))$.
}
, and hence, the total complexity of discretized incoherent BPQM is $\mathcal{O}(\mathrm{poly}(B)4^{\mathcal{N}(G)}\mathcal{N}(G)^2\abs{V})$.

For the rest of this section, our aim is to prove \Cref{lem:discretized_pno}.
The proof uses the following two intermediate results.
\begin{lemma}\label{lem:uniformly_controlled_gates}
    Consider a family of $m$-qubit unitaries $(U_i)_{i\in\ff_2^n}$, and assume that for each $i\in\ff_2^n$ we are given a unitary circuit that realizes $U_i$ with complexity $\leq C$.
    Then, there exists a circuit implementation of the unitary $\sum_{i\in\ff_2^n} U_i \otimes \proj{i}$ with complexity $\mathcal{O}(2^nnC)$.
\end{lemma}
\begin{proof}
    Apply the circuits for $U_1,\dots,U_{2^n}$ sequentially, with each gate replaced by an $n$-controlled version of itself.
\end{proof}
\begin{lemma}\label{lem:approx_state_prep}
    Fix some $B,n\in\mathbb{N}$.
    There exists a quantum circuit of complexity $\mathcal{O}(4^{n}n^2\mathrm{poly}(B))$ which acts as follows: For any $p\in\mathfrak{D}_{B,n}$ it maps
    \begin{equation}\label{eq:generalized_state_prep}
        \ket{p}\otimes \ket{0^n} \mapsto \ket{p}\otimes \left( \sum_{x\in\ff_2^n} \sqrt{\tilde{p}(x)}\ket{x} \right)
    \end{equation}
    for some $n$-bit probability distribution $\tilde{p}$ with $\norm{p-\tilde{p}}_1\leq \pi n2^{-B}$.
\end{lemma}
\begin{proof}
    Let's write the probability distribution as $p=p_{X_1,\dots,X_n}$ where $X_i$ are the $n$ involved bits.
    For a number $q\in[0,1]$, let us denote by $r(q)$ the closest (or one of the closest) number in $\mathfrak{Q}_B$.

    We begin by considering the somewhat simpler state preparation problem of realizing a circuit with the action $\ket{0^n}\mapsto \sum_x \sqrt{p(x)}\ket{x}$ for the given $n$-bit distribution $p$. 
    Möttönen \emph{et al.}\ proposed a method to perform this task in \cite{mottonen_statetrafo_2005}.
    The idea is to work sequentially, rotating each subsequent qubit by the correct probability distribution conditioned on all previous qubit values. 
    More precisely, the circuit proceeds by applying the sequence of gates $U_i=V_i\otimes \id^{\otimes n-i}$ for $i=1,\dots,n$, where $V_i$ is the uniformly controlled rotation gate 
    \begin{equation}
        V_i \coloneq \sum_{z\in\ff_2^{i-1}} \proj{z} \otimes \mathrm{R}_y(\alpha_{i|z_1,z_2,\dots,z_{i-1}})\,, 
    \end{equation}
    with 
    \begin{equation}
        \mathrm{R}_y(\theta)\coloneqq\begin{pmatrix}\cos\tfrac{\theta}{2} & -\sin\tfrac{\theta}{2} \\ \sin\tfrac{\theta}{2} & \cos\tfrac{\theta}{2}\end{pmatrix}
    \end{equation}
    and
    \begin{equation}
        \alpha_{i|z_1,\dots,z_{i-1}} \coloneq 2\arcsin\left(\sqrt{ p_{X_i|X_{1}=z_1,X_{2}=z_2,\dots,X_{i-1}=z_{i-1}}(1) }\right)
        \, .
    \end{equation}
    Thus, applying $U_1$ yields the state
    \begin{equation}
        U_1\ket{0^n}=\left( \sqrt{p_{X_1}(0)}\ket{0} + \sqrt{p_{X_1}(1)}\ket{1} \right) \otimes \ket{0^{n-1}}\,,
    \end{equation}
    and subsequently applying $U_2$ gives 
    \begin{equation}
        U_2U_1\ket{0^n}=\sum_{x\in\ff_2^2}\sqrt{p_{X_1,X_2}(x)}\ket{x} \otimes \ket{0^{n-2}}\,.
    \end{equation}
    A simple induction shows that the desired state is indeed produced after applying all $n$ uniformly-controlled gates $U_i$.

    We can now simply adapt this procedure for the more general task in \Cref{eq:generalized_state_prep}.
    Let us denote the two registers in \Cref{eq:generalized_state_prep} by $A$ and $C$.
    The procedure makes use of an additional $B$-qubit scratch register $E$ and proceeds as follows. For $i=1,\dots,n$: 
        \begin{enumerate}
            \item Coherently compute the angle $\theta =r(\tfrac{1}{\pi}\alpha_{i|z_1,\dots,z_{i-1}})$ from the register $A$, uniformly controlled on the values $z_1,\dots,z_{i-1}$ of the first $i-1$ qubits in $C$.
            Store the angle in $E$.
            \item Apply the gate $\sum_{\theta\in\mathfrak{Q}_B} \mathrm{R}_y(\pi\theta)_{C_i} \otimes \proj{\theta}_E$.
            \item Un-compute the register $E$ by repeating the first step.
        \end{enumerate}
    This procedure prepares the state $\ket{p}_A\otimes\sum_x\sqrt{\tilde{p}(x)}\ket{x}_C$, where $\tilde{p}$ is implicitly defined by
    \begin{equation}
        \frac{2}{\pi}\arcsin\left(\sqrt{ \tilde{p}_{X_i|X_{1}=z_1,X_{2}=z_2,\dots,X_{i-1}=z_{i-1}}(1) }\right)
        = 
        r\left( \frac{2}{\pi}\arcsin\left(\sqrt{ p_{X_i|X_{1}=z_1,X_{2}=z_2,\dots,X_{i-1}=z_{i-1}}(1) }\right) \right)
    \end{equation}
    for $i=1,\dots,n$.
    Since $f(x)=\sin(\tfrac{\pi}{2} x)^2$ is $(\pi /2)$-Lipschitz continuous, this directly implies that
    \begin{equation}
        \abs{
            \tilde{p}_{X_i|X_{1}=z_1,X_{2}=z_2,\dots,X_{i-1}=z_{i-1}}(z)
            - p_{X_i|X_{1}=z_1,X_{2}=z_2,\dots,X_{i-1}=z_{i-1}}(z)
        } \leq \frac{1}{2}\pi 2^{-B} \leq \pi2^{-B}
    \end{equation}
    for all $i=1,\dots,n$ and for all $z\in\ff_2^n$.
    Therefore, by the triangle inequality, 
    \begin{align}
        \norm{p-\tilde{p}}_1
        &= \sum_{x\in\ff_2^n} \abs{p_{X_1,\dots,X_n}(x_1,\dots,x_n) - \tilde{p}_{X_1,\dots,X_n}(x_1,\dots,x_n)} \\
        &= \sum_{x\in\ff_2^n} \abs{p_{X_1}(x_1)p_{X_2|X_1=x_1}(x_2)\cdots p_{X_n|X_{1}=x_{1}\dots X_{n-1}=x_{n-1}}(x_n) - \tilde{p}_{X_1}(x_1)\tilde{p}_{X_2|X_1=x_1}(x_2)\cdots \tilde{p}_{X_n|X_{1}=x_{1}\dots X_{n-1}=x_{n-1}}(x_n)} \\
        &\leq \sum_{x_1\in\ff_2}\abs{p_{X_1}(x_1)-\tilde{p}_{X_1}(x_1)}
        + \sum_{x_1,x_2\in\ff_2}\tilde{p}_{X_1}(x_1)\abs{p_{X_2|X_1=x_1}(x_2) - \tilde{p}_{X_2|X_1=x_1}(x_2) }
        + \cdots \\
        &\leq \pi n2^{-B} \, .
    \end{align}

    It remains to discuss the complexity of realizing the entire operation.
    The computation of the conditional marginal $P_{X_i|X_1=z_1,\dots,X_{i-1}=z_{i-1}}$ takes $\mathcal{O}(4^nn\mathrm{poly}(B))$ time: The $2^n$ additions followed by the division lead to a $2^n\mathrm{poly}(B)$ contribution, and the uniform control on the $i\leq n$ qubits in $C$ register contributes a factor of $2^nn$ (see \Cref{lem:uniformly_controlled_gates}).
    Subsequently, the computation of the arcsin of the square root can be done in complexity $\mathcal{O}(\mathrm{poly}(B))$.
    Note that the intermediate step of computing the conditional marginal may require an increase in precision, but no more than a constant increase in working precision will be needed. 

    Hence, the complexity of Step 1 and Step 3 is $\mathcal O(4^nn\mathrm{poly}(B))$. 
    Meanwhile, the operation in Step 2 can be realized through a sequence of $B$ controlled-$\mathrm{R}_y$ rotations which are controlled on $E_j$ and target $C_i$ for $j=1,\dots,B$.
    Its complexity is hence linear in $B$.
    The total cost is therefore $\mathcal{O}(4^nn^2\mathrm{poly}(B))$.
\end{proof}

\begin{proof}[Proof of \Cref{lem:discretized_pno}]
    As discussed in \Cref{def:pno}, the unitary $\pno(n,k,\ell,G,P)$ can be decomposed into two parts: $U_1$ and $U_2$.
    The unitary $U_1$ is static (i.e., it does not depend on the input distribution), so its generation can be done fully offline.
    The complexity of the circuit of $U_1$ is $\mathcal{O}(n^2)$~\cite{patel_synthesis_2008}.

    It remains to approximate the unitary
    \begin{equation}
        \sum_{s,y,\tilde{P}} \proj{y}_{Q_1}\otimes\proj{s}_{Q_2}\otimes V(y,s,\tilde{P})_{Q_3} \otimes \proj{\tilde{P}}_B\,, 
    \end{equation}
    where we denote by $V(y,s,\tilde{P})$ the unitary $V(y,s)$ from the definition of $U_2$ associated with the primitive node unitary with distribution $\tilde{P}$.
    We can approximate this unitary with the following steps:
    \begin{enumerate}
        \item Coherently compute $\mathfrak{r}_{B,n-k}(\tilde{P}_{A|Y=y,S=s})$, coherently conditioned on the system $Q_1$ and $Q_2$ being in the state $\ket{y}_{Q_1}\otimes\ket{s}_{Q_2}$, and store the result in an ancilla $2^{n-k}B$-qubit register $F$.
        \item Apply the reversed circuit from \Cref{lem:approx_state_prep} to $FQ_3$.
        \item Uncompute the ancilla register $F$ by repeating the first step.
    \end{enumerate}
    The complexity of the first and third step scales as $\mathcal{O}(4^nn\mathrm{poly}(B))$, where the uniform control on $Q_1Q_2$ contributes a factor $2^nn$ (see \Cref{lem:uniformly_controlled_gates}) and the computation of the marginal distribution contributes a factor $2^n\mathrm{poly}(B)$.

    Note that for each $y,s$, the unitary $V(y,s,\tilde{P})$ only depends on the conditional distributions $\tilde{P}_{A|Y=y,S=s}$.
    Due to the discretized storage of the marginal in $F$, as well as the approximation involved in \Cref{lem:approx_state_prep}, we do effectively apply the gate $V(y,s,\tilde{P})$ with an erroneous value of the conditional marginal, which we denote by $P'_{A|Y=y,S=s}$ and which fulfills
    \begin{equation}
        \forall y,s: \norm{\tilde{P}_{A|Y=y,s=s} - P'_{A|Y=y,S=s}}_1\leq n2^{-B}\pi + 2^{n-B} \leq (\pi +1)2^{n-B}
    \end{equation}
    due to a simple triangle inequality argument.
    We can complete these marginals $P'_{A|Y=y,S=s}$ to a complete joint distribution by defining it as $P'_{Y,S,A}(y,s,a)\coloneqq \tilde P_{S,Y}(s,y)P'_{A|Y=y,S=s}(a)$.
    Correspondingly, let $P'$ be the $n$-bit distributions such that $P'_{Y,S,A}(y,s,a)=P'(M^{-1}(y\cat s\cat a))$.
    We write $P'(\tilde P)$ to highlight that $P'$ depends on $\tilde P$.
    In summary, our circuit implements    
    \begin{equation}
        \sum_{s,y,\tilde{P}} \proj{y}_{Q_1}\otimes\proj{s}_{Q_2}\otimes V(y,s,P'(\tilde{P}))_{Q_3} \otimes \proj{\tilde{P}}_B\,, 
    \end{equation}
    and it holds that
    \begin{equation}
        \forall\tilde{P}: \norm{P'(\tilde{P}) - \tilde{P}}_1 = \norm{P'_{Y,S,A}-\tilde{P}_{Y,S,A}} = \sum_{y,s} \tilde{P}_{Y,S}(y,s) \norm{ P'_{A|Y=y,s=s} - \tilde{P}_{A|Y=y,S=s} }_1 \leq (\pi +1)2^{n-B}
        \, .
    \end{equation}
\end{proof}

\section{Detailed description of coherent BPQM}\label{app:coherent_bpqm}
In \Cref{def:coherent_bpqm}, we defined the coherent BPQM algorithm as the result of applying the deferred measurement principle to the discretized incoherent BPQM algorithm.
To be more explicit, this appendix provides a detailed step-by-step description of the algorithm.

\begin{definition}[Coherent BPQM node operation]\label{def:coherent_bpqm_nodeop}
Consider a node $v$ of some MPG and some fixed $B\in\mathbb{N}$.
The node operation of coherent BPQM is identical to \Cref{def:discretized_nodeop}, except for the following changes:
\begin{itemize}
    \item The classical registers storing the input distributions $P_{e_i^{\mathrm{out}}(v)}$ and the output distribution $P_{e^{\mathrm{in}}(v)}$ are replaced by $2^mB$-qubit quantum registers
    \item The computation of $\bar P$ in step 1 is done coherently, and the result is stored in a $2^mB$-qubit quantum register $F$
    \item Step 2 is realized with the reversed circuit from \Cref{lem:discretized_pno} applied to $\bar{Q}F$.
    \item The measurement in step 3 is omitted.
    \item The computation of $P'$ is done coherently and uniformly conditioned on the system $Q_2$. Denote the $2^{n_{e^{\mathrm{in}}}(v)}B$-qubit system storing the result by $E'$.
    \item At the end of the node operation, the register $F$ is uncomputed.
    \item The output of the node operation is given by $Q_1$ together with $E'$.
\end{itemize}
\end{definition}
The coherent BPQM algorithm is then obtained by chaining together these node operations on the MPG.
\begin{definition}[Coherent BPQM algorithm]
  The coherent BPQM algorithm is identical to \Cref{def:discretized_bpqm}, except for the following changes:
  \begin{itemize}
    \item In step 1, the message associated with the edge $e_i$ is fixed to be the quantum system $Q_i$ together with a $2^{n_{e_i}}B$-qubit register initialized in the computational state $\big|\mathfrak{r}_{B,n_{e_i}}(P_i)\big\rangle$
    \item In step 2, the node operation from \Cref{def:coherent_bpqm_nodeop} is used instead of the one from \Cref{def:discretized_nodeop}.
  \end{itemize}
\end{definition}
We now briefly discuss the complexity of coherent BPQM.
When comparing \Cref{def:coherent_bpqm_nodeop} with \Cref{def:discretized_nodeop}, the only step where the asymptotic complexity changes is the computation of $P'$, due to the uniform control on $Q_2$.
More precisely, the computation of $P_S(s)$ and the vector $P_{Y,S}(\cdot ,s)$ from $\bar{P}$ (which are needed to compute $P_{Y|S=s}$) must now be coherently controlled on the value $s$ of $Q_2$.
This increases the circuit complexity from $\mathcal{O}(2^{n_v}\mathrm{poly}(B))$ to $\mathcal{O}(4^{n_v}n_v\mathrm{poly}(B))$ (because of \Cref{lem:uniformly_controlled_gates}).

This change is asymptotically irrelevant, as the overall complexity of the node operation remains as $\mathcal{O}(\mathrm{poly}(B)4^{n_v}n_v^2)$.
Therefore, the total complexity of coherent BPQM is the same as that of discretized incoherent BPQM, i.e., it is $\mathcal{O}(\mathrm{poly}(B)4^{N(\mathcal{G})}N(\mathcal{G})^2\abs{V})$.

\section{Uniformly-controlled BPQM algorithm}\label{app:formalization_coherent_bpqm}
In \Cref{sec:bpqm_coherent}, we introduced the notion of \emph{coherent BPQM} by applying the deferred measurement principle to incoherent BPQM.
As discussed in that section, this only provides us with a well-defined quantum algorithm if we consider the \emph{discretized} version of incoherent BPQM, as otherwise, the quantum register storing the distributions would need to be infinite dimensional.
This puts us in a slightly uncomfortable situation that we do not have a precise notion of what an idealized version of coherent BPQM would even mean.

In this appendix, we address this problem by introducing yet another quantum algorithm, which we call \emph{uniformly-controlled BPQM} (UC-BPQM).
UC-BPQM is a well-defined quantum algorithm (i.e.\ acting only on finite-dimensional spaces) and can be viewed as a formalization of what computation an ``idealized coherent BPQM algorithm'' (i.e.\ without any sort of discretization errors) should perform.
As we will see, it achieves both bit-optimal and block-optimal decoding, though it is not an efficient algorithm.
Instead, it will purely serve as a theoretical tool, as we will show that UC-BPQM can be efficiently approximated using (discretized) coherent BPQM.

We start by defining an incoherent version of UC-BPQM.
\begin{definition}[BPQM with lookup table]\label{def:bpqm_lookup}
    Consider the incoherent BPQM algorithm as described in \Cref{def:incoherent_bpqm} and the unitary $\pno(n_v,k_v,\ell_v,G_v,\bar P)$ that is performed as part of the node operation of some node $v$.
    Notice that all the parameters, except $\bar P$, are fixed by the MPG, and are therefore identical for every execution of incoherent BPQM on every possible input.
    Furthermore, notice that for a given set of distributions input to the leaves of the MPG, the value of $\bar P$ is entirely determined by the measurement outcomes (see step 2 in \Cref{def:nodeop}) incurred in all the preceding node operations (i.e., from node operations of nodes $v'$ which succeed $v$ in the MPG).
    We jointly denote these measurement outcomes by $m\in\ff_2^{p_v}$ where $p_v\coloneqq \sum_{v' \text{ succeeds } v} (k_v-\ell_v)$.

    As such, it is possible to remove the classical part of the messages in the incoherent BPQM algorithm, and instead, make all the primitive node unitaries classically controlled on all the preceding measurement outcomes.
    More precisely, for every node $v$, we pre-compute a lookup table of all possible primitive node operation unitaries $(U^{(v)}_m)_m$ for every possible combination $m$ of preceding measurement outcomes.
    Instead of choosing a primitive node operation based on the incoming messages, we instead simply pick the appropriate unitary from this lookup table.

    The resulting algorithm is called BPQM with lookup table and it also solves the subspace decoding task.
\end{definition}
Clearly, BPQM with lookup table and incoherent BPQM perform the identical quantum computation on the quantum system that is fed as input into the algorithm.
The difference purely lies in the fact that we now do not keep track of the distribution information on-the-fly.
Hence, BPQM with lookup table can also optimally solve the subspace decoding task.
The downside is that we now have to pre-compute exponentially many possible node unitaries in advance to fill up the lookup tables; this requires both exponential time and memory.

\begin{definition}[Uniformly-controlled BPQM algorithm]\label{def:ubpqm}
    Consider BPQM with lookup table from \Cref{def:bpqm_lookup}.
    This algorithm contains two different types of measurements: some in the node operations, and one final measurement of the final message.
    We can remove the former using the deferred measurement principle, analogous to the definition of coherent BPQM in \Cref{def:coherent_bpqm}.
    More precisely, for the node $v$, we replace the classically-controlled $U^{(v)}_m$ gate by the uniformly controlled quantum gate $\sum_m U^{(v)}_m\otimes\proj{m}$.

    The resulting algorithm is called uniformly-controlled BPQM (UC-BPQM) and it also solves the subspace decoding task.
\end{definition}
Just as BPQM with lookup table, UC-BPQM is also an inefficient algorithm, as describing and applying the involved uniformly controlled gates is exponentially hard.
However, UC-BPQM inherits its optimality in solving the subspace decoding task due to the deferred measurement principle.
We briefly note that UC-BPQM corresponds to the description of BPQM in the original paper that introduced it~\cite{renes_belief_2017}.

Note that UC-BPQM does not require any additional ancilla qubits in its computation, and hence, the action of UC-BPQM can be described by an $n$-qubit unitary $U_{\mathrm{UC-BPQM}}$ followed by the measurement of the first $\ell$ qubits in the conjugate basis, which provides the output of the algorithm.
The action of $U_{\mathrm{UC-BPQM}}$ is easy to describe.
\begin{lemma}\label{lem:ubpqm_action}
    Consider an instance $\sdt(n,k,\ell,G,P)$ where the associated channel $W$ can be written as $W=\mathcal{W}[\mathcal{G},\mathcal{P}]$ for some MPG $\mathcal{G}$ and a sequence of probability distributions $\mathcal{P}=(P_i)_{i=1,\dots,m}$.
    Let us write the message ensemble as $\mensemble_{\mathcal{G}}^{\mathcal{P}}=(p_s,D_s)_{s\in\ff_2^{k-\ell}}$.
    The unitary $U_{\mathrm{UC-BPQM}}$ acts as follows
    \begin{equation}
        \spsc[P]((x\cat r)^TG) \mapsto
        \sum_{s\in\ff_2^{k-\ell}}
        \sqrt{p_s}(-1)^{s\cdot Ar}
        \spsc[D_s](x)_{Q_1} \otimes \ket{s}_{Q_2} \otimes \ket{0^{n-k}}_{Q_3}\,, 
    \end{equation}
    where $x\in\ff_2^{\ell},r\in\ff_2^{k-\ell}$, $A\in\ff_2^{(k-\ell)\times(k-\ell)}$ is some invertible matrix and $Q_1,Q_2,Q_3$ are $\ell$-qubit, $(k-\ell)$-qubit and $(n-k)$-qubit registers.
\end{lemma}
\begin{proof}
    Consider the channel $F[\mathcal{G}]$ which describes passing messages from the root to the leaves on the MPG $\mathcal{G}$.
    Recall that at every node $v$, the input string is concatenated with some random string $r_v\in\ff_2^{k_v-\ell_v}$ and then encoded according to $G_v$.
    Let us denote the concatenation of all the random strings by $\bar{r}\coloneqq \cat_{v\in V}r_v$, which is a $(k-\ell)$-bit string.
    So if we fix some particular value $\bar{r}\in \ff_2^{k-\ell}$, the encoding map described by the MPG becomes deterministic.
    Put differently, $\mathcal{G}$ induces a one-to-one mapping between pairs $(x,\bar{r})\in\ff_2^{\ell}\times\ff_2^{k-\ell}$ and codewords $c\in\mathcal{C}$.
    This map is linear, since all the ``local'' constituent encoders of $F[\mathcal{G}]$ are linear.
    Therefore, there exists a matrix $G'\in\ff_2^{k\times n}$ such that $F[\mathcal{G}]$ describes the encoding $x\mapsto (x\cat\bar{r})G'$.

    By iteratively applying \Cref{eq:nodeop_purestate} for each node operation, we see that $U_{\mathrm{UC-BPQM}}$ maps
    \begin{equation}
        \spsc[P]((x\cat \bar r)^TG') 
        \mapsto 
        \sum_{s\in\ff_2^{k-\ell}}
        \sqrt{p_s}(-1)^{\bar r\cdot s}
        \spsc[D_s](x)_{Q_1} \otimes \ket{s}_{Q_2} \otimes \ket{0^{n-k}}_{Q_3}
    \end{equation}

    Clearly, $G'$ must be a generator matrix for the same code as $G$ since $F[\mathcal{G}]=F_{\ell,G}$, and the first $\ell$ rows of the two generator matrices must be identical.
    As such, the remaining rows only differ by a basis change.
    Put differently, there exists an invertible matrix $A\in\ff_2^{(k-\ell)\times (k-\ell)}$ such that $(x\cat r)^TG = (x\cat Ar)^TG'$.
    This directly implies the desired statement.
\end{proof}

The fact that UC-BPQM requires no ancilla is particularly relevant for the task of block decoding, as discussed in \Cref{sec:block_optimality}.
Consider the following algorithm to perform block decoding using UC-BPQM.
It is identical to \Cref{def:bpqm_block_decoding}, except that coherent BPQM is replaced by UC-BPQM.
\begin{definition}[UC-BPQM block decoding]\label{def:ubpqm_block_decoding}
Consider an instance $\blockdec(\mathcal{C},p)$ for some $[n,k]$ binary linear code where $x_1,\dots,x_k$ are linearly independent codeword bits.
Let $\mathcal{G}_1,\dots,\mathcal{G}_k$ be MPG representations of $\mathcal{F}_{\mathcal{C},1},\dots,\mathcal{F}_{\mathcal{C},k}$.
\begin{problemdescr}
    \item[Input:] $n$-qubit quantum system $Q$ in the state $\ket{\Psi_x}$ for some $x\in\mathcal{C}$.
    \item[Output:] An estimate $\hat{x}$ of $x$.
    \item[Algorithm:]
    Denote the unitary associated to UC-BPQM w.r.t. $\mathcal{G}_i$ by $U_{\mathrm{UC-BPQM},i}$. 
    For $i=1,\dots,k$ perform the following actions
    \begin{enumerate}
        \item Apply $U_{\mathrm{UC-BPQM},i}$ to $Q$
        \item Measure the first qubit of $Q$ in the conjugate basis, set $\hat x_i$ to the measurement result.
        \item Apply $U_{\mathrm{UC-BPQM},i}^{\dagger}$ to $Q$
    \end{enumerate}
        Finally, extend $\hat{x}_1,\dots,\hat{x}_k$ to a valid codeword in $\mathcal{C}$.
\end{problemdescr}
\end{definition}
Due to the unitarity of BPQM, this algorithm implements a sequence of projective two-outcome measurements, which realize bit-optimal decoding.
Therefore, we can apply \Cref{prop:bitoptimal_implies_blockoptimal}, which states that UC-BPQM realizes block-optimal decoding.
\begin{corollary}\label{cor:ubpqm_block_optimal}
    The algorithm in \Cref{def:ubpqm_block_decoding} achieves block optimal decoding.
\end{corollary}

\section{Resilience of BPQM under discretization errors}\label{app:discretization}
In this section, we will prove \Cref{thm:bit_optimal_discretized} and \Cref{thm:block_optimal}.
We will assume that the reader has already familiarized themselves with \Cref{app:discretized_incoherent_bpqm,app:coherent_bpqm,app:formalization_coherent_bpqm}.
The two proofs essentially amount to showing that the discretization errors in BPQM are well-behaved, i.e., we can efficiently approximate incoherent BPQM with discretized incoherent BPQM, and we can efficiently approximate UC-BPQM with coherent BPQM.

Before we get to the proofs themselves, we need to provide some additional insights.
The basis for the BPQM with lookup table algorithm was the observation that in incoherent BPQM, the value of $\bar P$ in the node operation depends only on the measurement outcomes $m$ of the preceding node operations.
Hence, the primitive node operation for the node $v$ can only take a finite number of possible values $(U^{(v)}_m)_m$.

The same insight also applies to the \emph{discretized} incoherent BPQM algorithm, i.e., the applied node operation unitary also only depends on the preceding measurement outcomes.
Here, we denote the list of possible unitaries associated to the node $v$ by $(\tilde{U}^{(v)}_m)_m$.
The two unitaries $U^{(v)}_m$ and $\tilde{U}^{(v)}_m$ differ because of discretization errors in the classical part of the messages and because of the error in \Cref{lem:discretized_pno}.
As part of the analysis in this appendix, we will show that $\tilde{U}^{(v)}_m$ and $U^{(v)}_m$ are close to each other with high probability, provided that the number of discretization bits $B$ is large enough.
\begin{definition}[discretized BPQM with lookup table]\label{def:bpqm_lookup_discretized}
    We define \emph{discretized BPQM with lookup table} to be the identical algorithm as \Cref{def:bpqm_lookup}, except that we use the unitaries $(\tilde{U}^{(v)}_m)_m$ from discretized incoherent BPQM as a lookup table instead of the ideal unitaries $(U^{(v)}_m)_m$.
\end{definition}
In full analogy to the construction of UC-BPQM, we can also apply the deferred measurement principle here, and get a unitary version of this algorithm.
\begin{definition}[Discretized uniformly-controlled BPQM]
    Consider discretized BPQM with lookup table from \Cref{def:bpqm_lookup_discretized} and apply the deferred measurement principle to replace the classically conditioned unitaries with uniformly controlled quantum gates (analogous to \Cref{def:ubpqm}).
    The resulting algorithm is called \emph{discretized uniformly-controlled BPQM} (DUC-BPQM).
\end{definition}
DUC-BPQM, analogously to UC-BPQM, is neither an efficient algorithm, nor a message-passing algorithm.
Instead, it will purely serve as an analytical tool to prove \Cref{thm:bit_optimal_discretized} and \Cref{thm:block_optimal}.
The performance of DUC-BPQM in solving the subspace decoding task is equal to that of discretized BPQM with lookup table, which is in turn equivalent to that of discretized incoherent BPQM.
    Hence, to prove \Cref{thm:bit_optimal_discretized}, it suffices to characterize the success probability of DUC-BPQM.

Similarly, the action of DUC-BPQM and coherent BPQM on any given input state is equivalent, except that coherent BPQM involves ancilla systems to store the discretized distribution information. 
More formally, consider some instance $\sdt(n,k,\ell,G,P)$ and an appropriate MPG $\mathcal{G}$.
The action of the DUC-BPQM on an $n$-qubit input system $Q$ is described by some $n$-qubit unitary $U_{\mathrm{DUC-BPQM}}$ followed by a projective measurement $\{\Pi_j\}_{j\in\ff_2^{\ell}}$. 
If we separate the output $Q$ into the three systems $Q=Q_1Q_2Q_3$, as in \Cref{lem:ubpqm_action}, then $\{\Pi_j\}_j$ is simply the conjugate basis measurement on $Q_1$.
Meanwhile, the action of coherent BPQM is can be described by some $(n+A)$-qubit unitary $U_{\mathrm{cBPQM}}$, acting on the input $Q$ and an $A$-qubit ancilla system $E$ in the all-zero state, followed by the same conjugate basis measurement $\{\Pi_j\}_j$ on $Q_1$.
In fact, there exists an isometry $V_{Q_2\rightarrow Q_2E}$ from $Q_2$ to $Q_2E$ such that 
\begin{equation}\label{eq:iso_dubpqm_cbpqm}
    U_{\mathrm{cBPQM}}\left( \ket{\varphi}_Q\otimes \ket{0^A}_E \right) = V_{Q_2\rightarrow Q_2E}\cdot U_{\mathrm{DUC-BPQM}} \ket{\varphi}_Q
\end{equation}
for all $n$-qubit states $\ket{\varphi}$.
The isometry $V$ coherently computes the discretized distribution information given the joint systems $Q_2$ produced by all the node operations.
We formally prove this below in \Cref{lem:iso_cbpqm_dubpqm}.

Importantly, since $V$ and $\Pi_j$ act on different systems ($Q_2$ and $Q_1$ respectively), we can prove in the following that coherent BPQM and DUC-BPQM realize the identical projective measurement.
More precisely, coherent BPQM maps any state $\ket{\varphi}_Q\otimes\ket{0}_E$ to the post-measurement state
\begin{equation}
    \ket{\zeta_j}\coloneq \frac{1}{\sqrt{p_j}} \Pi_j U_{\mathrm{cBPQM}}\ket{\varphi}_Q\otimes \ket{0}_E \text{ with probability } p_j = \norm{\Pi_jU_{\mathrm{cBPQM}}\ket{\varphi}_Q\otimes\ket{0}_E}^2
    \, .
\end{equation}
Moreover, observe that $\ket{\zeta_j} = V_{Q_2\rightarrow Q_2E}\ket{\zeta_j'}$ where $\ket{\zeta_j'}\coloneq \tfrac{1}{\sqrt{p_j'}}\Pi_jU_{\mathrm{DUC-BPQM}}\ket{\varphi}$ is the post-measurement state of DUC-BPQM with probability $p_j'=\norm{\Pi_jU_{\mathrm{DUC-BPQM}}\ket{\varphi}}$.
Notice that
\begin{equation}
    p_j 
    = \norm{\Pi_j V_{Q_2\rightarrow Q_2E}U_{\mathrm{DUC-BPQM}}\ket{\varphi}}^2
    = \norm{V_{Q_2\rightarrow Q_2E} \Pi_j U_{\mathrm{DUC-BPQM}}\ket{\varphi}}^2
    = \norm{\Pi_j U_{\mathrm{DUC-BPQM}}\ket{\varphi}}^2
    = p_j'
    \, .
\end{equation}
Furthermore, after undoing the coherent BPQM node operations, we obtain the state
\begin{align}
    U_{\mathrm{cBPQM}}^{\dagger}\ket{\zeta_j}
    &= U_{\mathrm{cBPQM}}^{\dagger} V_{Q_2\rightarrow Q_2E} \ket{\zeta_j'} \\
    &= U_{\mathrm{cBPQM}}^{\dagger} V_{Q_2\rightarrow Q_2E} U_{\mathrm{DUC-BPQM}} (U_{\mathrm{DUC-BPQM}}^{\dagger}\ket{\zeta_j'}) \\
    &= U_{\mathrm{cBPQM}}^{\dagger} U_{\mathrm{cBPQM}} (U_{\mathrm{DUC-BPQM}}^{\dagger}\ket{\zeta_j'})\otimes\ket{0}_E \\
    &= (U_{\mathrm{DUC-BPQM}}^{\dagger}\ket{\zeta_j'})\otimes\ket{0}_E
\end{align}
which is identical to the state obtained after realizing DUC-BPQM and undoing $U_{\mathrm{DUC-BPQM}}$.

In summary, in order to study the performance of coherent BPQM in block decoding, we can equivalently study the performance of DUC-BPQM in block decoding (i.e., \Cref{def:ubpqm_block_decoding} where we replace UC-BPQM with DUC-BPQM).
We also briefly note that this argument proves \Cref{lem:coherent_bpqm_projection}.

\begin{lemma}\label{lem:iso_cbpqm_dubpqm}
    Consider an instance $T=\sdt(n,k,\ell,G,P)$ where the associated channel $W$ can be written as $W=\mathcal{W}[\mathcal{G},\mathcal{P}]$ for some MPG $\mathcal{G}$ with $m$ leaves and a sequence of probability distributions $\mathcal{P}=(P_i)_{i=1,\dots,m}$.
    The action of coherent BPQM and DUC-BPQM are related by \Cref{eq:iso_dubpqm_cbpqm} for some isometry $V_{Q_2\rightarrow Q_2E}$.
\end{lemma}
\begin{proof}
    Comparing the action of the coherent BPQM node operation and DUC-BPQM node operation, we observe that they are identical except that coherent BPQM produces additional ancilla registers containing ancilla information.
    More concretely, let us write the action of DUC-BPQM on some $n$-qubit state $\ket{\varphi}$ as follows
    \begin{equation}
        U_{\mathrm{DUC-BPQM}}\ket{\varphi}
        =
        \sum_{s} \ket{f_s}_{Q_1Q_3} \otimes \ket{s}_{Q_2}
        \, ,
    \end{equation}
    where $Q_1$ is the final message, $Q_2$ is the collection of all $S$ registers produced by all node operations, and $Q_3$ is the collection of all $A$ registers  produced by all the node operations.
    Here, $\ket{f_s}$ is a subnormalized quantum state.
    A simple (but verbose) induction argument shows that coherent BPQM produces the state
    \begin{equation}
        U_{\mathrm{cBPQM}}(\ket{\varphi}_Q\otimes \ket{0}_E)
        =
        \sum_{s} \ket{f_s}_{Q_1Q_3} \otimes \ket{s}_{Q_2} \otimes \ket{q_s}_{E}\,,
    \end{equation}
    where $E$ contains the distribution registers for all the edges in the graph.
    Moreover, $\ket{q_s}$ depends only on $s$.
    The desired isometry is therefore given by $V_{Q_2\rightarrow Q_2E}\coloneq \sum_s \proj{s}_{Q_2}\otimes \ket{q_s}_E$.
\end{proof}

\subsection{Characterizing the discretization error in the classical messages}
In this subsection, we characterize the impact of discretization errors on the classical part of the messages in discretized incoherent BPQM.
Let us consider some instance $\sdt(n,k,\ell,G,P)$ and an appropriate corresponding MPG $\mathcal{G}$.
Let us denote the root of $\mathcal{G}$ by $e^{\star}$.

In the absence of discretization errors, we know from \Cref{lem:incoherent_bpqm_intermediate_state} that the classical part of the final message over $e^{\star}$ is given by $D_m$ with probability $p_m$ where $(p_m,D_m)_m=\mensemble_{\mathcal{G}}^{\mathcal{P}}$.
Here, the index $m\in\ff_2^{k-\ell}$ captures the measurement outcomes of the $Q_2$ systems of all node operations.

If we run \emph{discretized} incoherent BPQM, the classical part of the message over the final edge is still fully determined by $m$, but now takes different values due to discretization errors.
We denote the possible values by $(\tilde{D}_m)_m$.
Naively, one might expect that $\tilde{D}_m$ is a very good approximation of $D_m$, provided that $B$ is chosen large enough.
Unfortunately this is is not true for all $m$.
However, it remains possible to make an average-case statement, which will end up being sufficient for our purposes.
The precise statement is formalized as follows.
\begin{lemma}\label{lem:discretization_distributions}
    Consider an instance $\sdt(n,k,\ell,G,P)$ where the associated channel $W$ can be written as $W=\mathcal{W}[\mathcal{G},\mathcal{P}]$ for some MPG $\mathcal{G}=(V,E)$ and a sequence of probability distributions $\mathcal{P}=(P_i)_{i=1,\dots,m}$.
    Denote the associated message ensemble $\mensemble_{\mathcal{G}}^{\mathcal{P}}=(p_m,D_m)_{m\in\ff_2^{k-\ell}}$.
    Consider the discretized incoherent BPQM algorithm with $B\in\mathbb{N}$ discretization bits, and denote the possible values of the classical part of the final message by $(\tilde{D}_m)_{m\in\ff_2^{k-\ell}}$.
    For $\Delta\coloneqq 2^{N(\mathcal{G}) - B}$, it holds that
    \begin{equation}
        \sum_m p_m \norm{D_m - \tilde D_m}_1 \leq (2N(\mathcal{G}) + 3)^{\abs{V}}\Delta \,. 
    \end{equation}
    \end{lemma}
The proof of this lemma proceeds by induction, by considering how the discretization errors accumulate at every node operation throughout the MPG.
This accumulated error at each node is captured by the following result.
\begin{lemma}\label{lem:error_classical_update}
    Consider a primitive node operation $U=\pno(n,k,\ell,G,D)$ and an $\ell$-bit probability distribution $\tilde D$.
    Let us write $f_U(D)=(p_s,D_s')_s$ and $f_U(\tilde D)=(\tilde p_s,\tilde D_s')_s$.
    Then, 
    \begin{equation}
        \sum_s p_s \norm{ D_s' - \tilde D_s' }_1 \leq 2\norm{D-\tilde D}_1
        \, .
    \end{equation}
\end{lemma}
\begin{proof}
    Let us denote $P_{Y,S,A}(s,y,a)\coloneqq D(M^{-1}(s\cat y\cat a))$ and $\tilde P_{Y,S,A}(s,y,a)\coloneqq \tilde D(M^{-1}(s\cat y\cat a))$ where $M$ is the matrix associated to the primitive node unitary.
    Then,
    \begin{align}
        \sum_s p_s \norm{ D_s' - \tilde D_s' }_1 
        &= \sum_s P_S(s) \norm{ P_{Y|S=s} - \tilde P_{Y|S=s} }_1 \\
        &\leq \norm{P_S - \tilde P_S}_1 + \sum_s \norm{P_S(s)P_{Y|S=s} - \tilde P_S(s)\tilde P_{Y|S=s}  }_1 \\
        &= \norm{P_S - \tilde P_S}_1 + \norm{ P_{S,Y} - \tilde P_{S,Y} }_1 \\
        &\leq \norm{P_{Y,S,A} - \tilde P_{Y,S,A} }_1 + \norm{ P_{Y,S,A} - \tilde P_{Y,S,A} }_1 \\
        &= 2\norm{ D - \tilde D }_1\,.
    \end{align}
    In the last step, we used that $P$ and $D$, respectively $\tilde P$ and $\tilde D$, only differ by some relabeling.
\end{proof}
\begin{proof}[Proof of \Cref{lem:discretization_distributions}]
    We prove our statement inductively over the size of the MPG.
    If the MPG has no nodes, then there is simply one element in the ensemble $\mensemble_{\mathcal{G}}^{\mathcal{P}} = (1,D)$ and the discretization error $\norm{D-\tilde{D}}_1$ is at most $\Delta$ by definition.

    If the MPG has at least one node, consider the node $v$ connected to the root edge.
    For $i=1,\dots,m_v$, denote the message ensembles associated to the incoming messages by $\mathcal{E}^{(i)}=(p_{s_i}^{(i)},D_{s_i}^{(i)})_{s_i}$ (i.e. $\mathcal{E}^{(i)}$ is the message ensemble of the sub-MPG for which $e_i^{\mathrm{out}}(v)$ is the root), and similarly, denote the associated distributions in incoherent BPQM by $\tilde D_{s_i}^{(i)}$.
    The joint message ensemble is $\otimes_{i=1}^{m_v}\mathcal{E}^{(i)}=(p_s,\bar D_s)$.
    Let us denote the distribution ensemble $f_U(D_s)=(p_{t|s},D_{s,t}')_t$ where $U$ is the primitive node unitary associated to $v$.
    Therefore, the output of non-discretized incoherent BPQM has classical part $D_{s,t}'$ with probability $p_sp_{t|s}$.

    Now consider \emph{discretized} incoherent BPQM.
    The incoming message over the $i$-th edge is given by $\tilde D_{s_i}$ with some probability $\tilde p_{s_i}$.
    From the induction hypothesis, we know that the joint message $\bar{\tilde{D}}_s\coloneqq \times_{i}\tilde{D}_{s_i}$ fulfills
    \begin{align}
        \sum_s p_s\norm{\bar D_s - \bar{\tilde{D}}_s}_1
        &\leq \sum_{i=1}^{m_v} \sum_{s_i} p_{s_i}\norm{D_{s_i} - \tilde{D}_{s_i}}_1 \\
        &\leq \sum_{i=1}^{m_v} (2N(\mathcal{G})+3)^{\abs{V}-1}\Delta \\
        &\leq N(\mathcal{G}) (2N(\mathcal{G})+3)^{\abs{V}-1}\Delta
        \, .
    \end{align}
    In the incoherent BPQM algorithm, $\bar{\tilde{D}}_s$ is computed and stored in its discretized form, which we denote $\bar{P}_s\coloneqq \mathfrak{r}_{B,n_v}(\tilde{D}_s)$ to mirror the notation in Step 1 from \Cref{def:discretized_nodeop}.
    This gives us
    \begin{equation}
        \sum_s p_s\norm{\bar D_s - \bar{P}_s}_1 \leq N(\mathcal{G}) (2N(\mathcal{G})+3)^{\abs{V}-1}\Delta + \Delta \, .
    \end{equation}
    Let us denote $f_U(\bar{P}_s) = (\tilde{p}_{t|s},P_{s,t}',)_t$.
    Next, we can apply \Cref{lem:error_classical_update} to obtain
    \begin{equation}
        \sum_{s,t} p_sp_{t|s} \norm{D_{s,t}' - P_{s,t}'}_1
        \leq 2\sum_{s} p_s \norm{\bar D_s - \bar P_s}_1
        \leq 2N(\mathcal{G})(2N(\mathcal{G})+3)^{\abs{V}-1}\Delta + 2\Delta
        \, .
    \end{equation}
    Finally, the resulting distribution $P_{s,t}'$ is again computed and stored in discretized fashion, so the output distribution of discretized incoherent BPQM is $P''_{s,t}\coloneqq \mathfrak{r}_{B,\ell_v}(P'_{s,t})$ (see Step 4 in \Cref{def:discretized_nodeop}).
    We hence obtain
    \begin{align}
        \sum_{s,t} p_sp_{t|s} \norm{D_{s,t}' - P_{s,t}''}_1
        &\leq \Delta + \sum_{s,t} p_sp_{t|s} \norm{D_{s,t}' - P_{s,t}'}_1 \\
        &\leq 2N(\mathcal{G}) (2N(\mathcal{G})+3)^{\abs{V}-1}\Delta + 3\Delta \\
        &\leq 2N(\mathcal{G}) (2N(\mathcal{G})+3)^{\abs{V}-1}\Delta + 3(2N(\mathcal{G})+3)^{\abs{V}-1}\Delta \\
        &= (2N(\mathcal{G})+3)^{\abs{V}}\,, 
    \end{align}
    which proves the induction step.
\end{proof}

\subsection{Characterizing the fidelity of DUC-BPQM}
The following lemma captures how imperfect knowledge of the incoming state impacts the action of the primitive node unitary.
\begin{lemma}\label{lem:error_wrong_distr}
    Consider a full-rank matrix $G\in\ff_2^{k\times n}$, $1\leq l\leq k\leq n$, and two probability distributions $D$ and $\tilde D$ over  $n$ bits. 
    Denote the two associated primitive node operations by $U=\pno(n,k,\ell,G,D)$ and $\tilde U=\pno(n,k,\ell,G,\tilde D)$.
    Next, consider some $n$-qubit state $\ket{\varphi_{x,r}}\coloneqq\spsc[D]((x\cat r)^TG)$ for some $x\in\ff_2^{\ell},r\in\ff_2^{k-\ell}$.
    Then, 
    \begin{equation}
      \braopket{\varphi_{x,r}}{U^{\dagger} \tilde U}{\varphi_{x,r}}
      \geq 1 - \norm{D - \tilde D}_1
      \, .
    \end{equation}
\end{lemma}
As a direct consequence, the trace distance between the output of the two node operations is bounded by
\begin{equation}
    \delta( U\ket{\varphi_{x,r}}, \tilde{U}\ket{\varphi_{x,r}} ) \leq \sqrt{2 \norm{D - \tilde D}_1}
\end{equation}
due to the Fuchs-van de Graaf inequality.
\begin{proof}
Following the notation in \Cref{def:pno}, consider the $U_1$, $U_2$, $P_{Y,S,A}$, $V(y,s)$, and $\ket{\xi_{y,s}}$ induced by $D$ as well as the $\tilde U_1$, $\tilde U_2$, $\tilde P_{Y,S,A}$, $\tilde V(y,s)$, $\ket{\tilde{\xi}_{y,s}}$  induced by $\tilde D$.
Notice that $U_1=\tilde U_1$, so only $U_2$ and $\tilde U_2$ differ.
Recall from the proof of \Cref{lem:node_operation} that
\begin{equation}
    U_2U_1\ket{\varphi_{x,r}} = \sum_{y,s}\sqrt{P_{Y,S}(y,s)}Z^x\ket{y}\otimes Z^r\ket{s}\otimes\ket{0}\,. 
\end{equation}
Similarly, we have
\begin{equation}
    \tilde U_2\tilde U_1\ket{\varphi_{x,r}} = \sum_{y,s}\sqrt{P_{Y,S}(y,s)} Z^x\ket{y}\otimes Z^r\ket{s}\otimes \left(\tilde V(y,s)\ket{\xi_{y,s}}\right) \, .
\end{equation}
The overlap between these states is therefore
\begin{align}
    \braopket{\varphi_{x,r}}{U^{\dagger}\tilde U}{\varphi_{x,r}}
    &= \sum_{y,s} P_{Y,S}(y,s) \braopket{0}{\tilde V(y,s)}{\xi_{y,s}} \\
    &= \sum_{y,s} P_{Y,S}(y,s) \braopket{\tilde \xi_{y,s}}{\tilde V(y,s)^{\dagger}\tilde V(y,s)}{\xi_{y,s}} \\
    &= \sum_{y,s} P_{Y,S}(y,s) \braket{\tilde \xi_{y,s}}{\xi_{y,s}} \\
    &= \sum_{y,s} P_{Y,S}(y,s) F(P_{A|YS}, \tilde P_{A|YS}) \\
    &\geq \sum_{y,s} P_{Y,S}(y,s) (1-\tfrac{1}{2}\norm{P_{A|YS}-\tilde P_{A|YS}}_1) \\
    & = 1 - \tfrac{1}{2}\sum_{y,s} P_{Y,S}(y,s) \norm{P_{A|YS} - \tilde P_{A|YS}}_1 \\
    & \geq 1 - \tfrac{1}{2}\left( \norm{P_{Y,S}-\tilde P_{Y,S}}_1 + \norm{P_{Y,S,A} - \tilde P_{Y,S,A}}_1 \right) \\
    & \geq 1 - \norm{P_{Y,S,A} - \tilde P_{Y,S,A}}_1 \\
    &= 1 - \norm{D -\tilde D}_1\,, 
\end{align}
where $F$ denotes the quantum fidelity (respectively the Bhattacharyya coefficient) and we used Fuchs-van de Graaf.
In the last step, we also used that $P_{Y,S,A}$ and $D$, respectively $\tilde P_{Y,S,A}$ and $\tilde D$, only differ by some relabeling.
\end{proof}
By iteratively applying \Cref{lem:error_wrong_distr}, we can now characterize the how close the output of DUC-BPQM approximates the output of UC-BPQM.
\begin{lemma}\label{lem:dubpqm_state}
    Consider an instance $\sdt(n,k,\ell,G,P)$ where the associated channel $W$ can be written as $W=\mathcal{W}[\mathcal{G},\mathcal{P}]$ for some MPG $\mathcal{G}=(V,E)$ and a sequence of probability distributions $\mathcal{P}=(P_i)_{i=1,\dots,m}$.
    Let us additionally fix a number of discretization bits $B\in\mathbb{N}$.
    Denote by $U_{\mathrm{UC-BPQM}}$ and $U_{\mathrm{DUC-BPQM}}$ the unitaries corresponding to the UC-BPQM and DUC-BPQM algorithms without the final measurement.
    For any $c\in\mathcal{C}$, where $\mathcal{C}$ is the code induced by $G$, the actions on the state $\ket{\varphi_{c}}\coloneqq\spsc[P](c)$ only differ by a trace distance error of
    \begin{equation}
        \delta(U_{\mathrm{UC-BPQM}}\ket{\varphi_{c}}, U_{\mathrm{DUC-BPQM}}\ket{\varphi_{c}}) \leq \abs{V}\beta\,, 
    \end{equation}
    where $\beta\coloneqq \sqrt{2\left((2N(\mathcal{G})+3)^{\abs{V}}+\pi + 1\right)\Delta}$ and $\Delta\coloneqq 2^{N(\mathcal{G})-B}$.
\end{lemma}
\begin{proof}
    Consider the channel $F[\mathcal{G}]$ described by the MPG.
    Recall that at every node $v$, the input string is concatenated with a random string $r_v\in\ff_2^{k_v-\ell_v}$ and then encoded according to $G_v$.
    Denote the concatenation of all the random strings by $\bar{r}\coloneqq \cat_{v\in V}r_v$, which is a $(k-\ell)$-bit string.
    For any particular value $\bar{r}$, the encoding map described by the MPG is deterministic.
    Put differently, $\mathcal{G}$ induces a one-to-one mapping between pairs $(x,\bar{r})\in\ff_2^{\ell\times k-\ell}$ and codewords $c\in\mathcal{C}$.
    From now on, let us denote by $x$ and $\bar r$ the two bit strings associated to the codeword $c$ in the lemma.

    We prove the desired statement inductively over the size $\abs{V}$ of the MPG.
    If the MPG is trivial (i.e. it has zero nodes), then both $U_{\mathrm{UC-BPQM}}$ and $U_{\mathrm{DUC-BPQM}}$ are simply the identity operation.
    It remains to show the induction step.

    So let us assume that the MPG is nontrivial and consider $v$ to be the node connected to the root edge.
    We can split $\bar r$ into $\bar r = r_1\cat r_2$ for $r_1\in\ff_2^{\ell_v-k_v}$ and $r_2\in\ff_2^{\ell-v-(\ell_v-k_v)}$ where $r_1$ corresponds to the random bit string that is associated to the node $v$, and $r_2$ to the concatenation of all the other random bit strings.

    Thanks to \Cref{lem:ubpqm_action}, we know that in the UC-BPQM algorithm applied to $\ket{\varphi_{x,r}}$, the state of the $n$-qubit system going into the last node operation is given by
    \begin{equation}
        \ket{\phi}\coloneqq
        \sum_{s\in\ff_2^{k-\ell-(k_v-\ell_v)}}
        \sqrt{p_s}(-1)^{r_2\cdot s}
        \spsc[D_s]((x\cat r_1)G_v)_{Q_1} \otimes \ket{s}_{Q_2} \otimes \ket{0^{n-k-(n_v-k_v)}}_{Q_3} \\
    \end{equation}
    where $(p_s,D_s)_s$ is the tensor product of all the message ensembles of the sub-MPGs where the edge $e_i^{\mathrm{out}}$ is the root edge, for $i=1,\dots,m_v$.
    Similarly, in the DUC-BPQM algorithm, the state before the final node operation is given by some $\ket{\tilde{\phi}}$ which fulfills 
    \begin{equation}
        \delta(\ket{\phi},\ket{\tilde{\phi}}) \leq (\abs{V}-1) \beta
    \end{equation}
    due to the induction hypothesis (because $\ket{\tilde\phi}$ is the tensor product of the DUC-BPQM output on smaller MPGs).

    The action of the final UC-BPQM node operation is given by the unitary
    \begin{equation}
        V = \sum_{s\in\ff_2^{k-\ell-(k_v-\ell_v)}} \pno(n_v,k_v,\ell_v,G_v,D_s)_{Q_1} \otimes \proj{s}_{Q_2} \otimes \id_{Q_3}
    \end{equation}
    and similarly, the action of the final DUC-BPQM node operation is given by
    \begin{equation}
        \tilde V = \sum_{s\in\ff_2^{k-\ell-(k_v-\ell_v)}} \pno(n_v,k_v,\ell_v,G_v,\tilde D_s)_{Q_1} \otimes \proj{s}_{Q_2} \otimes \id_{Q_3}
    \end{equation}
    for some $\tilde D_s$ fulfilling
    \begin{equation}
        \sum_s p_s \norm{D_s - \tilde D_s}_1 \leq (2N(\mathcal{G})+3)^{\abs{V}}\Delta + (\pi+1)\Delta = \tfrac{1}{2}\beta^2
    \end{equation}
    because of \Cref{lem:discretization_distributions,lem:discretized_pno}.

    By the triangle inequality, we have
    \begin{equation}
        \delta(V\ket{\phi}, \tilde V\ket{\tilde\phi})
        \leq
        \delta(\ket{\phi}, \ket{\tilde\phi})
        +
        \delta(V\ket{\phi}, \tilde V\ket{\phi})\,, 
    \end{equation}
    so it remains to characterize the trace distance between $V\ket{\phi}$ and $\tilde V\ket{\phi}$.
    We do this by considering the fidelity:
    \begin{align}
        \braopket{\phi}{V^{\dagger}\tilde V}{\phi}
        &= \sum_s p_s \braopket{\alpha_{s,x,r_1}}{\pno(n_v,k_v,\ell_v,G_v,D_s)^{\dagger}\pno(n_v,k_v,\ell_v,G_v,\tilde D_s)}{\alpha_{s,x,r_1}} \\
        &\geq 1 - \sum_s p_s \norm{D_s - \tilde D_s}
        \, ,
    \end{align}
    where $\ket{\alpha_{s,x,r_1}}\coloneqq\spsc[D_s]((x\cat r_1)^TG_v)$ and we applied \Cref{lem:error_wrong_distr}.
    From the Fuchs-van de Graaf inequality, we get
    \begin{equation}
        \delta(V\ket{\phi}, \tilde V\ket{\phi})
        \leq \beta\,, 
    \end{equation}
    which proves the induction step.
\end{proof}

\subsection{Putting it all together}
\begin{proof}[Proof of Theorem \ref{thm:bit_optimal_discretized}]
    As discussed in the beginning of this section, the success probability of discretized incoherent BPQM and DUC-BPQM in solving the subspace decoding task are identical.
    We know that UC-BPQM optimally solves the subspace decoding task, so it remains to show that DUC-BPQM and UC-BPQM output the same result with high probability.

    The difference in probability of obtaining the correct result with UC-BPQM and DUC-BPQM is upper bounded by the trace distance
    \begin{align}
        & \delta( U_{\mathrm{UC-BPQM}}W(x)U_{\mathrm{UC-BPQM}}^{\dagger}, U_{\mathrm{DUC-BPQM}}W(x)U_{\mathrm{DUC-BPQM}}^{\dagger} ) \\
        &\leq \frac{1}{2^{k-\ell}} \sum_r \delta( U_{\mathrm{UC-BPQM}}\proj{\Psi_{x,r}}U_{\mathrm{UC-BPQM}}^{\dagger}, U_{\mathrm{DUC-BPQM}}\proj{\Psi_{x,r}}U_{\mathrm{DUC-BPQM}}^{\dagger} ) \\
        &\leq \abs{V}\beta\,, 
    \end{align}
    where $\ket{\Psi_{x,r}}\coloneqq\spsc[D]((x\cat r)^TG)$ and we applied \Cref{lem:dubpqm_state}.
    Hence, it suffices to choose $B=\mathcal{O}(N(\mathcal{G}) + \log(1/\epsilon) + \abs{V}\log N(\mathcal{G}))$.
    By the discussion in \Cref{app:discretized_incoherent_bpqm}, we know that the total circuit complexity of discretized incoherent BPQM is $\mathcal{O}(\mathrm{poly}(B)4^{N(\mathcal{G})}N(\mathcal{G})^2\abs{V})$.
\end{proof}
\begin{proof}[Proof of Theorem \ref{thm:block_optimal}]
    In the beginning of this section, we argued that DUC-BPQM realizes the identical projection as coherent BPQM.
    Furthermore, we know from \Cref{cor:ubpqm_block_optimal} that the projection from UC-BPQM realizes block optimal decoding.
    It hence suffices to show that sequential DUC-BPQM projections are likely to produce the same measurement result as UC-BPQM.

    Following the notation in \Cref{sec:block_optimality}, let us denote by $\ket{\Psi_x}$ for $x\in\mathcal{C}$ the states that we want to distinguish, and let $U_{\mathrm{DUC-BPQM},i}$ and $U_{\mathrm{UC-BPQM},i}$ be the unitaries describing the node operations of UC-BPQM and DUC-BPQM for decoding the $i$-th codeword bit for $i=1,\dots k$.
    Our first goal will be to expand the result in \Cref{lem:dubpqm_state} to show that the two unitaries $U_{\mathrm{DUC-BPQM},i}$ and $U_{\mathrm{UC-BPQM},i}$ do not only have a similar action on the states $\ket{\Psi_x}$, but in fact, they act similarly on the entire subspace $V$.

    For this purpose, let us consider some vector (possibly unnormalized) $\ket{\phi}\in V$, which we can expand as $\ket{\phi}=\sum_x \alpha_x\ket{\Psi_x}$ for some expansion coefficients $\alpha_x\in\mathbb{C}$.
    For any $i=1,\dots,k$, we obtain
    \begin{equation}
        \norm{\left( U_{\mathrm{DUC-BPQM},i} - U_{\mathrm{UC-BPQM},i}\right) \ket{\phi} }
        \leq \sum_x \abs{\alpha_x} \norm{\left( U_{\mathrm{DUC-BPQM},i} - U_{\mathrm{UC-BPQM},i}\right) \ket{\Psi_x} }\,,
        \leq \sqrt{2}M\beta \sum_x \abs{\alpha_x}
    \end{equation}
    where we used \Cref{lem:dubpqm_state}, the fact that $\norm{\ket{f}-\ket{g}}\leq \sqrt{2}\delta(\ket{f},\ket{g})$ for any two vectors $\ket{f},\ket{g}$ with only real entries in the computationl basis, and we defined $\beta\coloneq \sqrt{2\left((2N+3)^{M}+\pi + 1\right)2^{N-B}}$.
    
    Next, let us define the $2^k\times 2^k$ Gram-matrix $G$ with entries $G_{x,y}\coloneq \braket{\Psi_x}{\Psi_y}$ for $x,y\in\mathcal{C}$.\footnote{As a slight abuse of notation, we index the rows and columns with elements of $\mathcal{C}$.}
    The matrix $G$ has the identical spectrum as $2^k\rho$ on $V$.\footnote{Consider a collection of linearly independent states $\ket{f_1},\dots,\ket{f_n}$ and $T\coloneq \sum_j\ketbra{f_j}{j}$.
    Then, the matrices $T^{\dagger}T=\sum_{j,k}\braket{f_j}{f_k}\ketbra{j}{k}$ and $TT^{\dagger}=\sum_j\proj{f_j}$ have the same eigenvalues.}
    We can write
    \begin{equation}
        \norm{ \ket{\phi} }^2 \\
        = \sum_{x,y} \bar\alpha_x\alpha_y \braket{\Psi_x}{\Psi_y}
        = \vec{\alpha}^{\dagger} G\vec{\alpha}
        \geq \lambda_{\mathrm{min}}(G)\norm{\vec{\alpha}}_2^2\,, 
    \end{equation}
    where $\vec{\alpha}$ is the column vector with the entries given by $\alpha_x, x\in\mathcal{C}$ and $\lambda_{\mathrm{min}}$ is the smallest non-zero eigenvalue.
    As such, we obtain
    \begin{align}
        \norm{\left( U_{\mathrm{DUC-BPQM},i} - U_{\mathrm{UC-BPQM},i}\right) \ket{\phi} }
        &\leq \sqrt{2}M\beta\norm{\vec{\alpha}}_1 \\
        &\leq \sqrt{2}M\beta\sqrt{2^k}\norm{\vec{\alpha}}_2 \\
        &\leq \sqrt{2}M\beta\sqrt{\frac{2^k}{\lambda_{\mathrm{min}}(G)}}\norm{\ket{\phi}} \\
        &= \sqrt{2}M\beta\frac{1}{\sqrt{\lambda_{\mathrm{min}}(\rho)}}\norm{\ket{\phi}}
        \, .
    \end{align}
    It remains to find a lower bound for the minimal non-zero eigenvalue of $\rho$.
    For this, we use $U=\pno(n,k,k,G,P)$ to diagonalize $\rho$ using \Cref{lem:node_operation}.
    \begin{align}
        U\rho U^{\dagger}
        &= \sum_{x\in\ff_2^k} \spsc[P_Y](x) \otimes \proj{0}^{\otimes (n-k)} \\
        &= \sum_{y\in\ff_2^k} P_Y(y) \proj{y} \otimes \proj{0}^{\otimes (n-k)}
        \, .
    \end{align}
    Hence $\lambda_{\min}(\rho)=\min_{y\in \ff_2^{k}}P_Y(y)$. 
    Recall that $P_{Y,S,A}(y,s,a)=P(M^{-1}(y\cat s\cat a))$ where $P=P_1\times \dots\times P_n$ where $P_i$ is a binary symmetric channel with error probability $p_i$.
    Since $P_Y(y)\geq \min_{s,a}P_{Y,S,A}(y,s,a)$, we thus have $P_Y(y)\geq \min_{z\in\ff_2^n}P(z)$.
    Since the smallest $P(z)$ is at least $\bar{p}^n$, where $\bar p\coloneqq\min_{i=1,\dots,n}p_i$, we have $P_Y(y)\geq \bar{p}^n$.

    In summary, we have
    \begin{equation}
        \norm{\left( U_{\mathrm{DUC-BPQM},i} - U_{\mathrm{UC-BPQM},i}\right) \ket{\phi} }
        \leq \sqrt{2}M\beta \bar p^{-n/2} \norm{\ket{\phi}}
        \, .
    \end{equation}

    Let us now turn our attention to comparing the block decoding performance of UC-BPQM and DUC-BPQM.
    The probability that UC-BPQM outputs the estimate $\hat{x}$ of $x$ is given by the norm of the (unnormalized) statevector
    \begin{equation}
        M_{\hat{x}_k}^{(k)} M_{\hat{x}_{k-1}}^{(k-1)} \dots M_{\hat{x}_1}^{(1)} \ket{\Psi_x}
    \end{equation}
    where the projetive measurement $\{M_0^{(i)}, M_1^{(i)}\}$ is defined by
    \begin{equation}
        M_j^{(i)} = U_{\mathrm{UC-BPQM},i}^{\dagger} \Pi_j U_{\mathrm{UC-BPQM},i}\,, 
    \end{equation}
    and $\{\Pi_0,\Pi_1\}$ is the projective measurement of the final qubit in the conjugate basis.
    Similarly, the probability that coherent BPQM outputs the estimate $\hat{x}$ of $x$ is given by the norm of
    \begin{equation}
        \tilde M_{\hat{x}_k}^{(k)} \tilde M_{\hat{x}_{k-1}}^{(k-1)} \dots \tilde M_{\hat{x}_1}^{(1)} \ket{\Psi_x}
    \end{equation}
    where the projective measurement $\{\tilde M_0^{(i)}, \tilde M_0^{(1)}\}$ is defined by
    \begin{equation}
        \tilde M_j^{(i)} = U_{\mathrm{DUC-BPQM},i}^{\dagger} \Pi_j U_{\mathrm{DUC-BPQM},i}
        \, .
    \end{equation}

    Observe that for UC-BPQM, every intermediate state
    \begin{equation}
        M_{\hat{x}_i}^{(i-1)} \dots M_{\hat{x}_1}^{(1)} \ket{\Psi_x}
    \end{equation}
    for $i=1,\dots,k$ is also an element of $V$, because of \Cref{prop:bitoptimal_implies_blockoptimal}.
    The same might not necessarily be true for DUC-BPQM.

    Observe that for any (possibly unnormalized) vector $\ket{\phi}\in V$, we have
    \begin{align}
        \norm{ M_j^{(i)}\ket{\phi} - \tilde M_j^{(i)}\ket{\phi} }
        &\leq
        \norm{ (U_{\mathrm{DUC-BPQM},i}^{\dagger} - U_{\mathrm{UC-BPQM},i}^{\dagger} ) \Pi_jU_{\mathrm{UC-BPQM},i}\ket{\phi} } \nonumber\\
        &\quad + \norm{ U_{\mathrm{DUC-BPQM},i}\ket{\phi} - U_{\mathrm{UC-BPQM},i}\ket{\phi} } 
        \, .
    \end{align}
    For the first term, observe that $\Pi_jU_{\mathrm{UC-BPQM},i}\ket{\phi} = U_{\mathrm{UC-BPQM},i}\ket{\varphi}$ for some $\ket{\varphi}\in V$ with $\norm{\ket{\varphi}}\leq\norm{\ket{\phi}}$, because we know that $M_{j}^{(i)}\ket{\phi}\in V$.
    Hence, we get
    \begin{equation}
        \norm{ (U_{\mathrm{DUC-BPQM},i}^{\dagger} - U_{\mathrm{UC-BPQM},i}^{\dagger} ) \Pi_jU_{\mathrm{UC-BPQM},i}\ket{\phi} } 
        \leq \norm{ U_{\mathrm{DUC-BPQM},i}\ket{\varphi} - U_{\mathrm{UC-BPQM},i}\ket{\varphi} }
        \, .
    \end{equation}
    In summary, we have 
    \begin{equation}
        \norm{ M_j^{(i)}\ket{\phi} - \tilde M_j^{(i)}\ket{\phi} }
        \leq
        2\sqrt{2}M\beta\bar p^{-n/2}\, \norm{\ket{\phi}}
        \, .
    \end{equation}    
    Chaining together this result $k$ times using the triangle inequality, we see that the probability of obtaining the outcome $\hat{x}$ with UC-BPQM and DUC-BPQM differs by
    \begin{align}
        & \abs{ \norm{M_{\hat{x}_k}^{(k)} M_{\hat{x}_{k-1}}^{(k-1)} \dots M_{\hat{x}_1}^{(1)} \ket{\Psi_x}} - \norm{\tilde M_{\hat{x}_k}^{(k)} \tilde M_{\hat{x}_{k-1}}^{(k-1)} \dots \tilde M_{\hat{x}_1}^{(1)} \ket{\Psi_x}} } \nonumber\\
        &\quad \leq \norm{ M_{\hat{x}_k}^{(k)} M_{\hat{x}_{k-1}}^{(k-1)} \dots M_{\hat{x}_1}^{(1)} \ket{\Psi_x} - \tilde M_{\hat{x}_k}^{(k)} \tilde M_{\hat{x}_{k-1}}^{(k-1)} \dots \tilde M_{\hat{x}_1}^{(1)} \ket{\Psi_x} } \\
        &\quad \leq k2\sqrt{2}M\beta\bar p^{-n/2}
        \, .
    \end{align}
    Therefore, it suffices to choose $B=\mathcal{O}(n\log(\frac{1}{\bar p}) + \log(1/\epsilon) + N + M\log(N) + \log(k))$.
    By the discussion in \Cref{app:discretized_incoherent_bpqm}, we know that the circuit complexity of discretized incoherent BPQM is $\mathcal{O}(\mathrm{poly}(B)4^{N}N^2M)$, which is repeated $k$ times.
\end{proof}

\printbibliography[heading=bibintoc,title={\large References}]

\end{document}